\documentclass[fleqn,final,authoryear,3p,times,twocolumn]{elsarticle}

\usepackage{amsmath, amssymb, amsfonts}
\usepackage{color, multirow, multicol, array, bm}

\usepackage{graphicx}
\usepackage{natbib}
\usepackage{charter}
\usepackage[english]{babel}

\usepackage{ulem}

\pdfoutput=1        

\newcommand{\flux}{\ensuremath{{\bf F}}}
\newcommand{\mdot}{\ensuremath{\dot M}}
\newcommand{\ldot}{\ensuremath{\dot L}}
\newcommand{\edot}{\ensuremath{\dot E}}
\newcommand{\msun}{\ensuremath{M_\odot}}
\newcommand{\rsun}{\ensuremath{R_\odot}}
\newcommand{\comment}[1]{} 

\journal{New Astronomy}    

\begin{document}

\begin{frontmatter}



\title{Properties of Strong and Weak Propellers from MHD Simulations}

\author[1]{M. M. Romanova}
\author[1]{A. A. Blinova}
\author[2]{G. V. Ustyugova}
\author[3]{A. V. Koldoba}
\author[1,4]{R. V. E. Lovelace}

\address[1]{Department of Astronomy, Cornell University, Ithaca, NY 14853, email:romanova@astro.cornell.edu,~ tel:
+1(607)255-6915, ~fax: +1(607)255-3433}
\address[2]{Keldysh Institute of Applied Mathematics RAS,
Miusskaya sq., 4, Moscow, 125047, Russia, email:
ustyugg@rambler.ru}
\address[3]{Moscow Institute of Physics and Technology, Dolgoprudnyy, Moscow region, 141700, Russia, email:
koldoba@rambler.ru}
\address[4]{Also Department of Applied and Eng. Physics}

\begin{abstract}

We investigate the properties of magnetized stars in the propeller
regime using axisymmetric numerical simulations. We were able to
model the propeller regime for stars with realistically large
magnetospheres ($5-7$ stellar radii) and relatively thin accretion
disks, $H/r\approx 0.15$, so that our results could be applied to
different types of magnetized stars, including Classical T Tauri
stars (CTTSs), cataclysmic variables (CVs), and accreting
millisecond pulsars (MSPs).
 A wide
 range of propeller strengths has been
studied, from very strong propellers (where the magnetosphere
rotates much more rapidly than the inner disk and most of the
inner disc matter is redirected into the wind) to very weak
propellers (where the magnetosphere rotates only slightly faster
than the inner disc, and only a small part of the inner disc
matter is redirected into the wind). In both the strong and weak
propellers, matter is accumulated at the inner disc for the
majority of the time, while episodes of accretion onto the star
and ejection into the wind are relatively brief. The efficiency of
the propeller, which characterizes the part of inner disk matter
flowing into the wind, strongly depends on the fastness parameter
$\omega_s$, which is the ratio of the stellar angular velocity to
the inner disc Keplerian velocity: propeller efficiency increases
with $\omega_s$. The properties of the winds are different in
strong and weak propellers. In the strong propellers, matter is
accelerated rapidly above the escape velocity and flows at a
relatively small opening angle of $40-45$ degrees. This matter
leaves the system, forming the large-scale outflows. In the weak
propellers (during episodes of ejection into the wind), matter may
flow faster or slower than the escape velocity and at a large
opening angle of $60-70$ degrees. Most of this matter is expected
to either fall back to the disk or form a magnetic turbulent
corona above the disk. A star-disk system loses energy and angular
momentum.
 A part of the rotational energy of
the star is ejected to the magnetically-dominated (Poynting flux)
jet, which is only present in the strong propellers. The other
part of the energy flows from the inner disk into a
propeller-driven wind. A star spins down partly due to the flow of
angular momentum from the star to the corona (in weak propellers)
or to the Poynting flux jet (in strong propellers) along the open
field lines, and partly due to the flow of angular momentum to the
inner disk along the closed field lines.

\end{abstract}

\begin{keyword}
accretion; accretion disks; MHD; stars: neutron; stars: magnetic;
magnetohydrodynamics
\end{keyword}

\end{frontmatter}


\section{Introduction}
\label{sec:introduction}

Different magnetized stars are expected to be in the propeller
regime if the magnetosphere rotates more rapidly than the inner
disk (e.g., \citealt{IllarionovSunyaev1975,LovelaceEtAl1999}).
This regime
is expected, e.g., when the accretion rate decreases and the
magnetosphere expands. If the inner disk matter penetrates through
the magnetosphere, then it
acquires angular momentum and can be ejected from the
disk-magnetosphere boundary in the form of a wind. Signs of the
propeller regime have been observed in Classical T Tauri stars
(CTTSs) (e.g.,
\citealt{DonatiEtAl2010,GrininEtAl2015,CodyEtAl2017}), in
cataclysmic variable AE Aqr (e.g.,
\citealt{Mauche2006,WynnEtAl1997}), and in a few accreting
millisecond pulsars (MSPs) at the ends of their outbursts, when
the accretion rate decreases and the disk moves away from the star
(e.g.,
\citealt{VanderklisEtAl2000,PatrunoEtAl2009,PatrunoDangelo2013,BultvanderKlis2014}).
Recently, transitional millisecond pulsars were discovered, where
a millisecond pulsar transits between the state of an accreting
MP, where the accretion disk moves close to the star, and
that of a radiopulsar, where the accretion disk moves to larger
distances from the star (e.g.,
\citealt{PapittoEtAl2013,FerrignoEtAl2014,Linares2014,PatrunoEtAl2014})
\footnote{Transitional MSPs were predicted long ago (e.g.,
\citealt{Bisnovatyi-KoganKomberg1974,AlparEtAl1982}), but were not
discovered until recently.}.
 In these types of stars, the
propeller regime is inevitable. In fact, different observational
properties of transitional MSPs may possibly be connected with the
propeller state, such as the highly variable X-ray radiation
(e.g.,
\citealt{FerrignoEtAl2014,PatrunoEtAl2014,ArchibaldEtAl2015}),
$\gamma-$ray flares  (e.g., \citealt{DeMartinoEtAl2010}) and
radiation in the radio band with a flat spectrum, which indicates
the presence of
outflows or jets (e.g.,
\citealt{BogdanovEtAl2015,DellerEtAll2015}). Many observational
properties of propeller candidate stars were not well-understood,
such as the accretion-induced pulsations observed at very low
accretion rates in some transitional MSPs and the CV AE Aqr.
According to theoretical estimates, at low accretion rates the
inner disk should be far away from the star and accretion should
be blocked by the centrifugal barrier of the propelling star
(e.g.,
\citealt{ArchibaldEtAl2015,PapittoTorres2015,PapittoEtAl2015}).
However, observations show that a small amount of matter accretes
in spite of the centrifugal barrier. This and other issues require
further understanding, so the propeller regime should be studied
in greater detail.



The propeller regime has been studied in a number of theoretical
works and numerical simulations. \citet{IllarionovSunyaev1975} and
\citet{LovelaceEtAl1999} investigated the strong propeller regime
analytically. They suggested that the propelling star ejects all
of the accreting matter into the wind, and no matter accretes onto
the star.

In other
analytical works and 1D numerical simulations it was suggested
that the magnetosphere rotates only slightly faster than the inner
disk, that is, the propeller is relatively weak, and there are no
outflows (e.g.,
\citealt{SunyaevShakura1977,SpruitTaam1993,DangeloSpruit2010,DangeloSpruit2012}).
In their models, the excess angular momentum
is transferred back to the disk, forming a dead disk, and matter
of the inner disk accretes onto the star quasi-periodically due to
the cyclic process of matter accumulation and accretion.



The propeller regime has been studied in a number of axisymmetric
(2.5D) simulations, where a magnetized, rapidly rotating star
interacts with an accretion disk (e.g.,
\citealt{RomanovaEtAl2005,UstyugovaEtAl2006,LiiEtAl2014}).
Simulations have shown that

\begin{enumerate}

\item In the propeller regime, both accretion and outflows are
present. Matter accretes onto the star in cycles.
For the major part of the cycle, matter accumulates in the inner
disk and slowly moves inward. Then, it partially accretes onto the
star and is partially ejected into the wind. Subsequently, the
magnetosphere expands. Therefore, both accretion and outflows
occur in brief episodes (spikes);

\item Accretion onto the star is typically accompanied by
outflow of matter from the inner disk. However, the outflows
may also be present at other times, such as when accretion is
blocked by the centrifugal barrier;

\item In strong propellers, a two-component outflow has been
observed: a relatively slow and dense, conically shaped inner disk
wind, which carries away most of the inner disk matter, and a
low-density, high-velocity collimated jet, which carries away
significant energy and angular momentum;

\item A star
spins down due to the outward flow of angular momentum along the
open and closed field lines.

\end{enumerate}

In earlier models, accretion from laminar $\alpha-$disks
\citep{ShakuraSunyaev1973} had been considered, where the
accretion rate in the disk was regulated
 by the $\alpha-$parameter of viscosity, $\alpha_v$, while the
rate of the field line diffusion through the disk was regulated
by a similar parameter, $\alpha_{\rm diff}$
\citep{RomanovaEtAl2004,RomanovaEtAl2005,UstyugovaEtAl2006,RomanovaEtAl2009}.
More recently, simulations were performed
for turbulent disks \citep{LiiEtAl2014}, where the turbulence is
driven by the magneto-rotational instability (MRI, e.g.,
\citealt{BalbusHawley1991}). Also, in contrast with the earlier
works, simulations were performed in both the top and bottom
hemispheres (no equatorial symmetry). These simulations show
similar
results to those obtained with the $\alpha-$disks. However, the
accretion funnels are not symmetric about the equatorial plane,
and the outflows are
typically one-sided.

In these earlier studies, only stars with relatively small
magnetospheres were modeled ($r_m\lesssim 3 R_\star$, where
$R_\star$ is the stellar radius)
 \footnote{Note that modeling the propeller regime is numerically
challenging and time-consuming, because the magnetic and velocity
gradients can be large compared with the cases of slowly-rotating
stars. It is somewhat easier to model stars with smaller
magnetospheres.}. However,  most of the propeller candidate stars
have larger magnetospheres, so the earlier models could only be
applied to a limited range of stars. This is why we adjusted the
model in such ways as to allow us to model the stars with larger
magnetospheres, $r_m\approx (5-7)R_\star$.

In addition, the earlier numerical simulations were mainly focused
on very strong propellers, where the magnetosphere rotates much
more rapidly than the inner disk,
(e.g.,
\citealt{RomanovaEtAl2005,RomanovaEtAl2009}).
However, propellers of lower strengths have not been
systematically studied. Some of the major questions are: (1) What
are the properties of outflows in propellers of different
strengths? In particular, (2) Which parts of the inner disk matter
flow into the winds? (3) What is the velocity of matter in the
wind? (4) What is the opening angle of the wind? (5) How much
energy flows into the inner disk wind and the Poynting flux jet?
(6) What is the rate of stellar spin-down? (7) How do these
properties depend on the strength of the propeller ?

To answer these questions, we performed a number of axisymmetric
simulations of propellers of different strengths, ranging from
very weak propellers to very strong propellers, and studied the
properties of matter, energy and angular momentum flow. As a base
case, we used a model with a turbulent disk similar to that used
by \citet{LiiEtAl2014}. However, compared with
\citet{LiiEtAl2014}, we (1) took the disk to be a few times
thinner, with an aspect ratio of $H/r\approx 0.15$, which is
closer to realistic (thin) disks; (2) considered magnetospheres of
larger sizes, $r_m\approx (5-7) R_\star$ (compared with
$r_m\approx 3 R_\star$ in \citet{LiiEtAl2014}), so that the model
could be applied to propelling stars with larger magnetospheres;
(3) suggested that the 3D instabilities are efficient at the
disk-magnetosphere boundary and added a diffusivity layer at
$r\leq 7R_\star$, where the diffusivity is high. The diffusivity
is very low (numerical) in the rest of the disk. \footnote{Note
that in \citet{LiiEtAl2014} the diffusivity has been very low
(numerical) in the entire simulation region, excluding a few test
cases where the diffusive layer was added, as in our current
simulations (see Appendix in \citealt{LiiEtAl2014}).} (4)
investigated the properties of propellers of different strengths.
Our simulations show that the properties of strong versus weak
propellers are qualitatively different, and are expected to
provide different observational properties.

The main goal of this new research was to develop a series of
models with parameter values similar to those expected in
propeller candidate stars, such as transitional millisecond
pulsars, intermediate polars, and Classical T Tauri stars. The
results of the simulations are presented in dimensionless form and
can be applied to all types of stars. We also provide convenient
formulae for the conversion of dimensionless values to dimensional
values in application to these
stars. We plan to apply the results of our models to particular
propeller candidate stars in future papers.

The plan of the paper is the following. In Sec.
\ref{sec:Theoretical} we discuss the theoretical background of the
problem. We describe our numerical model in Sec.
\ref{sec:Numerical model} and show the main results of simulations
and analysis in Sec. \ref{sec:Numerical
model}-\ref{sec:intervals-time}. In Sec. \ref{sec:applications} we
provide examples of applications and convenient formulae for
different types of stars. We conclude in Sec.
\ref{sec:Conclusions}. \ref{appen:numerical-model} and
\ref{appen:variation} provide the details of the numerical model
and the variation of different variables with time
for a number of representative models.

\section{Theoretical background}
\label{sec:Theoretical}

For investigation of propellers of different strengths, it is
important to find the main parameters which determine the
strengths of propellers and which determine the main properties of
propellers. In case of slowly-rotating (non-propelling) magnetized
stars, such a parameter is the fastness parameter $\omega_s$
(e.g., \citealt{Ghosh2007,BlinovaEtAl2016}). In the study of the
propeller regime, we also use the fastness parameter as the main
parameter of the problem.

\subsection{Fastness parameter $\omega_s$}

The fastness parameter is determined as the ratio between the
angular velocity of the star $\Omega_\star$ and the angular
velocity of the inner disk at the disk-magnetosphere boundary
$r=r_m$ (e.g., \citealt{Ghosh2007}):
\begin{equation}
\omega_s= \frac{\Omega_\star}{\Omega_K(r_m)}~, \label{eq:fastness}
\end{equation}
where,  $\Omega_K(r_m)$ is the Keplerian angular velocity of the
inner disk at $r=r_m$.



An importance of the fastness parameter can also be shown through
the simplified analysis of forces. In case of a thin (cold)
accretion disk, the matter pressure force is small, and the main
forces acting on the matter of the inner disk are the
gravitational, centrifugal and magnetic forces. In case of strong
propellers ($\omega_s>>1$) the centrifugal force is often the main
force driving matter to the outflows (e.g.,
\citealt{RomanovaEtAl2009,LiiEtAl2014}), so that
the total force acting to the unit mass of the disk is dominated
by the effective gravity:
$$g_{\rm eff}=g+g_c~ , $$
 where $g=-GM_\star/r^2$ and
$g_c={\Omega_\star}^2 r$ are the gravitational and centrifugal
acceleration, respectively. Taking into account that at the inner
disk, $r=r_m$,  $g(r_m)=-GM_\star/r_m^2=-\Omega_K(r_m)^2 r_m$, we
obtain
\begin{equation}
g_{\rm eff}=-\Omega_K^2(r_m) r_m + {\Omega_\star}^2 r_m =
\Omega_K^2(r_m) r_m (\omega_s^2-1)~. \label{eq:g-eff}
\end{equation}
One can see that in stars of the same mass $M_\star$ and the same
magnetospheric radius $r_m$, the main force acting onto the inner
disk matter depends only on the fastness parameter: $g_{\rm
eff}\sim (\omega_s-1)$. In cases of relatively strong propellers,
$\omega_s>>1$, the power-law dependence $g_{\rm eff}\sim
\omega_s^2$ is expected. We should note that the magnetic force
also contributes to driving and acceleration/collimation of matter
in the wind, so that the above power-law dependence on $\omega_s$
can be different in real situation.

Above analysis shows that the strength of propeller and processes
at the disk-magnetosphere boundary should depend on the fastness
parameter $\omega_s$. That is why we chose this parameter as the
main parameter of the model and investigate different properties
of propellers as a function of $\omega_s$.

\begin{figure}[ht!]
\centering
\includegraphics[width=8cm,clip]{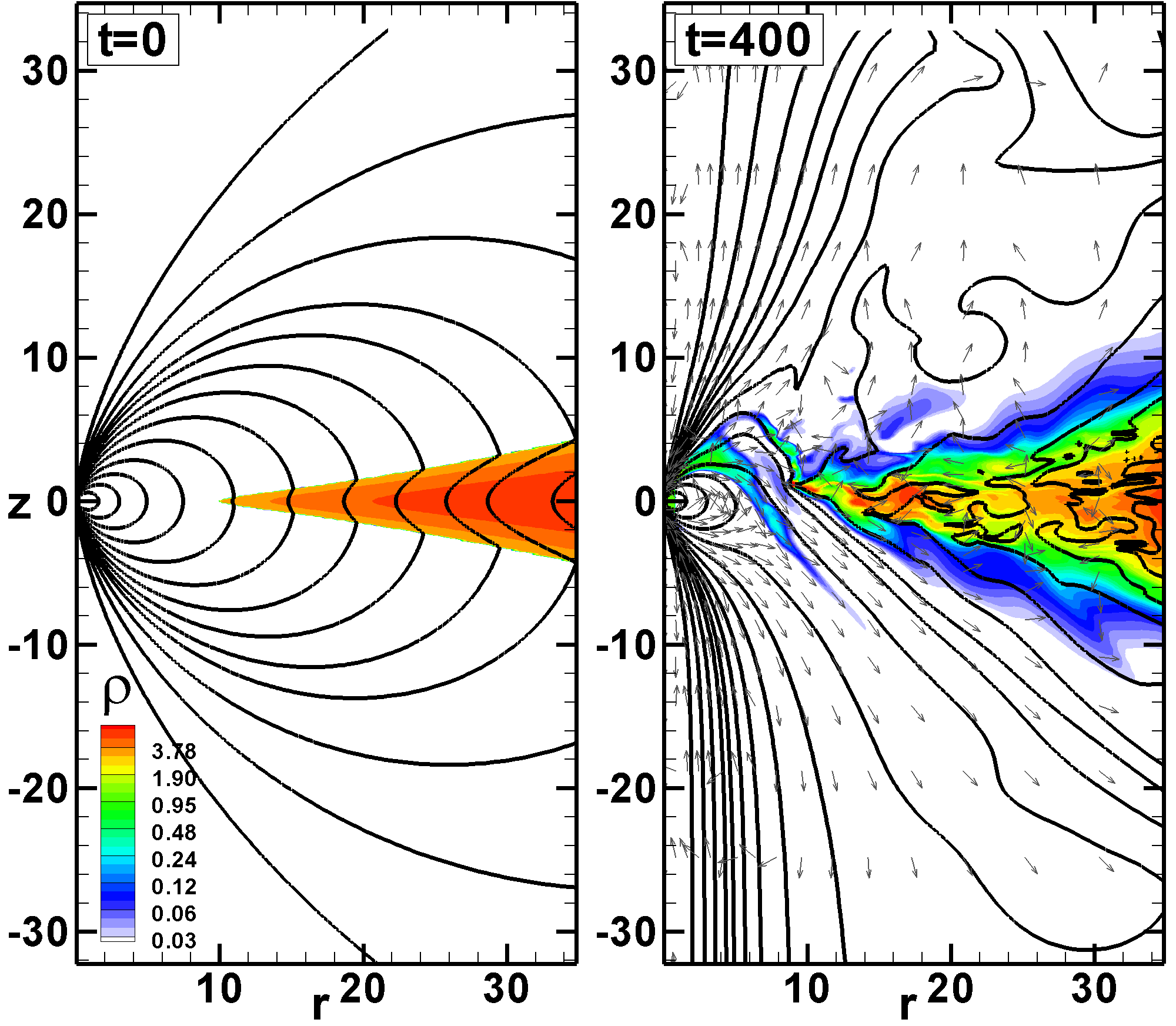}
\caption{\textit{Left panel:} Initial distribution of density and
sample magnetic field lines at $t=0$ in the model $\mu60c1.5$.
\textit{Right panel:} Same as left panel, but at $t=400$.}
\label{init-thin}
\end{figure}

\subsection{Convenient form for fastness parameter $\omega_s$ }


In the case of a Keplerian disk,
$\Omega_K(r_m)=(GM_\star/r_m^3)^{1/2}$, and the fastness parameter
can be presented in the form:
\begin{equation}
\omega_s = \bigg(\frac{r_m}{r_{\rm cor}}\bigg)^{\frac{3}{2}}~  ,
\label{eq:fastness}
\end{equation}
where $r_{\rm cor}$ is the corotation radius at which the angular
velocity of the star matches the Keplerian angular velocity in the
inner disk, $\Omega_\star=\Omega_K$:
\begin{equation}
r_{\rm cor}=\bigg(\frac{GM_\star}{\Omega_\star^2}\bigg)^{1/3} .
\label{eq:corotation-radius}
\end{equation}
The magnetospheric radius $r_m$ is the radius at which the
magnetic stress in the magnetosphere is equal to the matter stress
in the disk:
\begin{equation} \frac{B_p^2 + B_\phi^2}{8\pi}=\rho (v_p^2 + v_\phi^2) +
p~ . \label{eq:balance-stresses}
\end{equation}
Here, $\rho$ is density, $p$ is thermal pressure, $v_p$, $v_\phi$
and $B_p$, $B_\phi$ are the poloidal and azimuthal components of
velocity and the magnetic field, respectively.

In cases of slowly-rotating stars (not propellers), the
magnetospheric radius has been derived theoretically from the
balance of the largest components of the stresses : $B_{\rm
dip}^2/8\pi=\rho v_\phi^2$, where $B_{\rm dip}$ is the magnetic
field of the star which is suggested to be a dipole field, and
$v_\phi$ is the Keplerian angular velocity in the inner disk:
\begin{equation}
r_m = k \big[\mu_\star^4/(\dot{M}^2 GM_\star)\big]^{1/7},
~~~~~k\sim 1~, \label{eq:alfven}
\end{equation}
\noindent where $\mu_\star=B_\star R_\star^3$ is the magnetic
moment of the star with a surface field of $B_\star$, $\dot{M}$ is
the accretion rate in the disk, and $M_\star$  and $R_\star$ are
the mass and radius of the star, respectively (e.g.,
\citealt{LambEtAl1973}) \footnote{Comparisons of the
magnetospheric radii obtained in the axisymmetric simulations with
Eq. \ref{eq:alfven} provided the values $k\approx0.5$
\citep{LongEtAl2005} and $k\approx0.6$ \citep{ZanniFerreira2013}.
3D simulations of multiple cases have shown a slightly different
power (1/10 instead of 1/7) in Eq. \ref{eq:alfven} due to the
compression of the magnetosphere \citep{KulkarniRomanova2014}.}.

However, in the propeller regime, the magnetosphere departs from
the dipole shape and the poloidal velocity $v_p$ may become
comparable to or larger than the azimuthal velocity $v_\phi$. In
addition, the process of disk-magnetosphere interaction is
non-stationary, so all variables vary in time. This is why we
calculate the magnetospheric radius $r_m$ using the general
equation  for balance of stresses (Eq. \ref{eq:balance-stresses}),
where both poloidal and azimuthal components of velocity and
magnetic field are taken into account.

\begin{figure*}[ht!]
\centering
\includegraphics[width=14cm,clip]{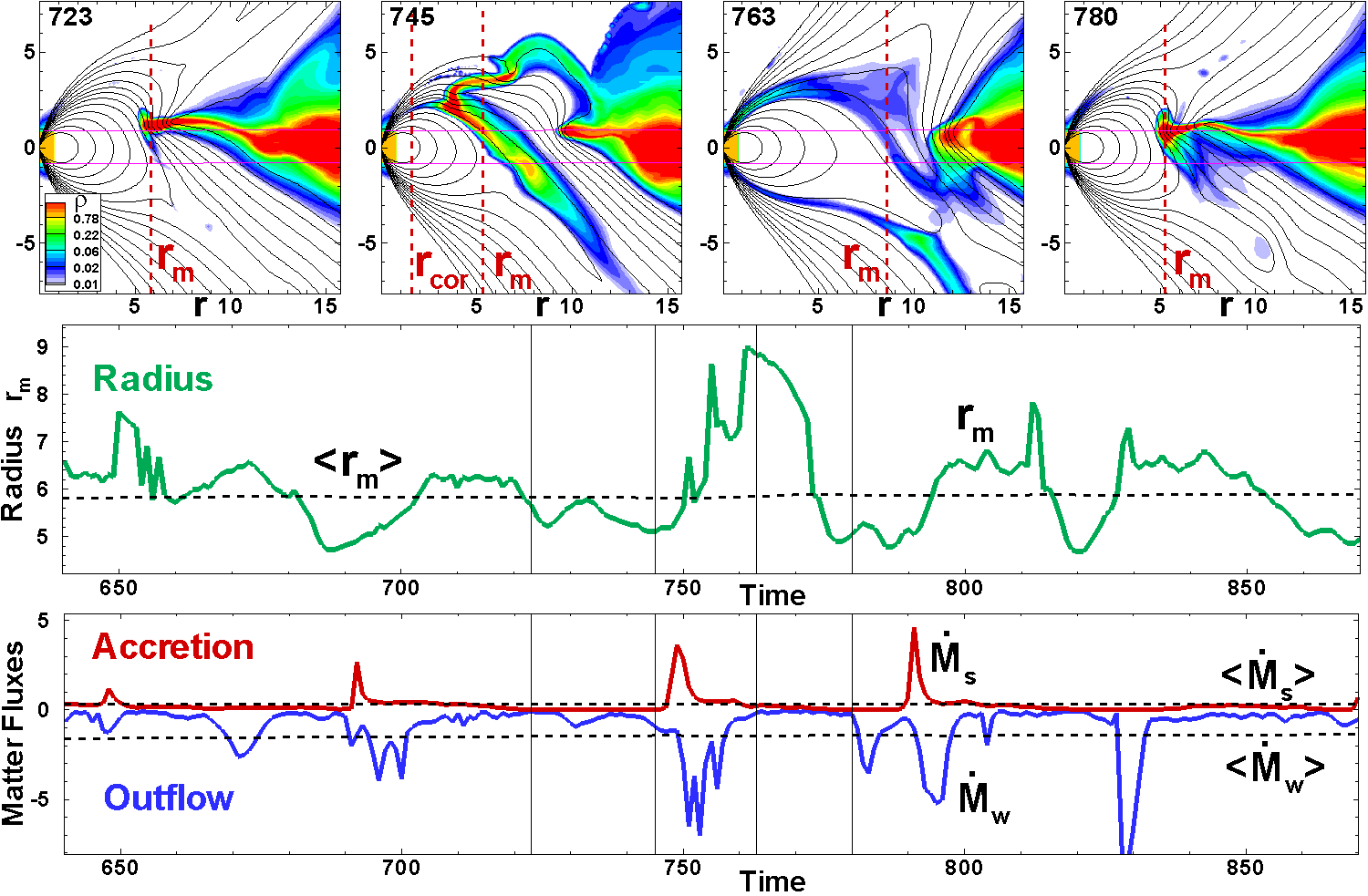}
\caption{ \textit{Top panels:} An example of the
accretion-ejection cycle
(model $\mu60c1.5$). The background shows density distribution and
the lines are sample field lines. The positions of the
magnetospheric and corotation radii are marked as $r_m$ and
$r_{\rm cor}$, respectively. \textit{Middle panel:} Variation of
the magnetospheric radius $r_m$. The thin vertical lines show
moments in time corresponding to the top panels.
 \textit{Bottom panel:} Variation of matter fluxes to the star,
 $\dot M_s$, and to the wind, $\dot M_w$. Dashed lines show the time-averaged
values: $\langle r_m\rangle$, $\langle \dot M_s\rangle$, and
$\langle \dot M_w\rangle$.}
 \label{2d-strong-expand}
\end{figure*}

\section{The numerical model}
\label{sec:Numerical model}

We performed axisymmetric simulations of disk accretion onto a
rotating magnetized star in the propeller regime. The model is
similar to that used in the simulations of \citet{LiiEtAl2014},
but with a few differences. Below, we briefly discuss the main
features of the numerical model and also the differences between
our model and that of \citet{LiiEtAl2014}. More technical details
of the model are described in \ref{appen:numerical-model}.

We consider accretion onto a magnetized star from a turbulent
accretion disk, where the turbulence is driven by the
magneto-rotational instability (MRI, e.g.,
\citealt{BalbusHawley1991}), which is initiated by a weak poloidal
magnetic field placed inside the disk (see Fig. \ref{init-thin}).
The accretion disk is cold and dense, while the corona is hot and
rarefied. The disk is 3,000 times cooler and denser
than the corona. The disk is geometrically thin, with an aspect
ratio of $h/r\approx 0.15$, where $h$ is the semi-thickness of the
disk. This disk is about 2.7 times thinner than that used in
\citet{LiiEtAl2014}. To achieve an accretion rate in the new thin
disk comparable with that in the \citet{LiiEtAl2014} thicker disk,
we increased the reference density in the disk by a factor of
three. \footnote{Note, that the first set of simulations was
performed for the thicker disk and only later was recalculated for
the thinner disk. Comparisons did not show any significant
differences between the results. However, the current paper is
based on the simulations of accretion from the thinner disk,
because a thinner disk is closer to realistic disks expected in
different accreting magnetized stars.}

\begin{figure*}[ht!]
\centering
\includegraphics[width=14cm,clip]{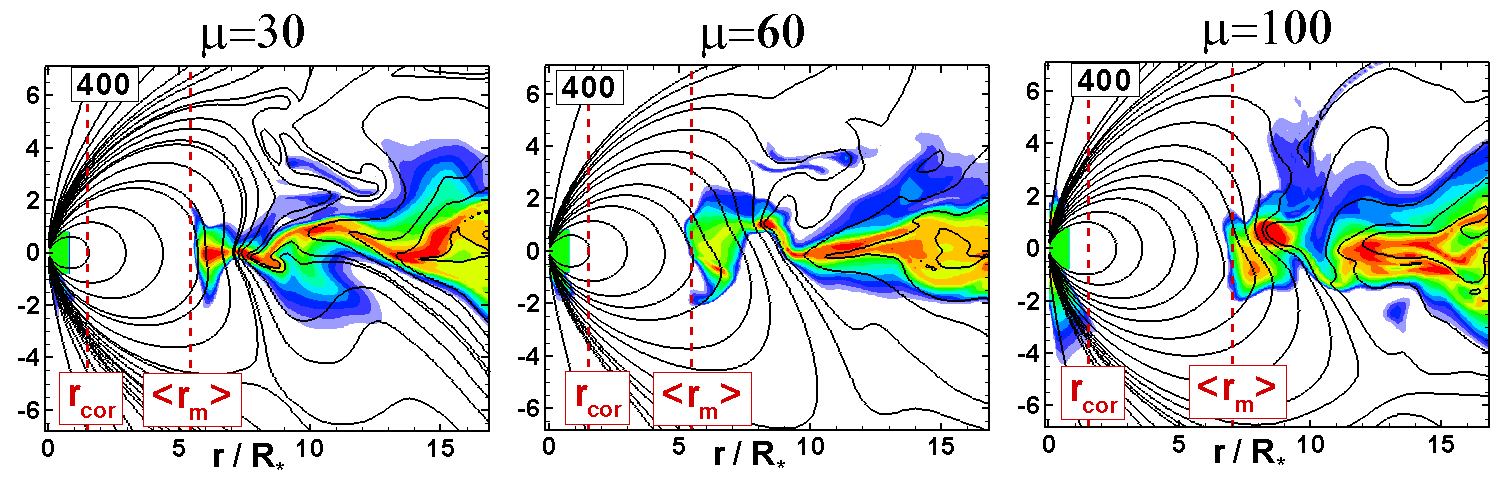}
\caption{An approximate position of the time-averaged
magnetospheric radius $\langle r_m\rangle$ in the cases of three
values of the magnetospheric parameter $\mu$: $\mu=30, 60, 100$.
The inner disk
oscillates. However, we chose the moments in time at which the
inner disk radius is approximately equal to the time-averaged
value $\langle r_m\rangle$. In all three models, the corotation
radius $r_{\rm cor}=1.5$. Note that the  magnetospheric radii in
our models, $\langle r_m\rangle\approx 5-7$,  are approximately
twice as large as those in the \citet{LiiEtAl2014} simulations,
where a smaller magnetospheric parameter, $\mu=10$, has been
taken.} \label{mu-diff-3}
\end{figure*}

A star with an aligned dipole magnetic field is placed at the
center of the coordinate system. The disk is placed at a distance
of 10 stellar radii from the center of the star, which
rotates slowly, with the Keplerian angular velocity corresponding
to corotation radius $r_{\rm cor}=10 R_\star$. Then, we gradually
spin up the star (over the period of 100 rotations of the inner
disk at $r=R_\star$) to a higher angular velocity, $\Omega_\star$,
corresponding to the propeller regime. The corresponding
corotation radius, $r_{\rm cor}$, is a parameter of the model. See
\ref{app:initial and boundary} for details of initial and boundary
conditions.

The disk-magnetosphere interaction in the propeller regime
requires some kind of diffusivity, so that matter of the inner
disk can penetrate through the rapidly-rotating layers of the
stellar magnetosphere.
This type of penetration may be connected with the magnetic
Rayleigh-Taylor (magnetic interchange) instability (e.g.,
\citealt{AronsLea1976}).
The magnetic interchange instability has been observed in 2D
simulations of propellers, performed in polar coordinates
\citep{WangRobertson1985}, as well as in global three-dimensional
(3D) simulations of accretion onto slowly-rotating stars (e.g.,
\citealt{KulkarniRomanova2008,RomanovaEtAl2008,BlinovaEtAl2016})
and in the local 3D simulations (e.g.,
\citealt{StoneGardiner2007a,StoneGardiner2007b}). However, in our
current 2.5D axisymmetric simulations, this instability is
suppressed by the axisymmetry of the problem. This is why we added
a diffusivity term into the code and suggested the presence of
large diffusivity at the disk-magnetosphere boundary. We
determined the coefficient of diffusivity in analogy with the
coefficient of viscosity in the $\alpha-$disk model:
$\eta_m=\alpha_{\rm diff} c_s^2/v_K$, where $c_s$ is the local
sound speed, and $\alpha_{\rm diff}$ is the $\alpha-$coefficient
of diffusivity. We chose the largest value, permitted by the
$\alpha-$disk theory, $\alpha_{\rm diff}=1$
\citep{ShakuraSunyaev1973} (see details in \ref{app:initial and
boundary}). This is different from the \citet{LiiEtAl2014}
simulations, where no diffusivity term was added in most of the
simulation runs, and where a small numerical diffusivity was
responsible for the disk-magnetosphere interaction. Test
simulation runs have shown that, when such a diffusive layer is
added, more matter is ejected into the outflows
(see Appendix in \citealt{LiiEtAl2014}). In the current paper, we
suggest high diffusivity in all simulation runs.

The equations were solved in dimensionless units,
so that the model could be applied to different types of stars,
from young stars to neutron stars (see \ref{app:reference units}
and Tab. \ref{tab-ref} for details of the dimensionalization
procedure). The results of simulations are shown in dimensionless
units, excluding those in Sec. \ref{sec:applications}.

One of the important dimensionless parameters is $\mu$, which
determines the typical size of the dimensionless magnetosphere,
$r_m/R_\star$.
Simulations of \citet{LiiEtAl2014} were performed at a relatively
small value of $\mu$ ($\mu=10$), which provided typical values of
the magnetospheric radius, $r_m\approx 3 R_\star$.
 \footnote{Note
that in application of the model to realistic stars,
\citet{LiiEtAl2014} suggested that the inner boundary
$R_0=2R_\star$, that is a star is located inside the inner
boundary. These provided the twice as larger efficient
magnetosphere of the star, $r_m/R_\star$. In current paper, stars
with larger magnetospheres are calculated, and we take
$R_0=R_\star$ during dimensionaliztion procedure.}. In the current
paper, we consider larger values of $\mu$: $\mu=30$, $60$ and
$100$, which provide larger sizes of the magnetosphere,
$r_m\approx (4-7) R_\star$. Larger magnetospheric sizes
are needed to model the propeller regime in different types of
stars, some of which may have relatively large magnetospheres.

A fine grid resolution is used, with grid compression in the
regions of the disk and the magnetosphere. A Godunov-type
numerical code in cylindrical coordinates has been developed by
\citet{KoldobaEtAl2016}, which incorporates an HLLD numerical
solver of \citet{MiyoshiKusano2005}.
See \ref{app:grid and code} for details of our numerical model.

\begin{table*}[ht!]
\centering
\begin{tabular}{llllll||lll||lll}

\\ Model & $\mu$ & $r_{\rm cor}$ & $\langle r_m \rangle$ & $\langle r_m \rangle/{r_{\rm cor}}$ & $\omega_s $ & $\langle\dot{M}_s\rangle$ & $\langle\dot{M}_w\rangle$  & $f_{\rm eff}$ & $\langle\dot{L}_{\rm sd}\rangle$ & $\langle\dot{E}_m\rangle$ & $\langle\dot{E}_f\rangle$ \\
\hline
$\mu30c1.3$  & 30 & 1.3 & 5.1 & 4.0 & 7.9 & 0.12 & 0.76 & 0.86 & 3.2 & 0.12 & 0.19 \\
$\mu30c1.5$  & 30 & 1.5 & 5.0 & 3.4 & 6.1 & 0.14 & 0.76 & 0.85 & 2.9 & 0.091 & 0.13 \\
$\mu30c2$    & 30 & 2 & 5.0 & 2.5 & 4.0 & 0.17 & 0.76 & 0.82 & 1.7 & 0.069 & 0.053 \\
$\mu30c3$    & 30 & 3 & 4.6 & 1.5 & 1.9 & 0.29 & 0.59 & 0.67 & 1.0 & 0.039 & 0.028 \\
$\mu30c4.2$  & 30 & 4.2 & 4.7 & 1.1 & 1.2 & 0.35 & 0.55 & 0.61 & 0.62 & 0.039 & 0.019 \\
\hline
$\mu60c1.3$  & 60 & 1.3 & 5.9 & 4.6 & 9.7 & 0.23 & 1.49 & 0.87 & 7.6 & 0.30 & 0.48 \\
$\mu60c1.5$  & 60 & 1.5 & 5.8 & 3.9 & 7.6 & 0.31 & 1.43 & 0.82 & 6.5 & 0.19 & 0.30 \\
$\mu60c2$    & 60 & 2 & 5.9 & 2.9 & 5.1 & 0.22 & 1.24 & 0.85 & 4.9 & 0.13 & 0.20 \\
$\mu60c3.1$  & 60 & 3.1 & 5.5 & 1.8 & 2.3 & 0.30 & 0.82 & 0.73 & 1.8 & 0.041 & 0.036 \\
$\mu60c3.7$  & 60 & 3.7 & 5.4 & 1.5 & 1.8 & 0.46 & 0.87 & 0.66 & 1.7 & 0.051 & 0.039 \\
$\mu60c4.2$  & 60 & 4.2 & 5.4 & 1.3 & 1.5 & 0.59 & 0.77 & 0.57 & 1.2 & 0.043 & 0.040 \\
$\mu60c5$    & 60 & 5 & 5.8 & 1.2 & 1.3 & 0.60 & 0.85 & 0.59 & 1.1 & 0.056 & 0.037 \\
\hline
$\mu100c1.5$  & 100 & 1.5 & 7.1 & 4.7 & 10.2 & 0.58 & 2.46 & 0.81 & 16.8 & 0.58 & 1.01 \\
$\mu100c2$    & 100 & 2 & 7.1 & 3.5 & 6.7 & 0.54 & 1.44 & 0.73 & 8.6 & 0.28 & 0.34 \\
$\mu100c2.5$  & 100 & 2.5 & 6.0 & 2.4 & 3.8 & 0.34 & 1.12 & 0.77 & 5.4 & 0.11 & 0.091 \\
$\mu100c3$    & 100 & 3 & 5.9 & 2.0 & 2.7 & 0.42 & 0.76 & 0.64 & 3.5 & 0.052 & 0.047 \\
$\mu100c3.7$  & 100 & 3.7 & 6.1 & 1.7 & 2.1 & 0.49 & 0.95 & 0.66 & 3.1 & 0.051 & 0.049 \\
$\mu100c4.3$  & 100 & 4.3 & 6.2 & 1.4 & 1.7 & 0.60 & 1.19 & 0.67 & 2.4 & 0.063 & 0.047 \\
$\mu100c5$    & 100 & 5 & 6.1 & 1.2 & 1.3 & 0.75 & 1.10 & 0.59 & 2.1 & 0.067 & 0.035 \\

\hline

\end{tabular}
\caption{Representative simulation models, calculated at
different values of magnetospheric parameter $\mu$ (which
determines the dimensionless size of the magnetosphere) and
different corotation radii $r_{\rm cor}$.  The time-averaged
magnetospheric radius $\langle r_m \rangle$ and matter fluxes to
the star $\langle\dot{M}_s\rangle$ and to the wind
$\langle\dot{M}_w\rangle$ are found from the simulations. The
fastness parameter $\omega_s$ is calculated using Eq.
\ref{eq:fastness}, while the propeller efficiency  $f_{\rm eff}$
is calculated using Eq. \ref{eq:prop-efficiency}. Matter fluxes to
the wind are calculated at the condition $v_p>0.1 v_{\rm esc}$.
$\langle\dot L_{\rm sd}\rangle$ is the angular momentum flux from
the surface of the star. $\langle\dot E_m\rangle$ and $\langle\dot
E_f\rangle$ are the energy fluxes through surface
$S(r=10,z=\pm10)$ carried by matter and magnetic field,
respectively.} \label{tab:models}
\end{table*}

\begin{figure}[ht!]
\centering
\includegraphics[width=8cm,clip]{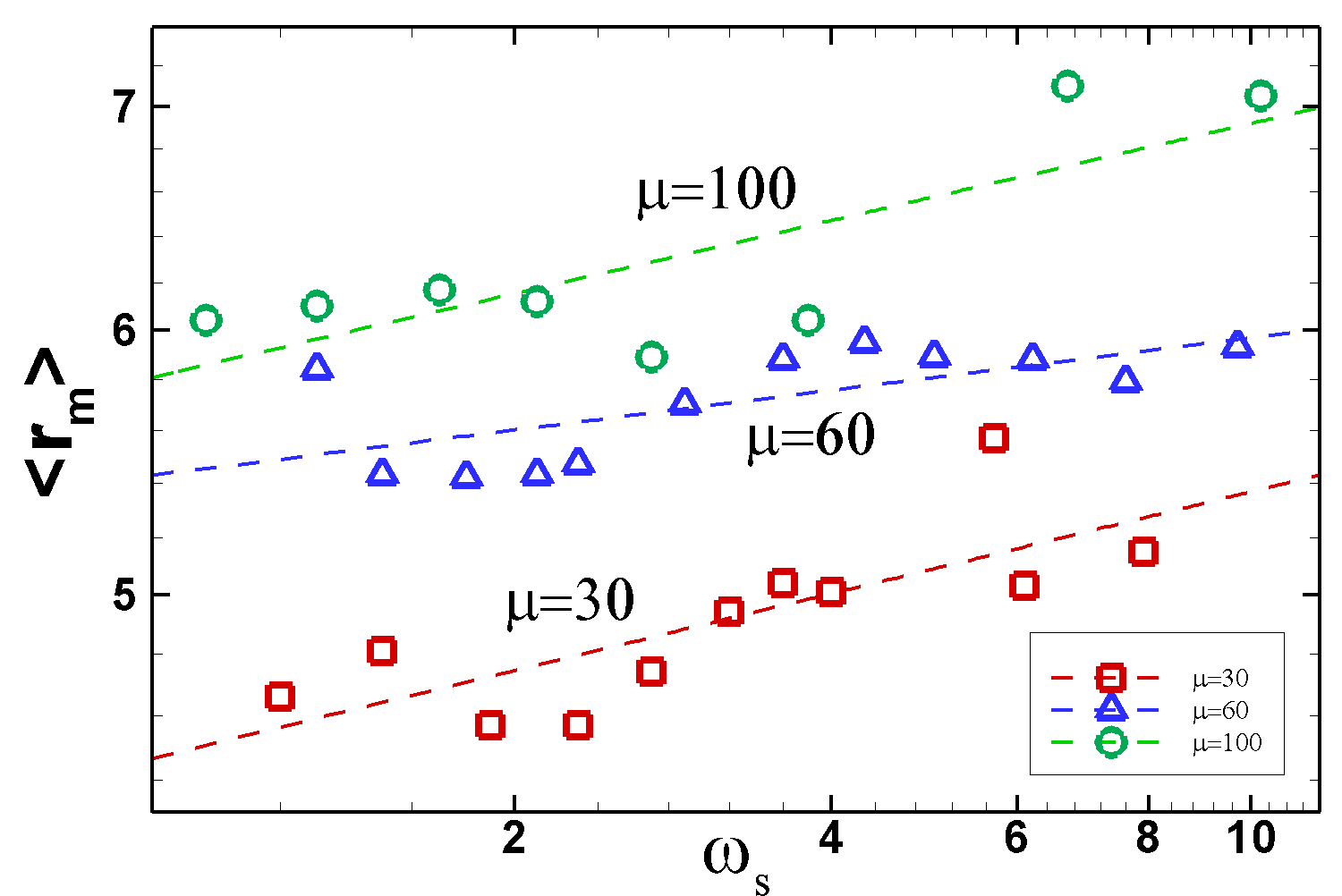}
\caption{The time-averaged magnetospheric radius $\langle
r_m\rangle$ obtained in models with different magnetospheric
parameters $\mu$ and different fastness parameters $\omega_s$.}
\label{rm-omega-s}
\end{figure}

\section{Variability and time-averaged values}
\label{sec:variabili-time-averaged}

We investigated the properties of propellers of different
strengths,
 from very weak propellers (in which the magnetosphere
rotates only slightly faster than the inner disk) to very strong
propellers (where the magnetosphere rotates much more rapidly than
the inner disk). To achieve different strengths of propeller, we
varied the corotation radius in the range of $r_{\rm cor}=1.3-6$.
We also varied the
magnetospheric parameter $\mu$ (which determines the dimensionless
size of the magnetosphere, $r_m/R_\star$), and performed
calculations for three values, $\mu=30, 60,$ and $100$. Table
\ref{tab:models} shows parameters for a number of representative
models and also results of simulations.


The disk-magnetosphere interaction in the propeller regime is a
non-stationary process, where the inner disk radius $r_m$
oscillates, and the matter fluxes to the star, $\dot M_s$, and to
the wind, $\dot M_w$, are also strongly variable.
 The variability is connected
with the cycle of matter accumulation, accretion/ejection, and
magnetosphere expansion.


\subsection{A cycle of accumulation-accretion/ejection-expansion}

To demonstrate the non-stationary nature of the propeller regime
and our procedure for time-averaging, we show the results obtained
for one of our models of a strong propeller, $\mu60c1.5$. The top
panels of Fig. \ref{2d-strong-expand} show a close-in view of the
inner part of the simulation region during one episode of the
accumulation-accretion/ejection-expansion cycle. (1) At $t=723$,
matter is accumulated at the inner disk, which gradually moves
inward (towards the star). The stellar field lines threading the
disk inflate and expand. Inflation is stronger in the part of the
magnetosphere that is below the equator. (2) At $t=745$, part of
the matter starts accreting onto the star above the magnetosphere,
while another part starts flowing away from the star below the
magnetosphere.
 It is very typical for matter
to accrete on one side of the magnetosphere while forming outflows
on the other side of the magnetosphere. The magnetosphere is
slightly compressed on the accreting side, while the field lines
strongly inflate on the outflowing side. One can see that a
significant part of the magnetic flux inflates in the direction
away from the star
below the equatorial plane.  
(3) After an accretion/ejection event, the magnetosphere expands
(see 3rd panel at $t=763$). (4) Subsequently, the inner disk
gradually moves inward (see 4th panel at $t=780$) and the process
repeats. At stage (2), accretion is possible because only a part
of the magnetosphere (where the field lines are closed) rotates
more rapidly than the inner disk and represents a centrifugal
barrier \footnote{ We should note that some matter is also ejected
into the outflows without significant accretion onto the star
(see, e.g., a burst to an outflow at $t\approx 830$ in Fig.
\ref{2d-strong-expand}). In this type of outflow, matter of the
inner disk penetrates through the magnetosphere, acquires
sufficient angular momentum and flows away from the star.}

\subsection{Magnetospheric radius $r_m$}

 We calculated the position of the inner disk (magnetospheric) radius $r_m$ using
Eq. \ref{eq:balance-stresses}. To calculate the magnetospheric
radius $r_m(t)$ at some moment $t$ in time, we take the values of
density, pressure and  components of velocity and magnetic field
in the equatorial plane from the simulations, and find the radius
$r_m(t)$, at which the balance of stresses (Eq.
\ref{eq:balance-stresses}) is satisfied. The magnetosphere is
often asymmetric about the equatorial plane. To take this issue
into account, we initially calculated the magnetospheric radius at
the surfaces $z=\pm R_\star$, which are above and below the
equatorial plane, and then took the half-averaged value
$r_m=[r_m(z=R_\star) + r_m(z=-R_\star)]/2$ as the main
magnetospheric radius in the model. We observed from the
simulations that at this radius, $r_m$, the density in the disk
drops from the high values in the disk down to very low values in
the magnetosphere, while the angular velocity changes from the
Keplerian angular velocity in the disk to the angular velocity of
the magnetosphere.


Simulations show that the magnetospheric radius varies in time.
 The middle panel
of Fig. \ref{2d-strong-expand} shows that it varies between
$r_m\approx 4.9$ and $r_m\approx 9$. To characterize the
magnetospheric radius in each model, we introduce the
time-averaged magnetospheric radius:
\begin{equation}
\langle r_m(t)\rangle = \frac{\int_{t_i}^{t} dt'
r_m(t')}{\int_{t_i}^{t} dt'} ~. \label{eqn:timeavg-rm}
\end{equation}
We show this time-averaged radius as a dashed line in Fig.
\ref{2d-strong-expand}. This radius slightly varies in time. For
consistency, we start averaging at moment $t=200$ in each
simulation run (so as to exclude the effects of the initial
conditions) and take this radius at moment $t=1,000$ for each
model. In the model shown in Fig. \ref{2d-strong-expand}, we
obtain $\langle r_m\rangle\approx 5.8$.

 The right panels of
Figs. \ref{app:fluxes-radii-d100}, \ref{app:fluxes-radii-d60} and
\ref{app:fluxes-radii-d30} show variation of the inner disk radius
in our representative models. The figures show that the amplitudes
of disk oscillations are larger in models with larger $\omega_s$
values (strongest propellers) and also increase with the
magnetospheric parameter $\mu$. In the weaker propellers, the
amplitude of the oscillations is much smaller, and the radius
varies only slightly. Tab. \ref{tab:models} shows the
time-averaged magnetospheric radii $\langle r_m\rangle$  for our
representative models.

Fig. \ref{mu-diff-3} shows the density distribution in three
models with the same corotation radius, $r_{\rm cor}=1.5$, but
different values of the magnetospheric parameter $\mu$: $\mu=30,
60, 100$, at the times when the magnetospheric radius is
approximately equal to the time-averaged radius. These radii are
approximately twice as large as the radii in the models of
\citet{LiiEtAl2014},

\subsection{Time-averaged fastness parameter}

 We use the time-averaged radius $\langle r_m\rangle$ to
calculate the time-averaged fastness parameter:
\begin{equation}
\langle\omega_s\rangle = \bigg(\frac{\langle r_m \rangle}{r_{\rm
cor}}\bigg)^{\frac{3}{2}}~ . \label{eq:fastness-rad-averaged}
\end{equation}
Subsequently, in this paper, we use this parameter, as the main
parameter of the model. For convenience, we remove the brackets
$\langle ~~\rangle$ and simply use the variable $\omega_s$. Tab.
\ref{tab:models} shows the values of the time-averaged fastness
parameter for our representative models. One can see that the
fastness parameter ranges from very low values, $\omega_s=1.2$, to
very high values, $\omega_s=10.2$.

Fig. \ref{rm-omega-s} shows the dependence of the time-averaged
magnetospheric radius, $\langle r_m\rangle$, calculated for all
models, on the time-averaged fastness parameter $\omega_s$. One
can see that in a set of models with the same parameter $\mu$, the
magnetospheric radius $\langle r_m\rangle$ slightly increases with
$\omega_s$. The dependencies are the following: (1) at $\mu=30$,
$\langle r_m\rangle\approx 4.5 \omega_s^{0.076}$; (2) at $\mu=60$,
$\langle r_m\rangle\approx 5.7\omega_s^{0.04}$; (3) at $\mu=100$,
$\langle r_m\rangle\approx 5.85\omega_s^{0.073}$. In each set, the
radii are larger at larger values of $\mu$. We took the dependence
on $\omega_s$ corresponding to $\mu=30$ and $\mu=100$, and found
an approximate general relationship for all models:
 \begin{equation}
 \langle r_m\rangle\approx 5.7
\mu_{60}^{0.21}\omega_s^{0.07}~,
\label{eq:rm-omega-s}\end{equation} where $\mu_{60}=\mu/60$. A
comparison of the radii obtained with this formula with the values
of $\langle r_m\rangle$ observed in the simulations shows that the
typical deviation of the observed radii
from those obtained with the formula is $\sim 5-10\%$.

\subsection{Why does matter accrete in the propeller regime?}

In the sample model shown in Fig. \ref{2d-strong-expand}, the
magnetospheric radius is always larger than the corotation radius,
$r_m > r_{\rm cor}$, and the time-averaged magnetospheric radius,
$\langle r_m\rangle\approx 5.8$, is also larger than the
corotation radius, $r_{\rm cor}=1.5$. In spite of this, matter
accretes onto the star. In all other models, matter also accretes
onto the star (see Tab. \ref{tab:models} for $\langle \dot
M_s\rangle$). This is different from the generally-accepted
definition that, in the propeller regime, accretion is possible if
$r_m<r_{\rm cor}$ and is completely forbidden otherwise.
Below, we describe the main reasons why accretion becomes possible
even in the cases of very strong propellers:

\begin{itemize}

\item In theoretical studies and one-dimensional models, it is
suggested that the centrifugal barrier is an infinite vertical
wall (e.g., \citealt{SunyaevShakura1977,SpruitTaam1993}). However,
two-dimensional simulations show that only the closed part of the
magnetosphere rotates more rapidly than the inner disk and
represents the centrifugal barrier. That is why, at favorable
conditions, matter can flow above or below the magnetosphere and
accrete onto the star at condition $r_m>r_{\rm cor}$.


\item The process is non-stationary. Most of the time, accretion
is blocked by the centrifugal barrier and matter does not accrete
onto the star. However, when the disk comes closer to the star,
conditions become favorable for accretion and matter accretes onto
the star in a brief episode. This explains why the time-averaged
matter flux to the star can be so low (much lower than the matter
flux to the wind).

\end{itemize}

These two-dimensional, non-stationary propeller models
help explain why
propellers can accrete a
part of the disk matter, in spite of the fact that their
magnetospheres rotate
more rapidly than their inner disks.

\subsection{Matter fluxes}

We also calculated the matter fluxes onto the star and to the
wind:
\begin{equation}
\dot M_s(t)=\int_{S_{\rm star}}{\rho {\bf v_p dS}}~, ~~~\dot
M_w(t)=\int_{S_{\rm wind}} {\bf \rho v_p dS} .
\end{equation}
The matter flux to the star has been calculated through the
stellar surface, $S_{\rm star}=S(r=R_\star,z=\pm R_\star)$ which
is a cylinder with radius $r=R_\star$ and height $z=\pm R_\star$
centered on the star, while the matter flux to the wind has been
calculated through  cylindrical surface $S_{\rm
wind}=S(r=10,z=\pm10)$  with radius $r=10$ and height $z=\pm 10$
\footnote{The surface $S(r=10,z=\pm10)$ is located relatively
close to the star, at a distance
 that is only slightly larger than the
time-averaged values of the magnetospheric radii in our models,
$\langle r_m \rangle \approx 4.6-7.1$ (see Tab. \ref{tab:models}).
It helps us select the propeller-driven wind from the
disk-magnetosphere boundary and deselect the slow winds from the
other parts of the disk. }. To exclude the slow motions in the
turbulent disk, we placed the condition that the poloidal velocity
in the wind should be larger than some minimum value $v_{\rm
min}=k v_{\rm esc}$, where $v_{\rm esc}=(2GM_\star/r)^{1/2}$ is
the local escape velocity, and $k\leq 1$. Our simulations show
that only in
strong propellers is matter ejected from the disk-magnetosphere
boundary with a velocity comparable to the local escape velocity.
In most cases, the initial outflow velocity
is low. It can be as low as $0.1 v_{\rm esc}$. In spite of that,
matter flows away from the simulation region, driven mainly by the
magnetic force of the inflating field lines. That is why we chose
the condition $v_{\rm min} = 0.1 v_{\rm esc}$ for the calculation
of the outflows. Using this condition, we take into account both
the fast and slow winds from the disk-magnetosphere boundary.

 The bottom panel of Fig. \ref{2d-strong-expand} shows the flux onto
 the star (in red) and the flux to the wind (in blue). One can see that
 the fluxes
 are ``spiky",
because most of the time accretion onto the star is stopped by the
centrifugal barrier of the propelling star.  The left panels of
Figs. \ref{app:fluxes-radii-d100}, \ref{app:fluxes-radii-d60} and
\ref{app:fluxes-radii-d30} show variation of the matter fluxes in
our representative models. To characterize fluxes in each model,
we introduced the time-averaged matter fluxes:
\begin{equation}
\langle\mdot_s(t)\rangle = \frac{\int_{t_i}^{t} dt'
\mdot_s(t')}{\int_{t_i}^{t} dt'} ~, ~~\langle\mdot_w(t)\rangle =
\frac{\int_{t_i}^{t} dt' \mdot_w(t')}{\int_{t_i}^{t} dt'}
~.\label{eqn_timeavg}
\end{equation}
The dashed lines in the bottom left panel of Fig.
\ref{2d-strong-expand} show the time-averaged matter fluxes onto
the star $\langle\dot M_s(t)\rangle$ and to the wind $\langle\dot
M_w(t)\rangle$.

We observed that the matter fluxes are
 affected by the initial conditions during the first
 $\sim$200 rotations, which is why we calculated the time-averaged
 values of the fluxes
 starting at $t=200$. We observed that the time-averaged fluxes vary only slightly in time.
 We chose a late moment in time,
$t=1,000$, which is near the end of most simulation runs
\footnote{Much longer simulation runs were performed in a few test
cases. However, they did not show new information compared with
shorter runs. That is why most of simulations were stopped shortly
after time $t=1,000$ (to save computing time). }, and took the
flux values at this moment to be the typical averaged fluxes for
any given model. The values of these averaged fluxes are
$\langle\mdot_s\rangle = 0.31$ and $ \langle\mdot_w\rangle = 0.82$
in the sample model shown in Fig. \ref{2d-strong-expand}.

 Left panels of Figures \ref{app:fluxes-radii-d100},
\ref{app:fluxes-radii-d60} and \ref{app:fluxes-radii-d30} of
\ref{appen:variation}) show examples of the fluxes in our
representative models. The dashed lines show their time-averaged
values.  Tab. \ref{tab:models} shows the time-averaged matter
fluxes for a number of calculated models.

\begin{figure}[ht!]
\centering
\includegraphics[width=7.0cm,clip]{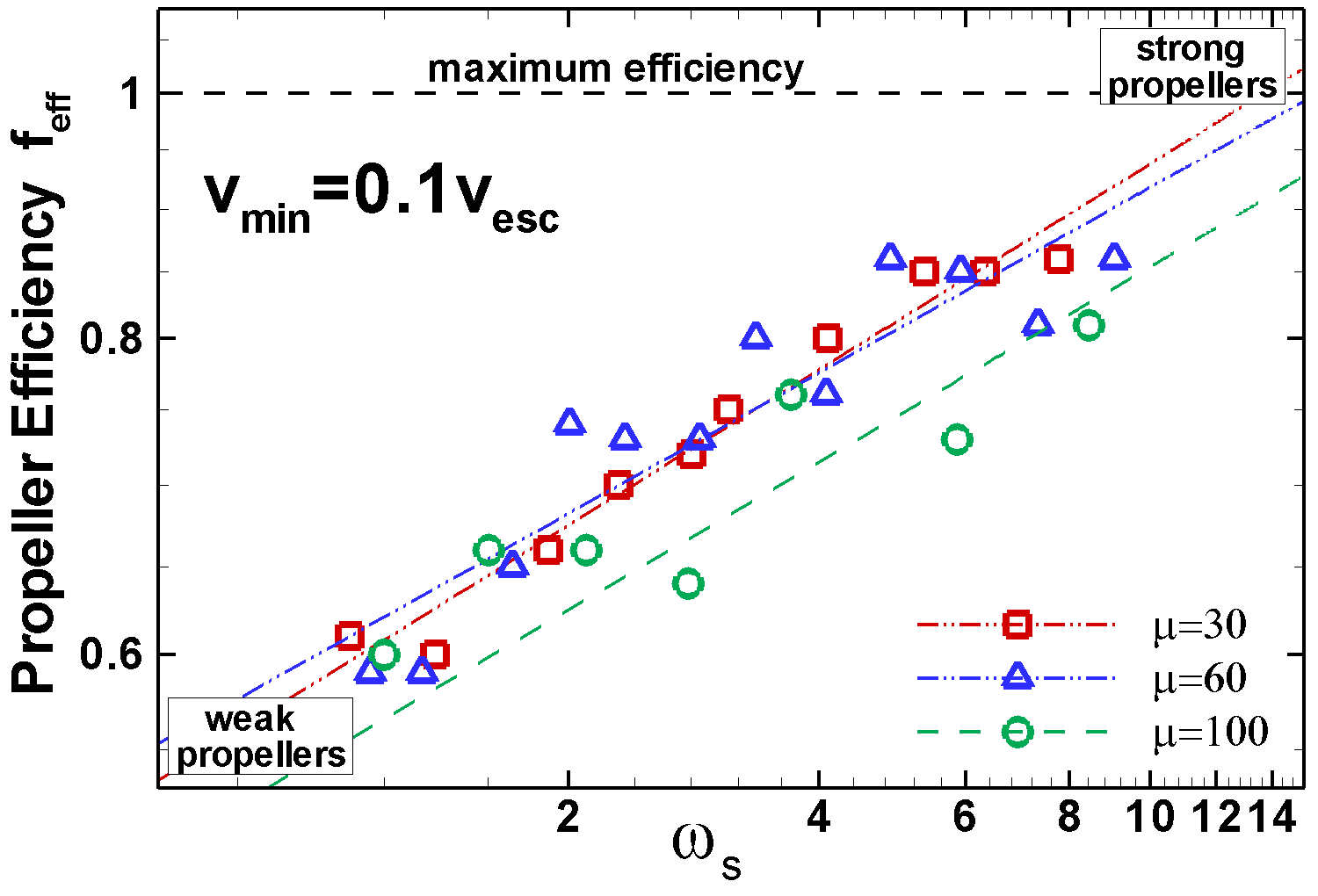}
\includegraphics[width=7.0cm,clip]{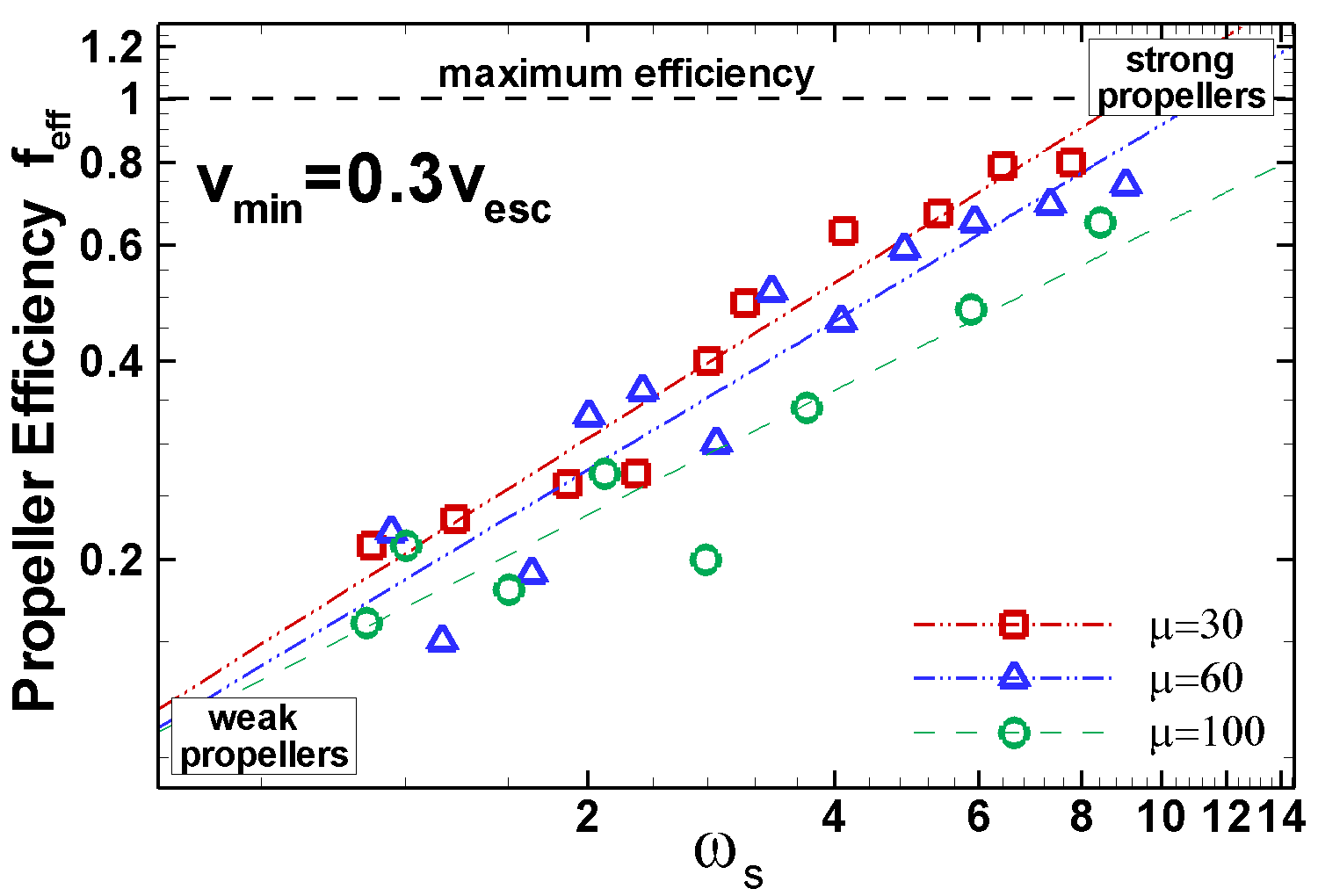}
\includegraphics[width=7.0cm,clip]{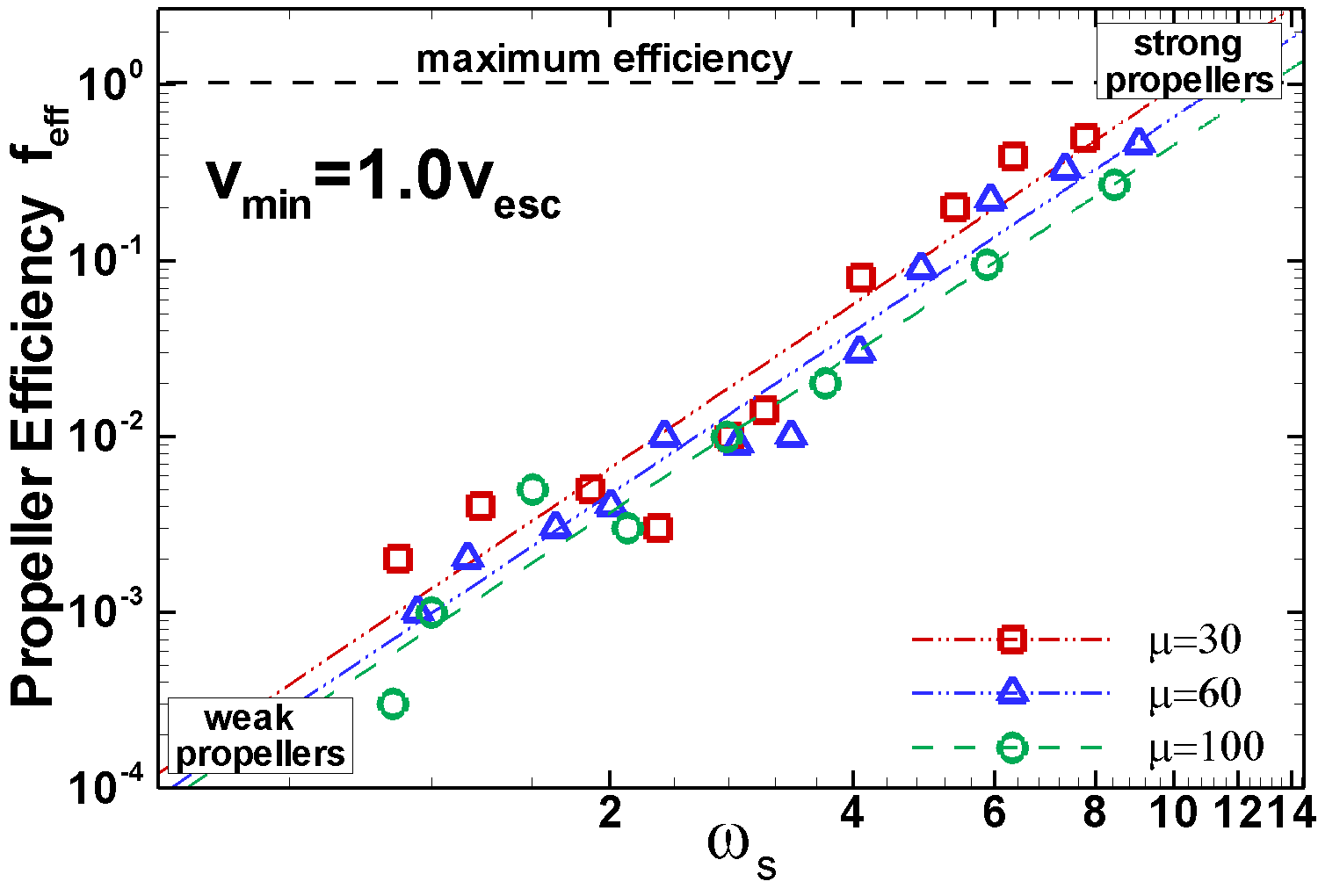}
\caption{Dependence of propeller efficiency $f_{\rm eff}$ on the
fastness parameter $\omega_s$, where the poloidal velocity in the
outflows (calculated through surface $S(r=10,z=\pm10)$) has been
restricted by the condition $v_p>v_{\rm min}$, where $v_{\rm
min}=k v_{\rm esc}$ is a part $k$ of the local escape velocity at
the same surface. \textit{Top panel:} $v_{\rm min}=0.1 v_{\rm
esc}$. \textit{Middle panel:} $v_{\rm min}=0.3 v_{\rm esc}$.
\textit{Bottom panel:} $v_{\rm min}=1.0 v_{\rm esc}$.}
\label{eff-thin-vesc01-03}
\end{figure}

\begin{table*}
\centering
\begin{tabular}{lllll}

\\ $v_{\rm min}/v_{\rm esc}$  & $\mu=30$ & $\mu=60$ & $\mu=100$ & Averaged\\
\hline
$0.1$ & $0.58 \omega_s^{0.32}$ & $0.59 \omega_s^{0.29}$ & $0.54 \omega_s^{0.28}$ & $f_{\rm eff}=0.57 \omega_s^{0.30}$ \\
\hline
$0.3$ & $0.15 \omega_s^{1.32}$ & $0.15 \omega_s^{1.16}$ & $0.15 \omega_s^{0.92}$ & $f_{\rm eff}=0.15 \omega_s^{0.92}$\\
\hline
$1$ & $0.0004 \omega_s^{5.14}$ & $0.0007 \omega_s^{4.35}$ & $0.0006 \omega_s^{4.01}$ & $f_{\rm eff}=0.0006 \omega_s^{4.01}$\\
\hline

\end{tabular}
\caption{Propeller efficiency $f_{\rm eff}$ as a function of
fastness parameter $\omega_s$ for different values of
magnetospheric parameter $\mu$ and different
minimum poloidal wind velocities: $v_{\rm min}=k v_{\rm esc}$,
where $k=0.1, 0.3, 1$.} \label{tab:models-feff}
\end{table*}

\subsection{Propeller Efficiency}
\label{subsec:propeller-efficiency}

Propellers of different strengths  eject different amounts of
matter into the wind. To characterize the relative matter flux
ejected into the wind, we introduce propeller efficiency:
\begin{equation}f_{\rm eff} =\frac{\langle\mdot_w\rangle}{\langle\mdot_s\rangle +
\langle\mdot_w\rangle},~~~ \label{eq:prop-efficiency}
\end{equation}
where $\langle\mdot_w\rangle$ and $\langle\mdot_s\rangle$ are the
time-averaged matter fluxes to the wind and to the star,
respectively.


For each model, we calculated the time-averaged matter fluxes and
the value of propeller efficiency $f_{\rm eff}$ using Eq.
\ref{eq:prop-efficiency}.
We also calculated the time-averaged value of the fastness
parameter for each model using eq. \ref{eq:fastness-rad-averaged}.

Fig. \ref{eff-thin-vesc01-03} shows the plot of efficiency $f_{\rm
eff}$ versus the averaged fastness, $\omega_s$, for all models,
where each point corresponds to a single model. The set of models
includes a wide variety of propeller strengths, from very weak
propellers (bottom left corners of the plots)  to very strong
propellers (top right corners of the plots), and different values
of magnetospheric parameter $\mu$, which correspond to different
magnetospheric sizes, $r_m/R_\star$, from relatively small
magnetospheres ($\mu=30$, marked as squares) to large
magnetospheres ($\mu=100$, marked as circles). The triangles show
models with intermediate magnetospheric sizes ($\mu=60$).

We calculated the propeller efficiency taking into account only
the faster component of the outflowing matter (to exclude the slow
motions in the inner disk), with poloidal velocities $v_p>v_{\rm
min}$, where  $v_{\rm min}=0.1v_{\rm esc}$, $v_{\rm min}=0.3v_{\rm
esc}$, and $v_{\rm min}=1.0v_{\rm esc}$ (see top, middle and
bottom panels of Fig. \ref{eff-thin-vesc01-03}).

The top panel of Fig. \ref{eff-thin-vesc01-03} shows that, at
condition $v_p>v_{\rm min}=0.1 v_{\rm esc}$, efficiency is high in
both the strong propellers, $f_{\rm eff}\approx 0.85$ (see top
right corner of the plot), and the weak propellers, $f_{\rm
eff}\approx 0.6$ (bottom left corner of the plot). This means
that, in propellers of different strengths, most of the inner disk
matter flows into the wind. This wind can be
slow in the cases of weak propellers and much faster in the strong
propellers (see Sec. \ref{subsec:velocity}). The middle panel of
Fig. \ref{eff-thin-vesc01-03} shows that, if we only include the
relatively fast outflows, with  $v_p>v_{\rm min}=0.3 v_{\rm esc}$,
then efficiency becomes lower, $f_{\rm eff}\approx 0.15-0.2$, in
the weak propellers. The bottom panel of Fig.
\ref{eff-thin-vesc01-03} shows that, if we only consider the
outflows with super-escape velocities ($v_p>v_{\rm min}=1.0 v_{\rm
esc}$), then, in the weak propellers, efficiency becomes very low,
$f_{\rm eff}\approx 10^{-3}$, but increases sharply with
$\omega_s$. One can see that, in all three cases, the dependency
$f_{\rm eff}$ on $\omega_s$ can be approximated as a power law:
 $f_{\rm eff}\approx K\omega_s^\alpha$.
Tab. \ref{tab:models-feff} shows these dependencies for different
$v_{\rm min}/v_{\rm esc}$ values and different values of $\mu$.
One can see that, at $v_{\rm min}=0.1 v_{\rm esc}$ and $v_{\rm
min}=0.3 v_{\rm esc}$, the efficiency is slightly lower for larger
magnetospheres ($\mu=100$) compared with the smaller
magnetospheres ($\mu=30$ and $\mu=60$).

The above analysis shows that, in propellers of different
strengths, a significant amount of the inner disk matter is lifted
above the disk plane and flows into the wind. However, the fate of
this wind is different in the cases of strong and weak propellers.
Below, we analyze the properties of the wind.

\begin{figure*}[ht!]
\centering
\includegraphics[width=16cm,clip]{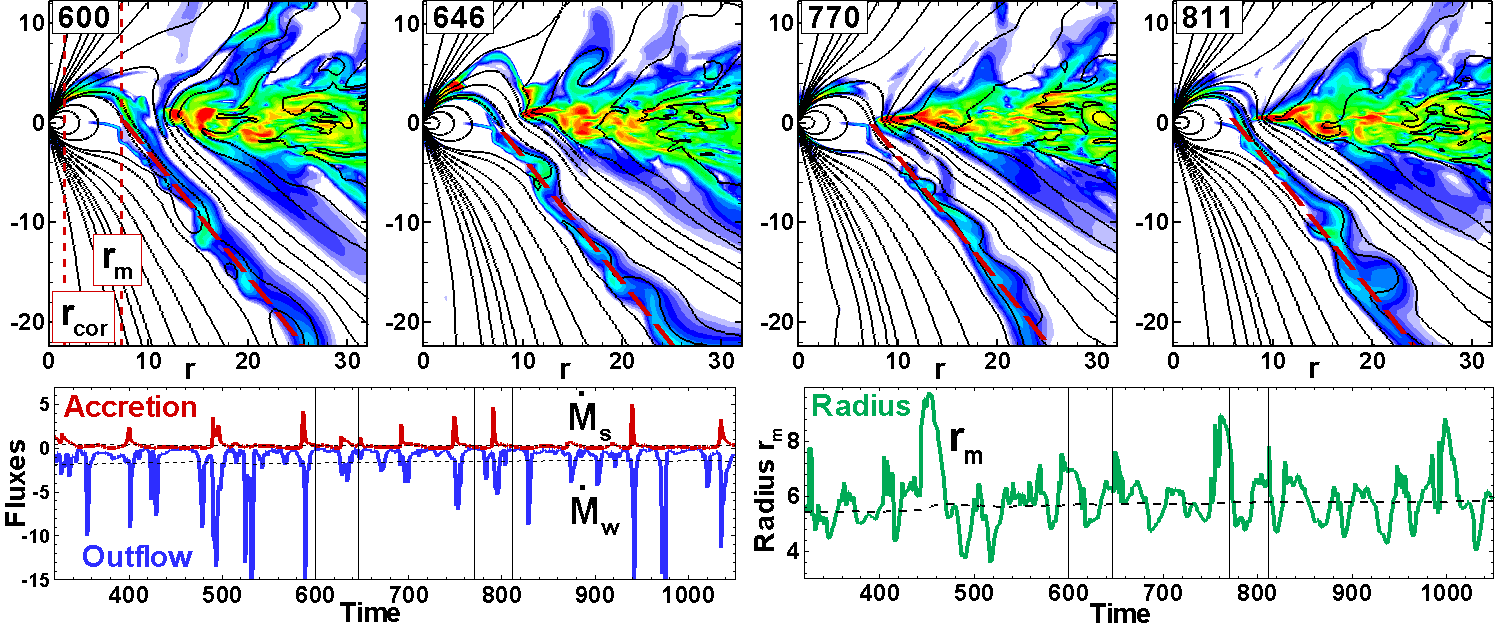}
\caption{\textit{Top panels:} Snapshots from a simulation run in
the strong propeller regime (model $\mu60c1.5$) at four different
moments in time. The color background shows matter flux density
$\rho \vert v_p\vert$, the lines are sample field lines. The thick
dashed line shows the approximate direction of the wind.
\textit{Bottom left panel:} Matter fluxes to the star (in red) and
to the wind (in blue), calculated at the condition $v > 0.1v_{\rm
esc}$. The vertical lines show the moments in time at which the
top panels are shown.
 \textit{Bottom right panel:}
Variation of the magnetospheric radius $r_m$ with time. The
long-dashed line shows the time-averaged value of $r_m$, $\langle
r_m \rangle$.} \label{2d-strong-6}
\end{figure*}

\begin{figure*}[ht!]
\centering
\includegraphics[width=16cm,clip]{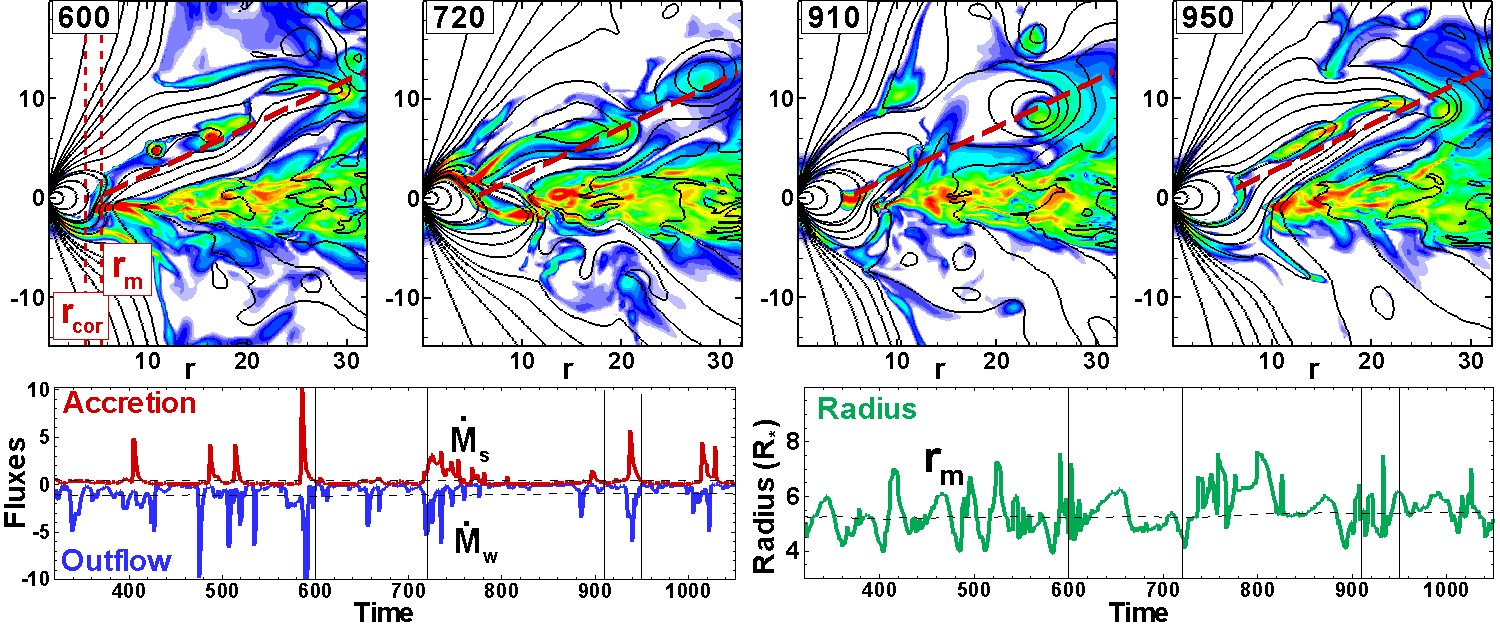}
\caption{Same as Fig. \ref{2d-strong-6}, but for a
star in the weak propeller regime (model $\mu60c3.7$).}
\label{2d-weak-6}
\end{figure*}

\section{Properties of propeller winds}
\label{subsec:strong-weak propeller}

\subsection{Matter flow in strong and weak propellers}

To demonstrate typical matter flow in strong and weak propellers,
we took two models ($\mu60c1.5$ and $\mu60c3.7$) with the same
magnetic moment, $\mu=60$, but different corotation radii $r_{\rm
cor}=1.5$ and $r_{\rm cor}=3.7$.

Fig. \ref{2d-strong-6} shows several snapshots of matter flow in
the strong propeller regime, taken during or after an episode of
matter outflow. The color background shows matter flux density
$\rho \vert v_p\vert$ and the lines are sample poloidal field
lines. One can see that most of the external field lines are open
and matter flows into conical-shaped wind at an angle of
$\Theta_{\rm wind}\approx 40^\circ-45^\circ$. The dashed red line
shows an approximate direction of the outflow.

The bottom left panel shows the matter fluxes to the star, $\dot
M_s$, and to the wind, $\dot M_w$, and their time-averaged values
(dashed lines), which were used to calculate the efficiency of the
propeller, $f_{\rm eff}\approx 0.82$. The bottom right panel shows
variation of the inner disk radius $r_m$ and its time-averaged
value $\langle r_m\rangle\approx 5.8$, which was used to calculate
the fastness parameter: $\omega_s=(\langle r_m\rangle/r_{\rm
cor})^{3/2}\approx 7.6$.

Fig. \ref{2d-weak-6} shows matter flow to the wind in a relatively
weak propeller (model $\mu60c3.7$) during several outbursts to the
wind. One can see that the magnetic field lines connecting the
inner disk with the star inflate and matter is ejected into the
wind at a larger opening angle, $\Theta_{\rm wind}\approx
60^\circ$, compared with the case of the stronger propeller. The
bottom left panel shows that the matter fluxes to the star and to
the wind look somewhat similar to those of the strong propeller
shown in Fig. \ref{2d-strong-6}. The efficiency of the weaker
propeller, $f_{\rm eff}\approx 0.66$, is only slightly lower than
that of the stronger propeller. This similarity is due to the fact
that, in both models, the outflows include any matter that flows
through surface $S(r=10,z=\pm10)$ with velocity $v>v_{\rm min}=0.1
v_{\rm esc}$. However, in the weak propeller, the velocity of
matter flow into the wind is much lower
than in the strong propeller (see Sec. \ref{subsec:velocity}). The
bottom right panel of Fig. \ref{2d-weak-6} shows that the
magnetospheric radius $r_m$ oscillates and the time-averaged value
$\langle r_m \rangle\approx 5.4$, which is only slightly smaller
than that in the above example
of the strong propeller \footnote{This approximate equality of the
radii $\langle r_m \rangle$ is due to the fact that the
magnetospheric radius is determined by the balance between the
magnetic and matter pressures, where $\rho v_\phi^2$ term
dominates over $\rho v_p^2$ term at the disk-magnetosphere
boundary.}. Note, however, that the fastness parameter,
$\omega_s\approx 1.8$, is much smaller than that of the strong
propeller.
The fastness parameter is one of the main factors contributing to
the differences in the properties of the winds
in strong versus weak propellers.

\begin{figure}[ht!]
\centering
\includegraphics[width=7.5cm,clip]{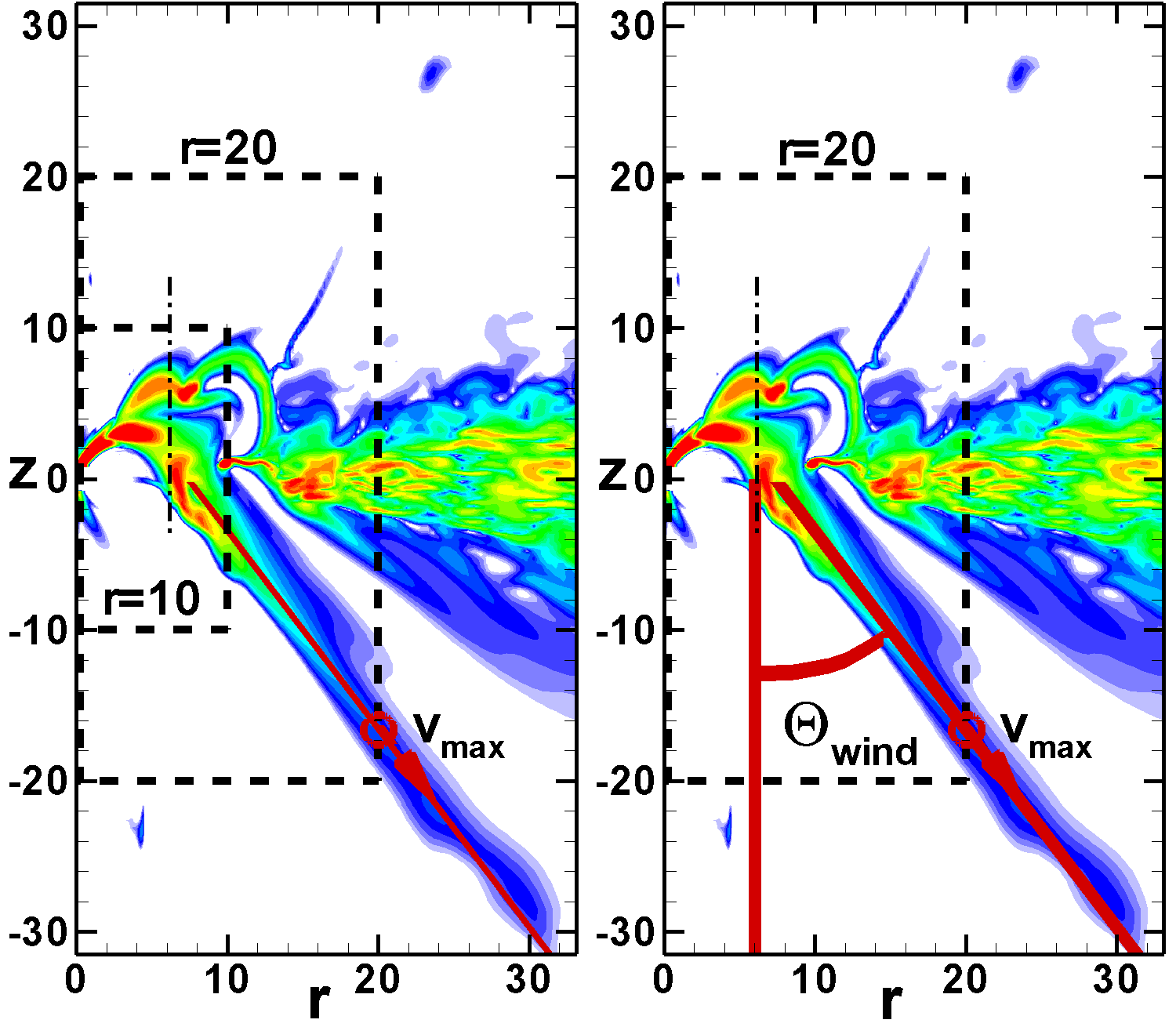}
\caption{Left panel shows the point on the outer box $(r=20,z=\pm
20)$ where the wind velocity
is at its maximum value, $v_{\rm max}$. Right panel shows the
opening angle of the wind, $\Theta_{\rm wind}$, which is
defined as the angle between the vertical line crossing the
magnetospheric radius $r_m$ and
the line connecting $r_m$ at $z=0$ to the $v_{\rm max}$ point.}
\label{2d-sketch}
\end{figure}


\begin{figure*}[ht!]
\centering
\includegraphics[width=16.0cm,clip]{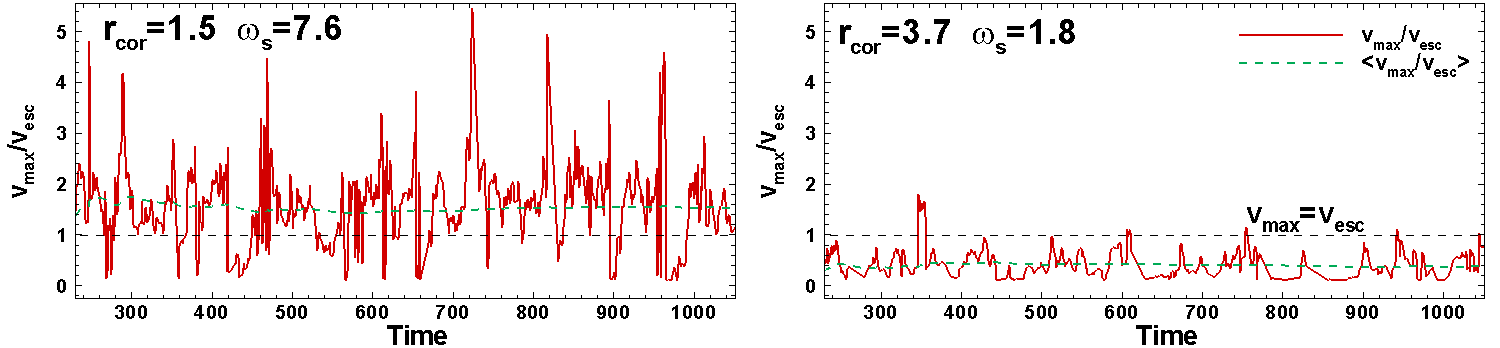}
\includegraphics[width=16.0cm,clip]{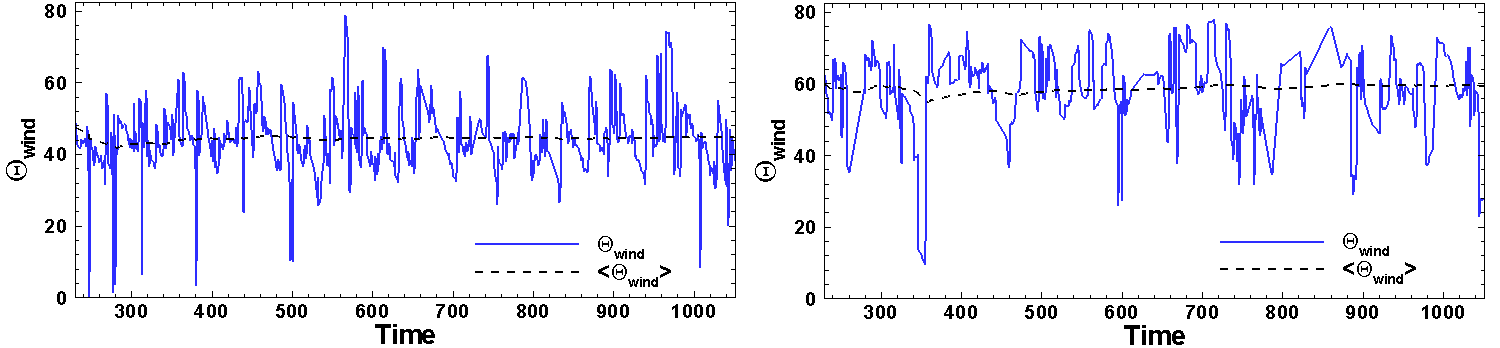}
\caption{\textit{Top panels:} Variation of the maximum wind
velocity
normalized to the local escape velocity, $v_{\rm max}/v_{\rm
esc}$, in the cases of a strong propeller (left panel, model
$\mu60c1.5$) and a weak propeller (right panel, model
$\mu60c3.7$). The dashed line shows the time-averaged value
$\langle v_{\rm max}/v_{\rm esc}\rangle$. \textit{Bottom panels:}
Same, but for the opening angle $\Theta_{\rm wind}$. }
\label{vmax-theta}
\end{figure*}

\begin{figure*}[ht!]
\centering
\includegraphics[width=8cm,clip]{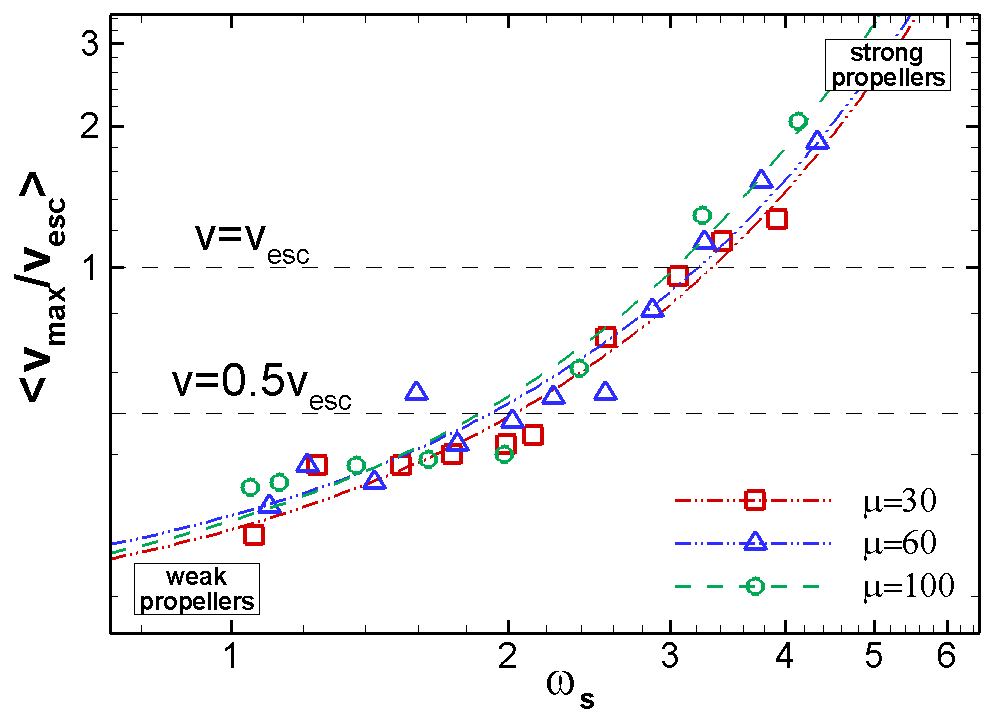}
\includegraphics[width=8cm,clip]{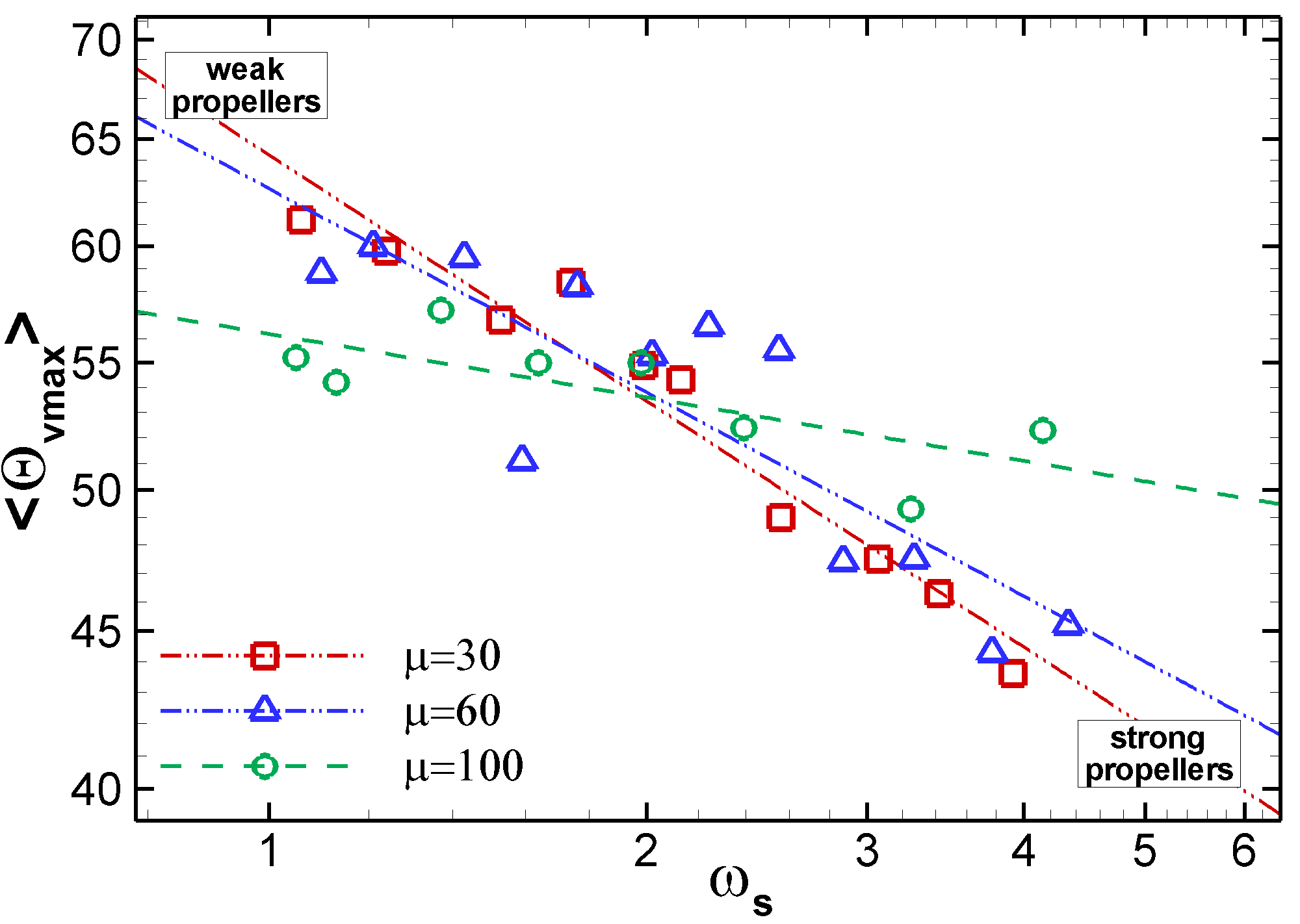}
\caption{\textit{Left panel:} The dependence of the time-averaged
maximum wind velocity
normalized to the local escape velocity, $\langle v_{\rm
max}/v_{\rm esc} \rangle$, on the fastness parameter $\omega_s$
for all simulation runs. Red squares, blue triangles and green
circles correspond to models with magnetospheric parameters
$\mu=30, 60$ and $100$, respectively.
\textit{Right panel:} Same, but for the time-averaged opening
angle $\langle \Theta_{\rm wind} \rangle$.} \label{vmax-theta-av}
\end{figure*}

\subsection{Velocities in the wind}

\label{subsec:velocity}

We investigated the velocities of matter in the wind component of
propellers of different strengths. Most of the matter flows from
the disk-magnetosphere boundary into the conical wind. To find the
typical velocity of matter flow in each model, we chose a surface
$S(r=20,z=\pm20)$,  which is a cylinder with dimensions $r=20$ and
$z=\pm 20$ (see Fig. \ref{2d-sketch}) \footnote{Note, that this
cylinder is larger than that used for the calculation of matter
fluxes, because matter in the wind often has low velocities at the
disk-magnetosphere boundary, but is accelerated at larger
distances from the star due to the magnetic force}, and searched
for the maximum poloidal velocity $v_{\rm max}$ at this surface.
The maximum velocity was calculated for the parts of the wind
where the density is not very low ($\rho>0.001$), so as to
deselect the regions of very low density and high velocity flow in
the axial regions. We also deselected the matter which moves at
low velocities in the disk by the condition  $v > v_{\rm min}=0.1
v_{\rm esc}$. We calculated the ratio $v_{\rm max}/v_{\rm esc}$,
which shows whether the maximum poloidal velocity in the wind is
larger or smaller than the local escape velocity $v_{\rm esc}$.

Fig. \ref{vmax-theta} (top panels) shows variation of the ratio
$v_{\rm max}/v_{\rm esc}$ with time in a strong (left panel) and
weak (right panel) propeller. One can see that in the case of a
strong propeller, this ratio varies in the range of $v_{\rm
max}/v_{\rm esc}\approx 2-5$ during the bursts. In the case of a
relatively weak propeller, the maximum velocity during the bursts
is either slightly larger or slightly smaller than the escape
velocity, so that $v_{\rm max}/v_{\rm esc} \approx 1$. Some of
this matter
escapes the star's gravity, while some of it returns back to the
star or falls onto the disk at some distance from the star. In
even weaker propellers, the maximum velocity in the wind is lower
than the escape velocity, so that matter will fall back onto the
star. Alternatively, it may contribute to the slowly-expanding
turbulent magnetic corona. The left panels of Figs.
\ref{app:vmax-theta-d100}, \ref{app:vmax-theta-d60}, and
\ref{app:vmax-theta-d30} {from \ref{appen:variation} show the
variation of $v_{\rm max}/v_{\rm esc}$ with time for some of the
representative models.

To characterize the velocity of the outflows in each model, we
calculated the time-averaged maximum velocity
\begin{equation}
\langle v_{\rm max}/v_{\rm esc}\rangle = \frac{\int_{t_i}^{t} dt'
v_{\rm max}(t')/v_{\rm esc}}{\int_{t_i}^{t} dt'} ~.
\label{eqn_timeavg-vmax}
\end{equation}
These averaged velocities are approximately 2-3 times lower than
the maximum velocities during the bursts to the wind (compare the
dashed lines in top panels of Fig. \ref{vmax-theta} with the
velocity maxima). For each model, we take the averaged velocity at
time $t=1,000$ and use it for comparisons with other models.


Fig. \ref{vmax-theta-av} (left panel) shows
 that the averaged
velocity increases with fastness exponentially (see Tab.
\ref{tab:velocity-angle} for dependencies). Note that the
dependencies are approximately the same for magnetospheres of
different sizes (different values of $\mu$).

\subsection{Opening Angle of the wind}
\label{subsec:opening-angle}

We also calculated the opening angle of the wind, $\Theta_{\rm
wind}$, which we determined as the angle between the line
connecting the inner disk to the point of maximum wind velocity
($v=v_{\rm max}$, located at the surface $S(r=20,z=\pm20)$) and
the vertical line crossing the inner disk, $r=r_m$ (see right
panel of Fig. \ref{2d-sketch}). Simulations show that, in both the
strong and weak propellers, the opening angle strongly oscillates
(see Fig. \ref{vmax-theta}, bottom panels). This angle is smaller
in the cases of stronger propellers. Figures
\ref{app:vmax-theta-d100}, \ref{app:vmax-theta-d60}, and
\ref{app:vmax-theta-d30} from \ref{appen:variation} show the
variation of $\Theta_{\rm wind}$ with time in our representative
models.

We calculated the time-averaged value of the opening angle for
each model:
\begin{equation}
\langle \Theta_{\rm wind}\rangle = \frac{\int_{t_i}^{t} dt'
\Theta_{\rm wind}}{\int_{t_i}^{t} dt'} ~. \label{eqn:Theta-avg}
\end{equation}
The dashed horizontal lines in Fig. \ref{vmax-theta} (bottom
panels) show the time-averaged opening angles in our sample cases
of strong and weak propellers. One can see that the time-averaged
angle $\langle \Theta_{\rm wind}\rangle$ is approximately
$45^\circ$ and $60^\circ$ in our examples of strong and weak
propellers, respectively.

Fig. \ref{vmax-theta-av} (right panel) shows the dependence of the
time-averaged opening angles, $\langle \Theta_{\rm wind}\rangle$,
taken for all models (at $t=1,000$), on the fastness parameter,
$\omega_s$. One can see that $\langle \Theta_{\rm wind}\rangle$
decreases with
$\omega_s$. The dependencies can be approximated by a power law
(see Tab. \ref{tab:velocity-angle}). One can see that the
dependencies are similar for $\mu=30$ and $\mu=60$. However, in
the models with the largest magnetospheres, $\mu=100$, the slope
is not as steep as in the other two cases.

The opening angle is large, $\langle \Theta_{\rm
wind}\rangle\approx 60^\circ-65^\circ$, in the weak propellers.
Velocities of outflows into the wind are also lower in the weak
propellers, and, during the outbursts, can be comparable to or
lower than the escape velocity. This wind matter may fall back to
the disk at some distance from the star. In the weak propeller
regime, a significant amount of matter may be recycled
through the process of ejection from the inner disk boundary, the
fall of this matter onto the disk at larger distances from the
star, and subsequent inward accretion in the disk.

\begin{table}
\centering
\begin{tabular}{llll}

\\         & $\mu=30$ & $\mu=60$ & $\mu=100$ \\
\hline
$\langle v_{\rm max}/v_{\rm esc}\rangle$ & $0.16 e^{0.55\omega_s}$ & $0.17 e^{0.55\omega_s}$ & $0.16 e^{0.61\omega_s}$  \\
\hline
$ \langle\Theta_{\rm wind}\rangle$       & $64.3 \omega_s^{-0.27}$ & $62.6 \omega_s^{-0.22}$ & $56.2 \omega_s^{-0.07}$  \\
\hline

\end{tabular}
\caption{Maximum velocity and
opening angle of the wind as a function of the fastness parameter,
$\omega_s$, for different values of magnetospheric parameter
$\mu$. The maximum velocity is calculated at the surface
$S(r=20,z=\pm20)$ and the condition $v_p > 0.1 v_{\rm esc}$. }
\label{tab:velocity-angle}
\end{table}

\begin{figure*}[ht!]
\centering
\includegraphics[width=16cm,clip]{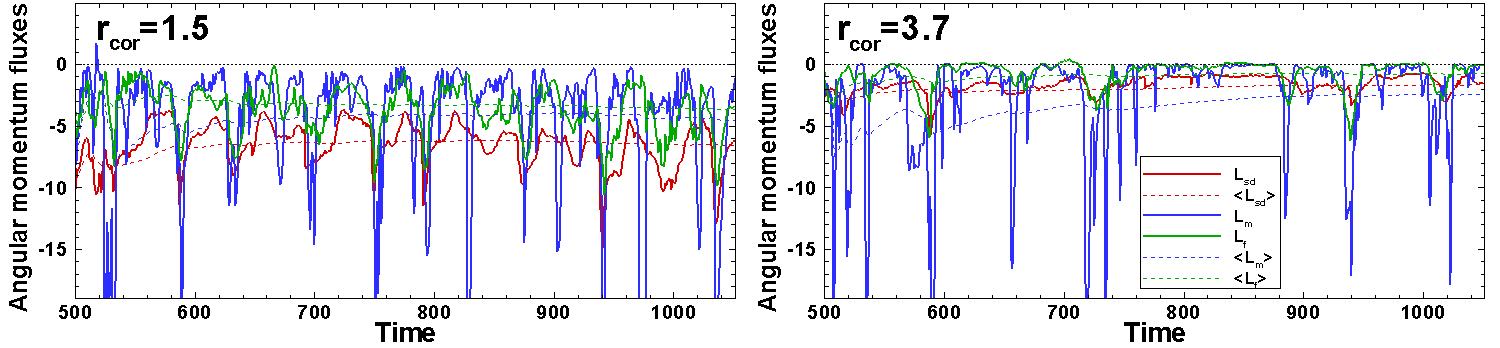}
\caption{Angular momentum fluxes carried by the magnetic field
from the star, $\dot L_{\rm sd}$ (in red), and to the outflows,
carried by matter, $\dot L_m$ (in blue), and by the field,  $\dot
L_f$ (in blue) in the cases of strong (model $\mu60c1.5$, left
panel) and weak (model $\mu60c3.7$, right panel) propeller
regimes. Outflows were calculated through surface
$S(r=10,z=\pm10)$ for velocities $v>0.1v_{\rm esc}$. The dashed
lines show the values of fluxes averaged in time.}
\label{angmom-fluxes-2}
\end{figure*}

\begin{figure*}[ht!]
\centering
\includegraphics[width=16cm,clip]{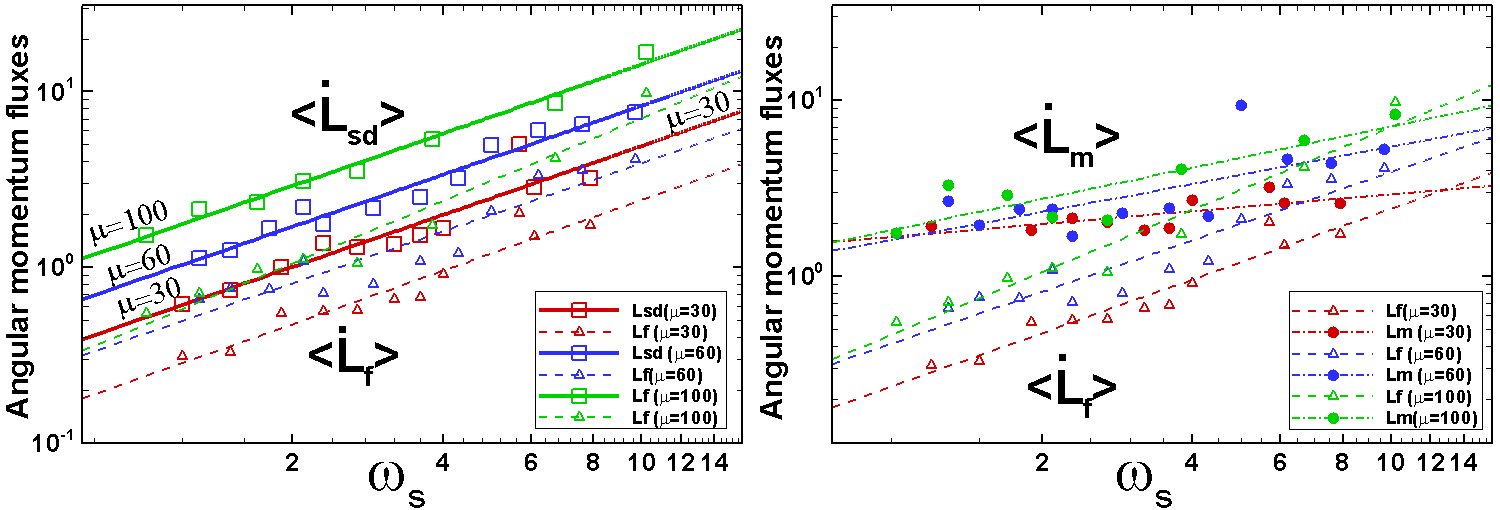}
\caption{\textit{Left panel:} Time-averaged angular momentum
fluxes carried by the magnetic field from the star, $\langle\dot
L_{\rm sd}\rangle$ and carried by the magnetic field $\dot L_f$
through the surface $S(r=10,z=\pm10)$. \textit{Right panel:}
Angular momentum flux carried by the matter to the wind,
$\langle\dot L_m\rangle$ through the same surface. Angular
momentum flux carried by magnetic field, $\langle\dot L_f\rangle$,
is also shown for reference. Red, blue and green symbols and lines
correspond to models with $\mu=30, 60$ and $100$, respectively.}
\label{fluxes-angmom-2}
\end{figure*}

\section{Angular momentum and energy}

In the propeller regime, a star-disk system loses angular momentum
and energy.

\subsection{Angular momentum flow and the spin-down rate}
\label{subsec:angmom-fluxes}

In the propeller regime, a star loses its angular momentum and
spins down (e.g., \citealt{LovelaceEtAl1999}). Angular momentum
flows from the surface of the star along the field lines
connecting the star with the disk and the corona. In addition,
angular momentum flows from the inner parts of the accretion disk
along the open field lines of the dipole, which have foot-points
at the disk.

 The angular momentum flux is
  calculated by
integrating the angular momentum flux densities through some
surface $S$:
\begin{equation}
\ldot = \ldot_{m} + \ldot_{f} = \int_S d{\mathbf S} \cdot
(\flux_{\rm Lm}+\flux_{\rm Lf}) ~,
\end{equation}
where $\flux_{\rm Lm}$ and $\flux_{\rm Lf}$ are the angular
momentum flux densities carried by matter and magnetic field,
respectively, and given by
\begin{eqnarray} \label{eqn_angmom}
\flux_{\rm Lm} &=&  r \rho v_\phi  {\bf v}_p~, \quad\quad
\flux_{\rm Lf}= - r \frac{B_\phi {\mathbf B}_p}{4 \pi}~.
\end{eqnarray}
Here, the normal vector to the surface $d{\mathbf S}$ points
inward towards the star. To estimate the rate of stellar
spin-down, $\dot L_{\rm sd}$, we calculated the angular momentum
flux through the surface of the star, $(r=R_\star,z=\pm R_\star)$
. We observed that the angular momentum flux is carried from the
stellar surface by the magnetic field \footnote{We forbid the
outflow of matter from the stellar surface and therefore the flux
carried by matter is zero.}. The red lines in Fig.
\ref{angmom-fluxes-2} show variation of this flux in the cases of
strong and weak propellers. We also calculated the fluxes carried
by matter, $\dot L_m$, and by the field, $\dot L_f$, through
surface $S(r=10, z=\pm10)$ \footnote{Here, we place the surface
$S$ close to the inner disk, so that to take into account angular
momentum which flows back to the disk along the closed field
lines.}. Fig. \ref{angmom-fluxes-2} shows these fluxes in blue and
green,
respectively. All fluxes strongly vary with time. Comparisons with
the matter fluxes
show that angular momentum fluxes are the largest during episodes
of matter outflow.

To compare the fluxes calculated for different models, we
calculated the time-averaged values using a formula similar to Eq.
\ref{eqn_timeavg}. We calculated separately the fluxes of angular
momentum carried from the surface of the star,
 $\langle \dot L_{\rm sd}\rangle$, and the fluxes carried  through surface
  $S(r=10,z=\pm10)$   by the magnetic field, $\langle
  \dot L_{f}\rangle$, and by matter, $\langle
  \dot L_{m}\rangle$. Fig. \ref{fluxes-angmom-2} shows these
  fluxes (taken at $t=1,000$) as a function of the fastness
  parameter, $\omega_s$. The dependencies can be
  approximately described by power laws
  (see solid lines in Fig. \ref{fluxes-angmom-2} and dependencies
  in Tab. \ref{tab:angmom}). One can see that the angular momentum
fluxes are larger at larger values of $\mu$. The left panel of
Fig. \ref{fluxes-angmom-2} and Tab. \ref{tab:angmom} show that, in
the models with the same values of $\mu$, the angular momentum
carried from the star, $\langle\dot L_{\rm sd}\rangle$, is
approximately twice as large as the angular momentum carried by
the field,  $\langle\dot L_{f}\rangle$, through surface
$S(r=10,z=\pm10)$. This means that only a
  part of the angular momentum flows from the star
  to the inflated field lines. The other part
  (approximately half of $\langle \dot L_{\rm sd}\rangle$) flows into the disk along the closed field lines \footnote{This result is in agreement with
  that obtained by \citet{UstyugovaEtAl2006}.}.

  In the weak propellers, magnetic angular momentum flux $\langle
  \dot L_{f}\rangle$ is associated with the inflation of the field lines and the outward propagation of magnetic flux.
  However, in the
  strong propellers, the magnetic flux acquires the form of a magnetic (Poynting flux) jet, where
  magnetic pressure accelerates the low-density plasma to high velocities inside the simulation region.
This jet is magnetically-driven and also magnetically-collimated.
The matter component of the flux through surface $S(r=10,z=\pm
10)$, $\langle
  \dot L_{m}\rangle$,
  is associated with the centrifugally-driven conical component of the wind coming from the
inner disk.  The right panel of Fig. \ref{fluxes-angmom-2} shows
that, at small values of $\omega_s$, the angular momentum flux
associated with matter,  $\langle
  \dot L_{m}\rangle$, is larger than that
associated with the field, $\langle
  \dot L_{f}\rangle$. However, at large values of $\omega_s$ they become
  comparable. In summary, the magnetic field carries angular momentum away from the star,
  while both matter and magnetic field carry angular momentum away from the star-disk system.

\begin{table}
\centering
\begin{tabular}{llll}

\\ Ang. mom. flux  & $\mu=30$ & $\mu=60$ & $\mu=100$ \\
\hline
$\langle\dot L_{\rm sd}\rangle$ & $0.51 \omega_s^{0.99}$ & $0.87 \omega_s^{0.98}$ & $1.47 \omega_s^{0.99}$  \\
\hline
$\langle\dot L_{f}\rangle$ & $0.24 \omega_s^{1.01}$ & $0.41 \omega_s^{0.98}$ & $0.47 \omega_s^{1.18}$ \\
\hline
$\langle\dot L_{m}\rangle$ & $1.68 \omega_s^{0.24}$ & $1.61 \omega_s^{0.53}$ & $1.84 \omega_s^{0.59}$\\
\hline

\end{tabular}
\caption{Angular momentum fluxes as a function of fastness,
$\omega_s$, for different values of magnetospheric parameter
$\mu$. $\langle\dot L_{\rm sd}\rangle$ is the time-averaged
angular momentum flux carried from  the surface of the star.
$\langle\dot L_m\rangle$ and $\langle\dot L_f\rangle$ are the
time-averaged angular momentum fluxes carried by matter and by the
field, respectively, through surface $S(r=10,z=\pm10)$ and
directed away from the star. } \label{tab:angmom}
\end{table}

\subsection{Energy fluxes}
\label{subsec:energy-fluxes}

Propeller-driven winds and jets also carry energy out of the
system.
We calculated the energy fluxes carried by matter and magnetic
field through surface $S(r=10,z=\pm10)$:

\begin{equation}
\dot  E = \dot E_m + \dot E_f = \int_S d{\mathbf S} \cdot
(\flux_{\rm Em}+\flux_{\rm Ef}) ~,
\end{equation}
where $\flux_{\rm Em}$ and $\flux_{\rm Ef}$ are the energy flux
densities carried by matter and magnetic field, given by
\begin{eqnarray} \label{eqn_energy}
\flux_{\rm Em} & = &  (\frac{\rho v^2}{2} + \frac{\gamma
p}{\gamma-1}
) {\bf v}_p~, ~\nonumber\\
\flux_{\rm Ef} & = & \frac{1}{4 \pi} \left( {\mathbf B}^2 {\bf
v}_p - ({\mathbf B} \cdot {\bf v}) {\mathbf B}_p \right)~.
\end{eqnarray}
Here, the normal vector to the surface $d{\mathbf S}$ points
inward towards the star. Fig. \ref{energy-fluxes-2} shows an
example of the temporal variations of energy fluxes in strong and
weak propellers. One can see that the energy fluxes strongly vary
in time, that is, energy is ejected into the outflows in the form
of bursts.

 Fig. \ref{energy-fluxes-all} shows the
time-averaged energy fluxes $\langle \dot E_m\rangle$ and $\langle
\dot E_f\rangle$ (taken at $t=1,000$) for all models. One can see
that the fluxes increase with $\omega_s$ and are larger at larger
values of $\mu$. Table \ref{tab:energy} shows the power law
dependencies for fluxes
at different values of $\mu$.

In both the strong and weak propellers, some energy is carried by
matter from the inner disk into a conically-shaped wind.
Additionally, in both cases, inflation of the field lines leads to
the flow of magnetic energy out of the star. However, in
the strong propellers, magnetic energy also flows into a
non-stationary, magnetically-driven and magnetically-collimated
low-density jet. Fig. \ref{2d-energy-2} shows the distribution of
energy flux density in a strong propeller (model $\mu60c1.5$). The
left panel shows that the energy carried by matter flows into the
conically-shaped wind. The right panel shows the energy flux
density associated with the magnetic field. One can see that this
energy flux is large and is more collimated than the matter energy
flux. This is the magnetically-dominated (Poynting flux) jet,
where matter is accelerated and collimated by the magnetic field.

\begin{figure*}[ht!]
\centering
\includegraphics[width=16cm,clip]{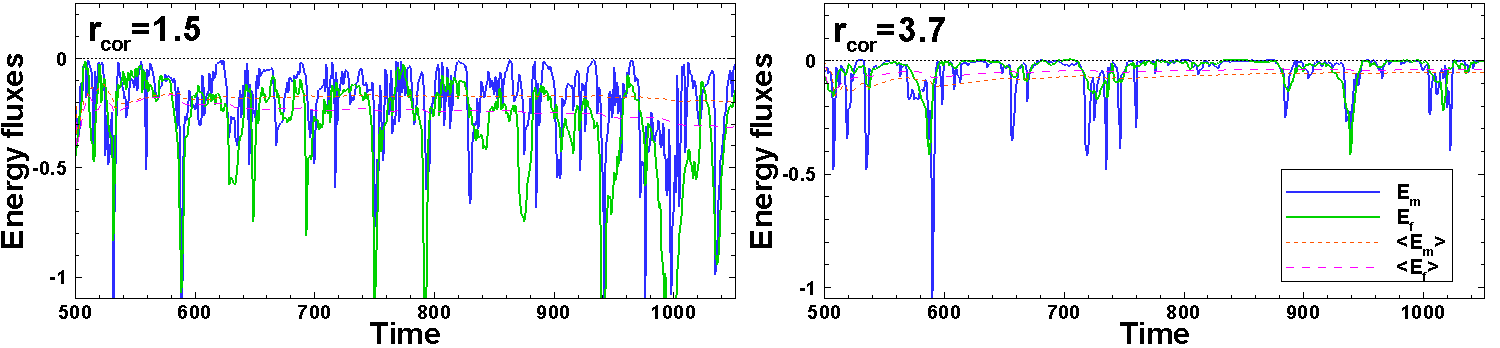}
\caption{Energy fluxes associated with matter (in green) and with
the field (in blue) in the cases of strong (model $\mu60c1.5$,
left panel) and weak (model $\mu60c3.7$, right panel) propeller
regimes. Outflows were calculated through surface $S(r=10,
z=\pm10)$ for velocities $v>0.1v_{\rm esc}$. Dashed lines show the
values of fluxes averaged in time.} \label{energy-fluxes-2}
\end{figure*}



\begin{table}
\centering
\begin{tabular}{llll}

\\ Energy flux & $\mu=30$ & $\mu=60$ & $\mu=100$ \\
\hline
$\langle \dot E_{\rm f}\rangle$ & $0.012e^{0.37\omega_s}$ & $0.021e^{0.35\omega_s}$ & $0.020e^{0.39\omega_s}$ \\
\hline
$\langle\dot E_{\rm m}\rangle$ & $0.029e^{0.19\omega_s}$ & $0.031e^{0.24\omega_s}$ & $0.033e^{0.29\omega_s}$\\
\hline

\end{tabular}
\caption{Energy fluxes through surface $S(r=10,z=\pm10)$ as a
function of fastness, $\omega_s$. $\langle\dot E_f\rangle$ and
$\langle\dot E_m\rangle$ are the energy fluxes carried by the
magnetic field and matter, respectively. Only the components of
energy directed away from the star are taken into account.}
\label{tab:energy}
\end{table}

\section{Summary of wind properties in strong and weak
propellers} \label{subsec:summary-weak-strong}

To summarize the specifics of matter flow in propellers of
different strengths, we show two sketches that demonstrate the
properties of strong and weak propellers during accretion/outburst
events (see left and right panels of Fig.
\ref{sketch-strong-weak}).

\begin{itemize}

\item In the strong propeller regime (left panel), matter flows
from the inner disk into the disk wind along the open field lines
connecting the disk with the corona. This wind has high
super-escape velocities and relatively small opening angles,
$\langle\Theta_{\rm wind}\rangle\approx 40^\circ-45^\circ$. Such a
wind
may flow to large distances from the star, forming large-scale
wind structures. Alternatively, it may be collimated by the
external medium, forming a jet. There is also a low-density,
high-velocity, magnetically-dominated and magnetically-driven
Poynting flux jet,
whose matter flows along the stellar
field lines.
This jet
carries a significant amount of
energy and angular momentum away from the star. In a typical case
of non-stationary ejections, shock waves are expected to form
along  the flow, where particles
may be accelerated to high energies. In summary, a strong,
two-component outflow is expected in strong propellers.

\item In the weak propeller regime (right panel), matter flows
from the inner disk into the low-velocity wind, where the maximum
velocity is comparable with or lower than the local escape
velocity. Matter flows into
the conical wind at large opening angles, $\langle\Theta_{\rm
wind}\rangle\approx 60^\circ-70^\circ$. This wind partly forms the
large-scale outflow structures, and partly falls back onto the
disk at some distance from the star. The fallen matter then
accretes back towards the disk-magnetosphere boundary and
is again ejected into the slow, conically-shaped wind.
Such recycling of inner disk matter is expected in many weak
propellers. In addition, inflation and reconnection of the field
lines lead to the formation of magnetic islands, which are ejected
at low, sub-escape velocities,
forming the slow, magnetically-dominated wind. Some of this matter
may accrete back to the star, driven by gravitational force.


\end{itemize}

\begin{figure}[ht!]
\centering
\includegraphics[width=8cm,clip]{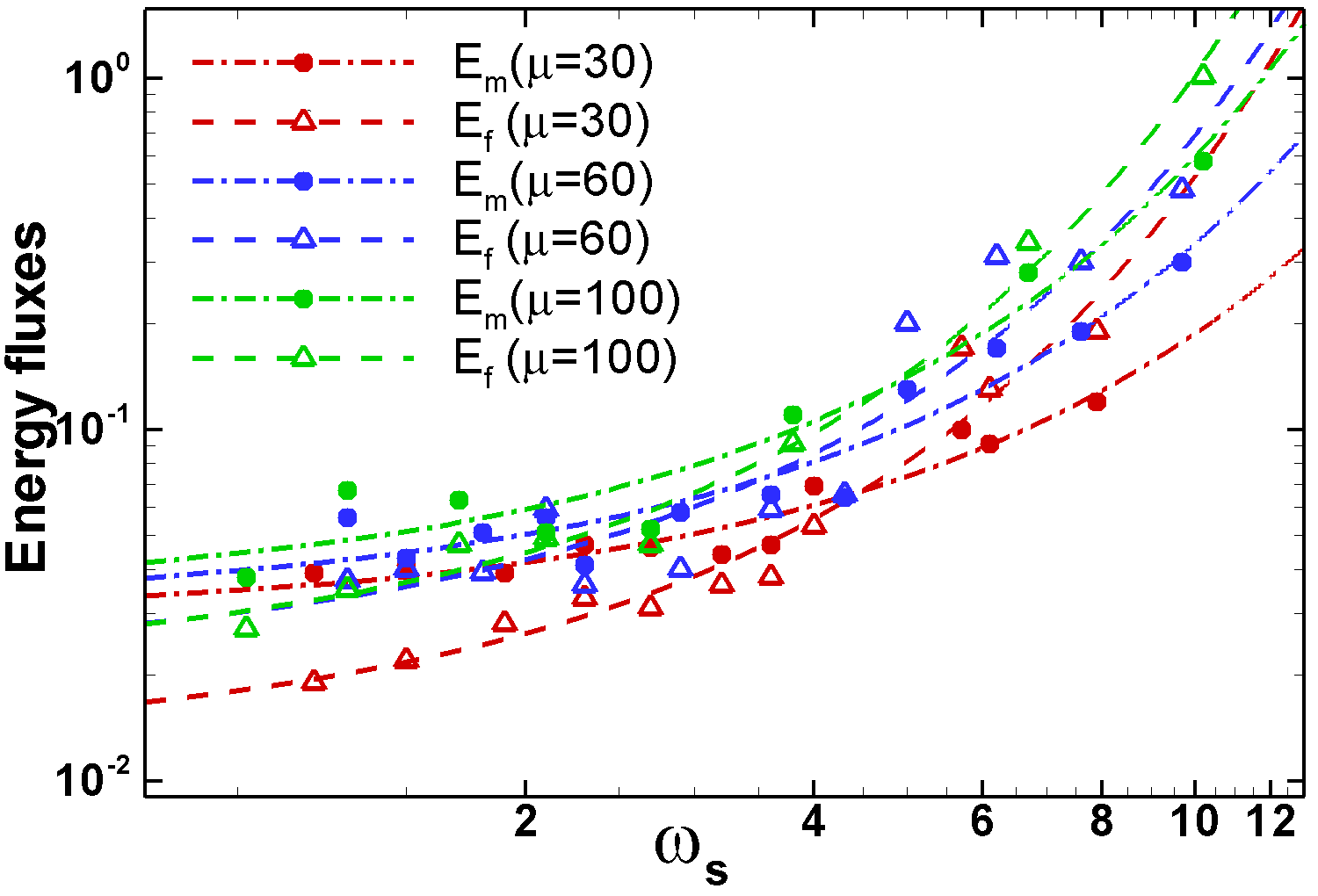}
\caption{Time-averaged energy fluxes carried away from the star by
matter, $\langle\dot E_m\rangle$, and by the magnetic field,
$\langle\dot E_f\rangle$, calculated through surface
$S(r=10,z=\pm10)$. Red, blue and green symbols and lines
correspond to models with $\mu=30, 60$ and $100$, respectively.}
\label{energy-fluxes-all}
\end{figure}

\begin{figure}[ht!]
\centering
\includegraphics[width=8cm,clip]{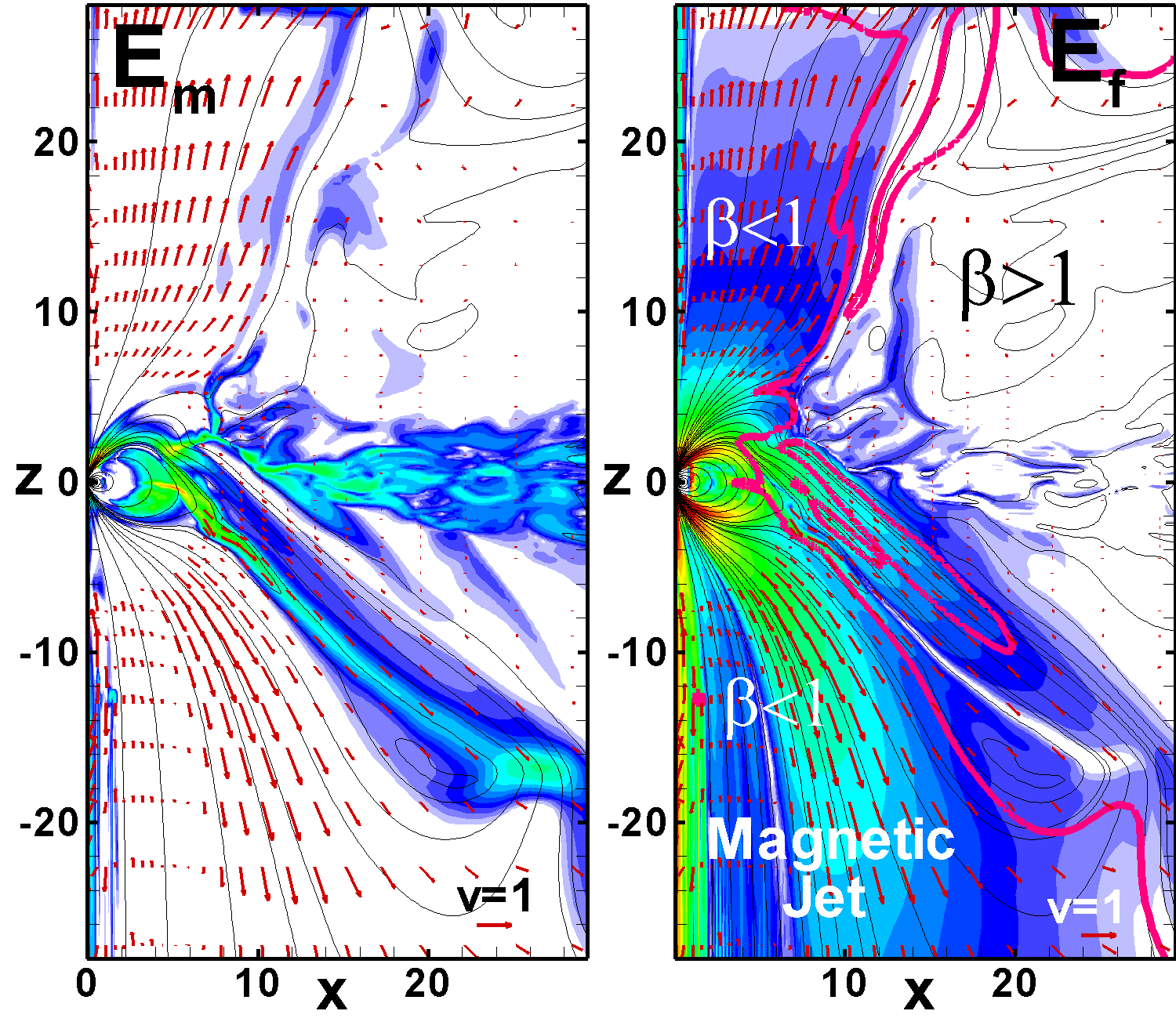}
\caption{Color background shows the energy flux densities
associated with matter (left panel) and the field (right panel) in
the model $\mu60c1.5$ at $t=598$. The thick red line shows the
$\beta=1$ line, which separates the magnetically-dominated region
($\beta<1$) from the matter-dominated region ($\beta>1$).}
\label{2d-energy-2}
\end{figure}


We should note that, in both the strong and weak propellers,
matter accretes (and is ejected) during brief episodes, and the
inner disk strongly oscillates. Therefore, in both cases, strong
variability in the light curves is expected.


\begin{figure*}[ht!]
\centering
\includegraphics[width=8cm,clip]{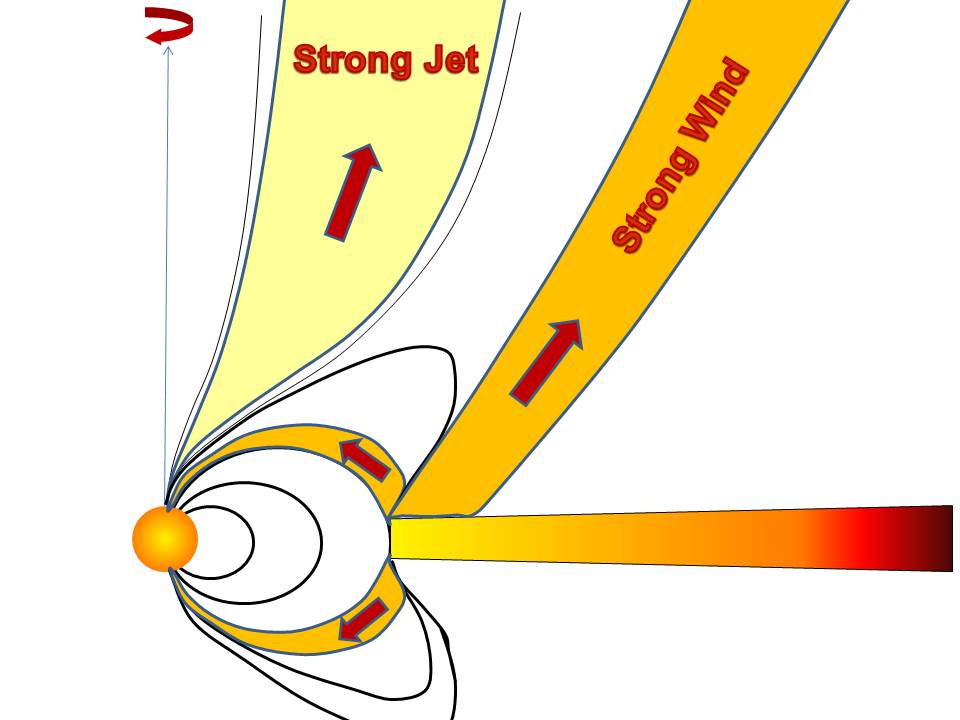}
\includegraphics[width=8cm,clip]{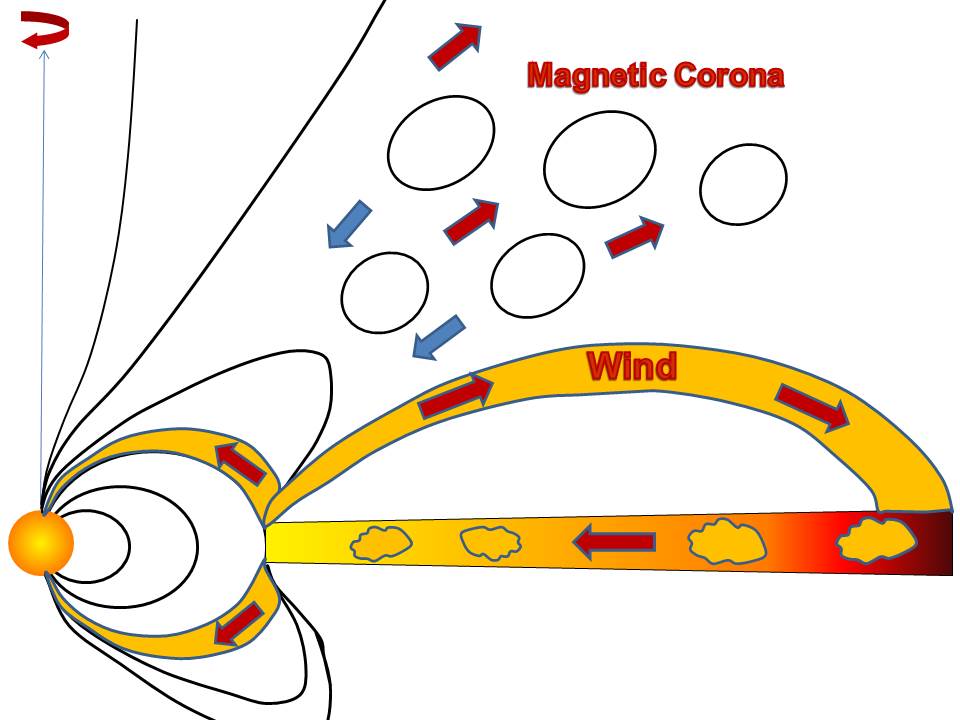}
\caption{\textit{Left panel:} A sketch of matter flow in the
strong propeller regime during an accretion/ejection event, where
matter (1) partly accretes onto the star, (2) partly flows into
the high-velocity wind, and (3) partly flows into the low-density,
high-velocity axial jet. \textit{Right panel:} A sketch of matter
flow in the weak propeller regime during an accretion/ejection
event, where (1) matter partly flows into the low-velocity,
high-opening angle wind which may fall back to the disk, and (2)
partly into the turbulent corona, where the magnetic islands form
as a result of the inflation/reconnection, but return back due to
gravity.} \label{sketch-strong-weak}
\end{figure*}

\section{Time intervals between accretion/ejection events}
\label{sec:intervals-time}

Here, we estimate the characteristic time intervals between
accretion/ejection events. Simulations show that
accretion/ejection events are typically associated with a cycle in
which (1) Matter accumulates at the inner disk and slowly moves
inward, (2) Matter of the inner disk penetrates through the
external regions of the magnetosphere and the field lines begin to
inflate,  (3) Matter partly accretes onto the star and partly
flows into the wind, (4) The magnetosphere expands and the cycle
repeats. An outflow becomes possible when the field lines inflate
and open.


Here, we consider two possible scenarios: (1)
The diffusivity at the disk-magnetosphere boundary is relatively
high, and matter penetrates through the disk-magnetosphere
boundary rapidly (we observe this in most of our simulation runs);
(2)
The diffusivity is low, so that the inner disk matter gradually
penetrates through the magnetosphere.

\subsection{High Diffusivity Scenario}

Let's suggest that, after an event of accretion/ejection, the
magnetosphere is ``empty" and the inner disk is located at some
radius $r_m$. Then, matter of the inner disk gradually penetrates
through the external layers of the magnetosphere due to some 3D
instabilities. The depth of the penetration is unknown. However,
we can suggest that this matter penetrates into the magnetosphere
up to some depth $\Delta r$ and then stops at some distance from
the star due to an even stronger centrifugal barrier of the
propelling star \footnote{This scenario is observed in most of our
simulation runs.} In parallel, new matter comes in from the disk
to its inner parts with an accretion rate $\dot M$. It carries the
angular momentum flux
\begin{equation} \dot L_{\rm m}=r_m^2 (\Omega_\star
- \Omega_d) \dot M . \label{eq:time-int-dotL_m}
\end{equation}


The dipole magnetic field of the star inflates due to the
difference between the angular velocity of the star, $\Omega_*$,
and the disk, $\Omega_d$. Angular momentum flows from a unit
length of the disk at radius $r$ to the inflating field lines. Its
value (per unit length) is:
\begin{equation}
L_f = \frac{B_m^2 r_m^2} {(\Omega_* - \Omega_d)}~,
\label{eq:time-int-L_f}
\end{equation}
where $B_m$ is the magnetic field at $r=r_m$.
Inflation occurs when the angular momentum of matter in the disk
is larger than the angular momentum required for inflation:
\begin{equation}
r_m^2 (\Omega_\star - \Omega_d) \dot M \Delta t > \frac{B_m^2
r_m^2} {(\Omega_* - \Omega_d)}
\Delta r~. \label{eq:time-int-L_m-eq-Lf}
\end{equation}
Therefore, the next episode of inflation will occur after an
interval of time $\Delta t$, if
\begin{equation}
\Delta t > \frac{B_m^2 \Delta r}{\dot M (\Omega_\star -
\Omega_d)^2} ~. \label{eq:Delta-t-high-diff}
\end{equation}
Taking into account the fact that $B_m=B_\star (R_\star/r_m)^3$
and definition of fastness, $\omega_s=\Omega_s/\Omega_d$, we
obtain:
\begin{equation}
\Delta t > \frac{\mu_\star^2 \Delta r}{\dot M r_m^3 G M_\star
(\omega_s - 1)^2} ~. \label{eq:Delta-t-high-diff}
\end{equation}

To find the
characteristic time interval between inflation events in our
simulations, we re-write this condition in dimensionless form
using the dimensionalization procedure from  \ref{app:reference
units}: $\Delta r = R_0 \Delta \widetilde{r}$, $\Delta t = P_0
\widetilde{\Delta t}$ (where $P_0=2\pi t_0$ is the period of
rotation at $r=R_0$), etc., and obtain Eq.
\ref{eq:Delta-t-high-diff} in dimensionless form:
\begin{equation}
\Delta t > \frac{\mu^2 \Delta r}{2\pi \dot M r_m^3 (\omega_s -
1)^2} ~. \label{eq:Delta-t-high-diff-dim}
\end{equation}
Here, we take into account the fact that
$B_\star/B_0=\widetilde{\mu} (R_\star/R_0)^3=\widetilde{\mu}$ and
remove all tildes above the dimensionless variables.


\begin{figure}[ht!]
\centering
\includegraphics[width=8cm,clip]{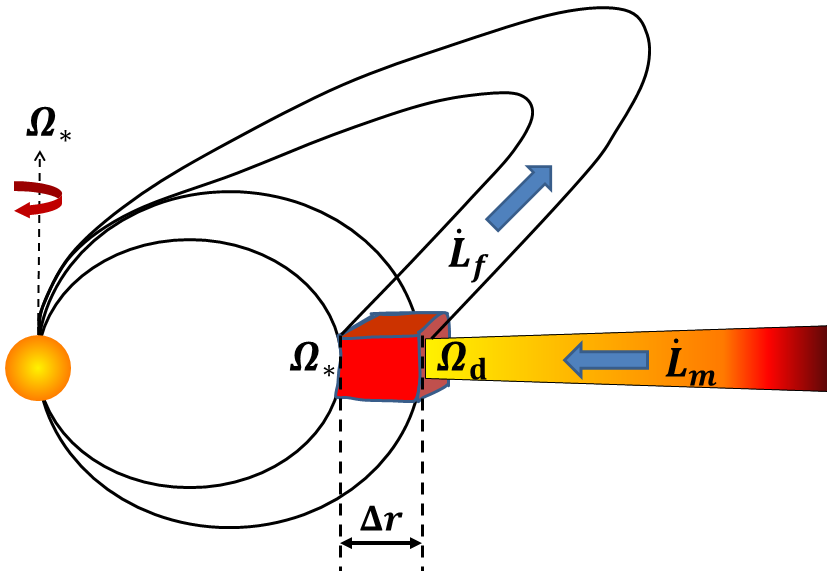}
\caption{The sketch shows inflation of the field lines threading a
ring with the radial width of $\Delta r$. A star and its
magnetosphere rotate with angular velocity $\Omega_\star$. Matter
of the inner disk rotates with angular velocity $\Omega_d$. Matter
of the inner disk brings in angular momentum
at a rate of $\dot L_m$. The inflating field lines carry angular
momentum $\dot L_f$ away from the star.} \label{sketch-funnel}
\end{figure}

\subsection{Low Diffusivity Scenario}

At a low diffusivity rate, matter of the inner disk slowly
penetrates through the external layers of the magnetosphere (in
the direction of the star),
and the depth of penetration is proportional to the time interval
$\Delta t$:  $\Delta r = \sqrt{\eta_m \Delta t}$, where $\eta_m$
is the diffusivity coefficient. During this time interval $\Delta
t$, matter accretes towards the inner disk and is accumulated in
the amount of $\dot M \Delta t$. This matter carries the angular
momentum flux described
by Eq. \ref{eq:time-int-dotL_m}, and the inflating field lines
(which thread the ring with width $\Delta r$) carry away the
angular momentum flux described by Eq. \ref{eq:time-int-L_f}.
Inflation becomes possible when the angular momentum flux carried
by matter becomes larger than the angular momentum flux carried by
the field:
\begin{equation}
r_m^2 (\Omega_\star - \Omega_d) \dot M \Delta t > \frac{B_m^2
r_m^2} {(\Omega_\star - \Omega_d)} \sqrt{\eta_m \Delta t}~.
\end{equation}
Therefore, the next episode of inflation
occurs after an interval of time $\Delta t$, if
\begin{equation}
\Delta t > \eta_m \frac{\mu^4 }{2\pi\dot M^2 r_m^6 (\omega_s -
1)^4} ~. \label{eq:Delta-t-low-diff-dim}
\end{equation}
Here, we have already converted the time interval to dimensionless
units and removed the tilde's.

One can see that the time interval is proportional to the
diffusivity coefficient, $\eta_m$, and all the variables that
$\Delta t$ is dependent on have coefficient powers that are twice
as high as those in the high-diffusivity scenario (see Eq.
\ref{eq:Delta-t-high-diff-dim}).

\subsection{\textbf{Comparison with simulations}}

In our model, the diffusivity is high, so we take Eq.
\ref{eq:Delta-t-high-diff-dim} (for the high diffusivity scenario)
and compare the time intervals obtained with this formula with the
time intervals between episodes of accretion obtained in our
simulations.

Eq. \ref{eq:Delta-t-high-diff-dim} shows that in stars with a
larger magnetospheric parameter $\mu$ the time interval $\Delta t$
should be larger. To check the dependence on $\mu$, we compare the
time intervals between the bursts in models with different values
of parameter $\mu$ (see left panels of Figures
\ref{app:fluxes-radii-d100}, \ref{app:fluxes-radii-d60} and
\ref{app:fluxes-radii-d30} from \ref{appen:variation}). One can
see that, in the models with comparable values of $\omega_s$, the
time intervals between the main bursts are larger at $\mu=100$
than at $\mu=60$ and $\mu=30$. This trend is even more clear if we
compare the variation of the inner disk radius (compare the right
panels of the same figures).

 Eq. \ref{eq:Delta-t-high-diff-dim} also shows that the time interval
$\Delta t$ should increase when the difference $(\omega_s-1)$
becomes small. Therefore, the time intervals between bursts of
accretion should be larger in the weaker propellers.  The left
panels of Figures \ref{app:fluxes-radii-d100},
\ref{app:fluxes-radii-d60} and \ref{app:fluxes-radii-d30} show
that, for each value of $\mu$, the time interval $\Delta t$
between accretion/ejection events increases  when the fastness
parameter $\omega_s$ decreases.
Therefore,
 Eq. \ref{eq:Delta-t-high-diff-dim} describes the dependence
 on these two parameters
 correctly.  Eq. \ref{eq:Delta-t-high-diff-dim} also shows that the time interval
 $\Delta t$ increases when the accretion rate $\dot M$ decreases. In our simulations, we did not vary
 the initial accretion rate in the disk. However, we observed that in some long simulation runs the accretion rate decreases
 and the time interval increases, which is consistent with the theoretical dependence.




\section{Applications to Different Types of Stars}
\label{sec:applications}

In this section, we provide convenient estimates and formulae for
the application of our model to different types of stars.

\subsection{Application to accreting millisecond pulsars}

For accreting millisecond pulsars, we take the mass and radius of
the star to be $M_\star=1.4 M_\odot$ and $R_\star = 10$km,
respectively, and the magnetic field to be $B_\star=10^8$G. Using
equations \ref{eq:app-matter flux}, \ref{eq:app-angmom flux} and
\ref{eq:app-energy flux}, we obtain the reference values for
matter, angular momentum and energy fluxes in the following form:
\begin{equation}
\noindent \mdot_0  = \rho_0 v_0 R_0^2 =
 3.22\times 10^{-12}{\mu_{60}}^{-2}{B_8}^2 R_6^{5/2}
M_{1.4}^{-1/2} {M_\odot}/{\rm yr} ~, \label{eq:matter-flux-ns}
\end{equation}
\begin{equation}
\noindent \ldot_0  = \mdot_0 v_0 R_0 =
 2.80\times 10^{30}{\mu_{60}}^{-2}{B_8}^2 R_6^3 {\rm ergs} ~,
\label{eq:angmom-flux-ns}
\end{equation}
\begin{equation}
\noindent \edot_0  = \mdot_0 v_0^2 =
 3.83\times 10^{34}{\mu_{60}}^{-2}{B_8}^2 R_6^{3/2}M_{1.4}^{1/2} {\rm ergs}/{\rm s}
 ~,
\label{eq:energy-flux-ns}
\end{equation}
where $M_{1.4}=M_\star/{1.4 M_\odot}$, $R_6=R_\star/10^6{\rm cm}$,
$B_8=B_\star/10^8$G, and $\mu_{60}=\mu/60$.
  For example, to obtain the dimensional matter fluxes to
the star, $\dot M_s$, and to the wind, $\dot M_w$, one should take
the dimensionless values  $\langle\dot M_s\rangle$ and
$\langle\dot M_w\rangle$ from Tab. \ref{tab:models} and multiply
them by $\dot M_0$:
\begin{equation}
\dot M_s = \dot M_0 \langle\dot M_s\rangle~, ~~ \dot M_w = \dot
M_0 \langle\dot M_w\rangle~.
\end{equation}
Analogously, we can find the energy fluxes to the wind/jet
associated with matter and magnetic field:
\begin{equation}
\dot E_m = \dot E_0 \langle\dot E_m\rangle~, ~~\dot E_f = \dot E_0
\langle\dot E_f\rangle~.
\end{equation}
A star in the propeller regime spins down. The spin-down energy
flux (spin-down luminosity) is:
\begin{equation}
\dot E_{\rm sd}=\dot L_{\rm sd} \Omega_\star =\dot L_0 \langle
\dot L_{\rm sd}\rangle \Omega_\star=\dot L_0\langle\dot L_{\rm
sd}\rangle 2\pi/P_\star = \nonumber
\end{equation}
\begin{equation}
=1.76\times 10^{34}{\mu_{60}}^{-2}{B_8}^2 R_6^3 \langle{\dot
L}_{\rm sd}\rangle P_{-3}^{-1}  {\rm ergs/s} , \end{equation}
where $P_{-3}$ is the period of a neutron star in milliseconds.
The spin-down time scale can be estimated as
\begin{equation}
t_{\rm sd} = \frac{L_\star}{\ldot_{\rm sd}}  = \frac{I
\Omega_\star}{\dot L_0 \langle\ldot_{\rm sd}\rangle} ~,
\label{eqn:spindown-brief}
\end{equation}
where $L_\star=I\Omega_\star$ is the angular momentum of the star,
$I=k M_\star R_\star^2\approx 1.12\times 10^{45} k_{0.4} M_{1.4}
R_6^2~ \rm{g cm^2}$ is the star's moment of inertia,
$k_{0.4}=k/0.4$. Substituting in $\ldot_0$, $\Omega_\star=\Omega_0
\tilde\Omega_\star$, and taking $\Omega_0$ from Tab.
\ref{tab-ref}, we obtain:
\begin{equation}
t_{\rm sd} \approx 1.73\times 10^{11} k_{0.4}
M_{1.4}^{3/2}R_6^{-5/2} {B_8}^{-2} \mu_{60}^2
\frac{\tilde\Omega_\star} {\langle\ldot_{\rm sd}\rangle}~{\rm
yr}~. \label{eqn:spindown-ns-brief}
\end{equation}
 We can re-write the last term of Eq.
\ref{eqn:spindown-ns-brief}, $\tilde\Omega_\star/\langle\ldot_{\rm
sd}\rangle$, in the following way.  From Tab. \ref{tab:angmom} we
note that the spin-down flux $\langle\ldot_{\rm sd}\rangle$ is
proportional to the fastness parameter, $\omega_s$, and there is
also an approximately linear dependence on $\mu$:
$\langle\ldot_{\rm sd}\rangle\approx 0.83 \mu_{60} \omega_s$. On
the other hand, using the definition of the fastness parameter,
$\omega_s=\Omega_\star/\Omega_K
(r_m)=\tilde\Omega_\star/\tilde\Omega_K
(r_m)\approx\tilde\Omega_\star/\langle r_m\rangle^{-3/2}$, we can
re-write $\tilde\Omega_\star$ as $\tilde\Omega_\star=\langle r_m
\rangle ^{-3/2} \omega_s$ and obtain the following relationship:
\begin{equation}\frac{\tilde\Omega_\star}{\langle\dot L_{\rm
sd}\rangle}\approx 0.10 \bigg(\frac{\langle
r_m\rangle}{5}\bigg)^{-3/2} \mu_{60}^{-1}~,
\end{equation}
and the time scale in the form of
\begin{equation}
t_{\rm sd} \approx 1.74\times 10^{10} k_{0.4}
M_{1.4}^{3/2}R_6^{-5/2} {B_8}^{-2} \mu_{60} \left(\frac{\langle
r_m \rangle}{5}\right)^{-3/2}~{\rm yr}~.
\label{eqn:spindown-ns-final}
\end{equation}
The spin down time-scale does not depend on the angular velocity
of the star, $\Omega_\star$, because the faster rotators have
larger spin-down rates but also
 a larger amount of initial angular momentum.
   Eq. \ref{eqn:spindown-ns-final} shows that in stars with
the same mass, radius and magnetic field, the spin-down time scale
 is roughly the same in models with the same magnetospheric
radius $\langle r_m\rangle$.

Using the values of $\langle r_m \rangle$ from Table
\ref{tab:models}, we obtain the time scales in the range of
$t_{\rm sd}=(1.0-2.0)\times 10^{10}$ yrs.


In application to millisecond pulsars, the rate of spin-down is
often measured as the rate of variation of frequency $\nu$ with
time, $\dot \nu = d\nu/dt$ (in Hz/s). Taking into account the fact
that the angular momentum of the star
$L_\star=I_\star\Omega_\star$ and the angular momentum flux from
the star $\dot L_{\rm sd}=I_\star \dot\Omega_\star$, we obtain:
\begin{equation}
d\nu/dt = (1/2\pi)\dot\Omega_\star = \Omega_\star/(2\pi) (\dot
L_{\rm sd}/L_\star) = \nu_\star/t_{\rm sd} =
\nonumber
\end{equation}
\begin{equation}
=1.82\times 10^{-15} \nu_3 \frac{R_6^{5/2}{B_8}^2 ({\langle r_m
\rangle/5})^{3/2}}{k_{0.4} M_{1.4}^{3/2}\mu_{60}} {\rm Hz/s} ,
\end{equation}
where $\nu_3=\nu/1000$ Hz.

The time interval between accretion/ejection events varies in the
range of $\Delta \tilde{t}=30-200$ in dimensionless units. To
convert to dimensional units of time, we multiply this value by
the reference period of rotation, $P_0=0.46$s, and obtain $\Delta
t\approx (14-92)$ ms.

\subsection{Application to cataclysmic variables}

For cataclysmic variables, we take the mass and radius of the star
to be $M_\star=M_\odot$ and $R_\star = 5000$km, respectively, and
the magnetic field to be $B_\star=10^6$G. We obtain the reference
values for matter, angular momentum and energy fluxes in the
following form:
\begin{equation}
\noindent \mdot_0  =
 2.11\times 10^{-9}{\mu_{60}}^{-2}{B_6}^2 R_{5000}^{5/2}
M_{\odot}^{-1/2} {M_\odot}/{\rm yr} ~, \label{eq:matter-flux-wd}
\end{equation}
\begin{equation}
\noindent \ldot_0  =
 3.46\times 10^{34}{\mu_{60}}^{-2}{B_6}^2 R_{5000}^3 {\rm ergs} ~,
\label{eq:angmom-flux-wd}
\end{equation}
\begin{equation}
\noindent \edot_0  =
 3.57\times 10^{34}{\mu_{60}}^{-2}{B_6}^2 R_{5000}^{3/2}M_{1}^{1/2} {\rm ergs}/{\rm s}
 ~,
\label{eq:energy-flux-wd}
\end{equation}
where where  $R_{5000}=R_\star/5,000 {\rm km}$, and
$B_6=B_\star/10^6$G.

We can obtain the spin-down luminosity of the star in a convenient
form:
\begin{equation}
\dot E_{\rm sd}=\dot L_{\rm sd} \Omega_\star =\dot L_0\langle\dot
L_{\rm sd}\rangle 2\pi/P_\star =
\nonumber
\end{equation}
\begin{equation}
= 2.2\times 10^{35}{\mu_{60}}^{-2}{B_6}^2 R_{5000}^3 \langle\dot
L_{\rm sd}\rangle P^{-1}~{\rm ergs/s} ~, \label{eq:angmom-flux-wd}
\end{equation}
where $P$ is the period of the star in seconds.

Using Eq. \ref{eqn:spindown-brief} and the value for the
moment of inertia for white dwarf,  $I=k M_\star
R_\star^2=2.0\times 10^{50} k_{0.4} M_\odot R_{5000}^2
 \rm{g cm^2}$, we obtain the spin-down time scale in a form similar to that obtained for
 millisecond pulsars:
\begin{equation}
t_{\rm sd} \approx 1.89\times 10^7 k_{0.4}
M_{\odot}^{3/2}R_{5000}^{-5/2} {B_6}^{-2} \mu_{60} ({\langle r_m
\rangle/5})^{-3/2}~{\rm yr}~. \label{eqn:spindown-cv-brief}
\end{equation}
Using the values of $\langle r_m \rangle$ from Table
\ref{tab:models}, we obtain the time scales in the range of
$t_{\rm sd}=(1.47-1.68)\times 10^7$ yrs.

The time interval between accretion/ejection events varies in the
range of $\Delta \tilde{t}=30-200$ in dimensionless units. To
convert to dimensional units of time, we multiply this value by
the reference period of rotation, $P_0=6.08$s, and obtain $\Delta
t\approx (180-1220)$ s.

\subsection{Application to CTTSs}

In application to Classical T Tauri stars,  we take the mass and
radius of the star to be $M_\star=0.8 M_\odot$ and $R_\star =
2R_\odot$, respectively, and the magnetic field to be
$B_\star=10^3$G.

We obtain the reference values for matter, angular momentum and
energy fluxes in the following form:
\begin{equation}
\noindent \mdot_0  =
 2.1\times 10^{-9}{\mu_{60}}^{-2}{B_3}^2 R_{2R_\odot}^{5/2}
M_{0.8}^{-1/2} ~ {{\rm M}_\odot}/{\rm yr} ~,
\label{eq:matter-flux-wd}
\end{equation}
\begin{equation}\noindent \ldot_0  =
 3.5\times 10^{34}{\mu_{60}}^{-2}{B_3}^2 R_{2R_\odot}^3 ~ {\rm ergs} ~,
\label{eq:angmom-flux-wd}
\end{equation}
\begin{equation}
\noindent \edot_0  =
 3.6\times 10^{34}{\mu_{60}}^{-2}{B_3}^2 R_{2R_\odot}^{3/2}M_{0.8}^{1/2} ~ {\rm ergs}/{\rm s} ~,
\label{eq:energy-flux-wd}
\end{equation}
where  $M_{0.8}=M_\star/{0.8 M_\odot}$, $R_{2R_\odot}=R/2R_\odot$,
and $B_3=B_\star/10^3$G.


Using Eq. \ref{eqn:spindown-brief} and the value for the moment of
inertia of CTTSs,  $I=k M_\star R_\star^2\approx 1.25\times
10^{55} k_{0.4} M_{0.8} R_{2R_\odot}^2
 \rm{g cm^2}$, we obtain the spin-down time scale in a form similar to that obtained for
 millisecond pulsars and CVs:
\begin{equation}
t_{\rm sd} \approx 1.03\times 10^7  k_{0.4} {M_{0.8}}^{1.5}
 R_{2\rsun}^{-5/2}  {B_3}^{-2} \mu_{60} \bigg(\frac{\langle
r_m\rangle}{5}\bigg)^{-3/2}~{\rm yr}~.
\label{eqn:spindown-ctts-final}
\end{equation}
Using the values of $\langle r_m \rangle$ from Table
\ref{tab:models}, we obtain the time scales in the range of
$t_{\rm sd}=(5.3\times 10^6 - 1.3\times 10^7)$ yrs.
These time scales are in agreement
with the observations of CTTSs, which show that CTTSs are already
slow rotators after 1-10 million years.

The time interval between accretion/ejection events varies in the
range of $\Delta \tilde{t}=30-200$ in dimensionless units. To
convert to dimensional units of time, we multiply these values by
the reference period of rotation, $P_0=0.37$ days, and obtain
$\Delta t\approx (11-74)$ days. A recent analysis of the
light-curves of accreting young stars in the $\rho$ Oph and Upper
Sco regions of star formation (obtained with the K-2 Kepler
mission) has shown that bursts of accretion
occur every 3-80 days \citep{CodyEtAl2017}. Stars with infrequent
bursts may be in the propeller regime.

\section{Conclusions and Discussions}
\label{sec:Conclusions}

We performed axisymmetric simulations of accretion onto rotating
magnetized stars in the propeller regime, ranging from very weak
to very strong propellers. We used the fastness parameter
$\omega_s$ to characterize the strength of the propellers. We
observed that many properties of the propellers depend on the
fastness parameter.

\subsection{Main conclusions}

The main conclusions are the following:

\textbf{1.} Both accretion and outflows are observed in propellers
of different strengths. The relative amount of matter ejected into
the outflows (propeller efficiency, $f_{\rm eff}$, see Eq.
\ref{eq:prop-efficiency}) increases with $\omega_s$ as a power
law.

\textbf{2.} The accretion/ejection cycle is observed at different
propeller strengths. In this cycle: (a) Matter of the inner disk
slowly moves inward and penetrates through the field lines of the
external magnetosphere, (b) The magnetic field lines inflate and
open. Matter partly accretes onto the star and is partly ejected
into the outflows along the inflated field lines. (c) The
magnetosphere expands and the cycle repeats. Most of the time
matter accumulates in the inner disk, while the accretion/ejection
events occur during brief intervals of time.

\textbf{3.} The inner disk oscillates.
The time-averaged inner disk (magnetospheric) radius $\langle r_m
\rangle$ is larger than the corotation radius $r_{\rm cor}$. In
spite of this, matter accretes onto the star. Accretion is
possible due to the fact that (a) only the closed part of the
magnetosphere represents the centrifugal barrier, (b) the process
is non-stationary: accretion occurs
in brief episodes as the inner disk moves closer to the star.

\textbf{4.} The velocity of matter ejected into the wind is
different in propellers of different strengths: (a) In strong
propellers, the maximum velocity of ejecting matter is a few times
larger than the local escape velocity; (b) In weak propellers, the
maximum velocity is slightly larger or smaller than the escape
velocity; (c) In very weak propellers, matter is ejected at
sub-escape velocities, forming a turbulent corona above the disk.
The time-averaged velocity of matter ejected into the wind
increases with the fastness parameter ($\omega_s$) exponentially.


\textbf{5.} The time-averaged opening angle of the wind $\langle
\Theta_{\rm wind} \rangle$ is also different in propellers of
different strengths: (a) In strong propellers, this angle is
relatively small, $\langle\Theta_{\rm wind}\rangle\approx
40^\circ-45^\circ$. (b) In weak propellers, it is larger,
$\langle\Theta_{\rm wind}\rangle \approx 60^o$. The opening angle
decreases with $\omega_s$ as a power law.


\textbf{6.} A star in the propeller regime spins down due to the
outward angular momentum flow along the field lines. Approximately
half of the angular momentum flows to the disk along the closed
field lines. The other half flows along the open field lines
connecting the star with the corona.

\textbf{7.} A star-disk system loses mass, angular momentum and
energy. Most of the matter flows from the inner disk into a
conically-shaped wind, which carries the energy and angular
momentum associated with that matter. In addition, the inflating
field lines carry angular momentum and energy associated with the
magnetic field. In the
strong propellers, the field lines originating at the star wind up
rapidly and form a magnetically-dominated and magnetically-driven
(Poynting flux) jet, which accelerates a small amount of matter to
high velocities. This jet
takes a significant amount of angular momentum out of the star. In
addition, it carries angular momentum and energy out of the
system. Ejections to the conical wind and Poynting flux jet are
strongly non-stationary, so the formation of shocks is expected at
some distances from the star.




\subsection{Application to propeller candidate stars}
 \label{sec:Discussions}


Our research shows that our models of propellers can explain the
different observational properties of propeller candidate stars:

\begin{itemize}

\item Strong variability in the light-curves, which can be
associated with  (1) variable accretion rate onto the star, (2)
variable ejection rate to the wind, (3) oscillations of the inner
disk.

\item Accretion of matter onto the stellar surface in the
low-luminosity (low accretion rate) regime, when the
magnetospheric radius $r_m$ is larger than the corotation radius
$r_{\rm cor}$. Our models show that a small amount of matter
accretes onto a star even in the strongest propeller regime.

\item Outflows from propeller candidates stars. These outflows can
be associated with conical winds and
more collimated magnetic jets.

\item Flares of high-energy radiation (e.g., the gamma-ray flares
observed in some transitional MSPs) can be associated with
acceleration of particles in shocks, which form during
non-stationary ejections to jets and winds in the strong propeller
regime.

\end{itemize}

In future studies, we plan to model the propeller candidate stars
individually (using the known stellar parameters) and to compare
our models with observations in detail.

\subsection{Comparisons with other models}

Our model is somewhat similar to the model of
\citet{AlyKuijpers1990}, who suggested that the field lines
connecting the star and the disk should inflate and reconnect
quasi-periodically. This model
was developed for accreting (non-propelling) stars. However,
differential rotation between the foot-points of the field lines
and their inflation is expected in both regimes (see also
\citealt{NewmanEtAl1992,LovelaceEtAl1995,UzdenskyEtAl2002})
\footnote{Inflation of the field lines has been observed, e.g., in
simulations by \citet{MillerStone1997}. The signs of such
inflation were observed in CTTS AA Tau \citep{BouvierEtAl2007}.}.
Axisymmetric MHD numerical simulations by
\citealt{GoodsonEtAl1997,GoodsonEtAl1999} confirmed this type of
instability.
In their simulations, they observed several cycles in which matter
accumulated, the field lines inflated and subsequently
reconnected, matter accreted onto the star and then was ejected
into the winds, and the magnetosphere expanded. A similar cycle
has been observed in axisymmetric simulations of the propeller
regime \citep{RomanovaEtAl2004,ZanniFerreira2013}. However, both
types of simulations (for slowly and rapidly-rotating stars) have
only been performed in the top part of the simulation region
(above the equatorial plane).
In these models, reconnection of the field lines has been
necessary for the subsequent accretion of matter onto the star.
More recent simulations by \citet{LiiEtAl2014} have shown that
modeling the entire simulation region (above and below the
equator) leads to a new
phenomenon: the magnetic flux inflates in one direction (above or
below the disk), but matter accretes onto the star on the opposite
side
of the equator \footnote{This phenomenon has been initially
observed in simulations by \citealt{LovelaceEtAl2010}, where
accretion onto stars with complex fields has been modeled.}. This
phenomenon leads to the fact that reconnection is not required for
accretion: matter of the inner disk accretes above the
magnetosphere (on the opposite side of inflation relative to the
equator), where the magnetic flux does not block its path (see,
e.g., Fig. \ref{2d-strong-expand}). In our current studies of the
propeller regime, we observed a similar phenomenon in the models
with larger magnetospheres and thinner disks.
We should note that, in the models of slowly-rotating stars (e.g.,
\citealt{GoodsonEtAl1997}), accretion is blocked by the magnetic
flux of the inflated field lines, while in the models of
propellers the centrifugal barrier of the rapidly-rotating star is
a more important factor in blocking accretion.


Our model also has some similarities with the ``dead disk" model,
proposed by \citet{SunyaevShakura1977,SpruitTaam1993} and further
developed by \citet{DangeloSpruit2010,DangeloSpruit2012}: in these
models, matter of the inner disk is blocked by the centrifugal
barrier
for some interval of time, and the periods of matter accumulation
alternate with episodes of matter accretion onto the star.
However, compared with their models, our model is two-dimensional
and
takes into account (1) inflation of the field lines, (2) formation
of outflows and jets, which can be driven by both centrifugal and
magnetic forces, and (3) the possibility of accretion above or
below the centrifugal barrier (which has the shape of a closed
magnetosphere). Also, in \citet{DangeloSpruit2010}, it is
suggested that the magnetospheric radius should be near the
corotation radius. In our models, the position of the
magnetospheric radius does not depend much on the rotation of the
star, but is instead determined by the balance of magnetic and
matter stresses, while the position of the corotation radius is
determined by the period of the star. We modeled propellers with
different ratios of these two radii, which are in the range of
$\langle r_m\rangle/r_{\rm cor}=1.1-4.7$ (see Tab.
\ref{tab:models}). Moreover, in each model, the magnetospheric
radius
typically varies strongly. In spite of these differences,
cyclic 
accretion is also observed in our models. However, in our models,
we observe several time-scales
associated with more complex processes of disk-magnetosphere
interaction.


\subsection{Restrictions of the model and future work}

Current simulations are axisymmetric. This restricts us from
modeling instabilities at the disk-magnetosphere boundary, which
determine the rate of matter penetration through the external
magnetosphere. 3D instabilities are shown to be effective in cases
of slowly-rotating magnetized stars (e.g.,
\citealt{KulkarniRomanova2005,RomanovaEtAl2008,BlinovaEtAl2016}).
 In this paper, we suggested that
similar instabilities may also operate and provide an effective
diffusivity at the disk-magnetosphere boundary. We used the
$\alpha-$diffusivity approach and took the maximum possible value
of $\alpha_{\rm diff}=1$ (acting only inside the spherical radius
$R=7$, which typically includes the disk-magnetosphere boundary).
We observed that this diffusivity provides rapid penetration of
matter through the external layers of the magnetosphere. However,
the effective diffusivity may depend, for example, on the fastness
parameter $\omega_s$, and can be high at some values of $\omega_s$
and low at other values (as in the cases of slowly-rotating stars,
see \citealt{BlinovaEtAl2016}).

Fortunately, the results of the propeller model do not depend too
much on the value of diffusivity. Our earlier studies of
propellers, performed at different values of the diffusivity
parameter $\alpha_{\rm diff}$ (see Appendix B in
\citealt{LiiEtAl2014}), have shown that the process of
disk-magnetosphere interaction is similar in the cases of high and
low diffusivity.  However, at very low diffusivity, $\alpha_{\rm
diff}=0.01$, matter is accumulated at the disk-magnetosphere
boundary for longer time before it accretes onto the star. In this
case, accretion is more ``spiky" (see left panel of Fig. B2 of
\citealt{LiiEtAl2014}). In the opposite
scenario, when the diffusivity is high, $\alpha_{\rm diff}=1$,
matter of the inner disk penetrates more rapidly through the
external layers of the magnetosphere, acquires angular momentum
and is ejected into the winds (see right panel of Fig. B2 of
\citealt{LiiEtAl2014}). In this case, the accretion rate is
smaller. As a result, efficiency is higher
at higher diffusivity values. However, the difference in not very
large: $f_{\rm eff}=0.70$ in the low-diffusivity case versus
$f_{\rm eff}=0.86$ in the high-diffusivity case. Overall, the
results of \citet{LiiEtAl2014} obtained at a very low diffusivity
\footnote{Most of results in \citet{LiiEtAl2014} simulations were
obtained using ideal MHD code with no diffusivity term added, and
where only a small numerical diffusivity determined penetration of
the disk through the magnetosphere. The numerical diffusivity in
the code corresponded to $\alpha_{\rm diff}\approx 0.01-0.003$.}
do not differ qualitatively from the results obtained in the
current paper. The issue of diffusivity should be further studied
in 3D simulations.

\smallskip

On the other hand, in three dimensions, the magnetic axis of the
dipole can be tilted about the rotational axis of the disk. 3D MHD
simulations of accreting stars have shown that the magnetospheric
radius $r_m$ is approximately the same in stars with different
tilts of the magnetic axis \citep{RomanovaEtAl2003}, and therefore
the centrifugal barrier
should be located at the same distance as in the 2D simulations.
However, the centrifugal barrier will have a slightly different
shape, which may be against accretion. On the other hand, the
tilted dipole is more favorable for accretion. Therefore, the
efficiency of the propeller may be somewhat different compared
with the axisymmetric case. Global 3D simulations should be done
to determine the difference between 2D and 3D simulations.

\section*{Acknowledgments}
Resources supporting this work were provided by the NASA High-End
Computing (HEC) Program through the NASA Advanced Supercomputing
(NAS) Division at the NASA Ames Research Center and the NASA
Center for Computational Sciences (NCCS) at Goddard Space Flight
Center. The research was supported by NASA grant NNX12AI85G. AVK
was supported by the Russian academic excellence project
``5top100". We also acknowledge the International Space Science
Institute (ISSI), which funded and hosted an international team
devoted to the study of transitional millisecond pulsars, and we
thank all the members of the team for fruitful discussions.

{}

\appendix

\section{Description of Numerical Model}
 \label{appen:numerical-model}

\subsection{Initial and boundary conditions}
\label{app:initial and boundary}

\paragraph*{Initial Conditions:} In this work, the initial conditions for the hydrodynamic variables are similar
 to those
 used in our previous works (e.g., \citealt{RomanovaEtAl2009, LiiEtAl2014}), where
 the initial density and entropy distributions
 were calculated
 by balancing the gravitational, centrifugal and
pressure forces. The disk is
 initially cold and dense, with temperature $T_d$ and density $\rho_d$. The corona
 is hot and rarified, with temperature $T_c = 3\times 10^3 T_d$ and density $\rho_c = 3.3\times 10^{-4} \rho_d$.
 In the beginning of
 the simulations,
 the inner edge of the disk is placed
 at $r_d$ = 10, and the star rotates with $\Omega_i$ = 0.032 (corresponding to $r_{\rm
cor}$ = 10), so that the
  magnetosphere and the inner disk initially corotate. This condition helps to ensure that
  the magnetosphere and the disk are initially in near-equilibrium at the disk-magnetosphere
  boundary.
  The star is gradually spun up from $\Omega_i$ to the final
  state with angular velocity $\Omega_\star$, corresponding to $r_{\rm cor}$ (given in Table \ref{tab:models}).
  The initial pressure distribution in the simulation is determined from the
Bernoulli equation:
\begin{equation}
F(p) + \Phi + \Phi_c = B_0 = {\rm constant},
\end{equation}
where $\Phi = -GM_\star/(r^2+z^2)^{1/2}$ is the gravitational
potential, $\Phi_c=-kGM_\star/r$ is the centrifugal potential, $k$
is a Keplerian parameter\footnote{We take $k$ slightly greater
than unity to balance the disk pressure gradient (k=1+0.003).} and
\begin{equation}
F(p) =
\begin{cases}
{\cal R} T_d \ln(p/p_b), & \mbox{if } p > p_b \mbox{ and } r > r_d, \\
{\cal R} T_c \ln(p/p_b), & \mbox{if } p \leq p_b \mbox{ or } r \leq r_d, \\
\end{cases}\end{equation}
where $p_b$ is the pressure at the boundary that separates the
disk from the corona. We assume the system to be initially
barotropic, and determine the density from the pressure:
\begin{equation}
\rho(p) =
\begin{cases}
p/{\cal R} T_d, & \mbox{if } p > p_b \mbox{ and } r > r_d, \\
p/{\cal R} T_d, & \mbox{if } p \leq p_b \mbox{ or } r \leq r_d.
\end{cases}\end{equation}
To initialize the MRI, 5\% velocity perturbations are added to
$v_\phi$ inside the disk.

\paragraph*{Initial magnetic field configuration:} Initially, the disk is threaded
by the dipole magnetic field of the star. We also add a small
``tapered'' poloidal field inside the disk (see left panel in Fig.
\ref{init-thin}), which is given by
$$
\Psi=\frac{B_0 r^2}{2}\cos\bigg(\pi\frac{z}{2h}\bigg), ~~
h=\sqrt{\bigg(\frac{GM_*}{\Phi_c(r)-E}\bigg)^2 - r^2},
$$
where $h$ is the half-thickness of the disk and $E$ is a constant
of integration in the initial equilibrium equation (see
\citealt{RomanovaEtAl2002, RomanovaEtAl2011}). This tapered field
helps initialize the MRI in the disk and has the same polarity as
the stellar field at the disk-magnetosphere boundary.

\paragraph*{Boundary Conditions:}
{\it Stellar surface:} all the variables on the surface of the
star have ``free'' boundary conditions, such that $\partial
(...)/\partial n=0$ along the entire surface. We
do not allow for the outflow of matter from the star
(i.e. we prohibit stellar winds), and adjust the matter velocity
vectors to be parallel to the magnetic field vectors. This models
the frozen-in condition on the star.

\noindent \textit{Top and bottom boundaries:} all variables have
free boundary conditions along the top and bottom boundaries. In
addition, we implement outflow boundary conditions on velocity to
prohibit matter from flowing back into the simulation region once
it leaves.

\noindent \textit{Outer side boundary:} the side boundary is
divided into a ``disk region'' ($|z| < z_{\rm disk}$) and a
``coronal region" ($|z| > z_{\rm disk}$), with
$$
z_{\rm disk} = h(R_{\rm out}) =
\sqrt{\left(\frac{GM_*}{\Phi_c(R_{\rm out})-E}\right)^2 - R_{\rm
out}^2},
$$
where $R_{\rm out}$ is the external simulation radius. The matter
along the disk boundary ($|z| < z_{\rm disk}$) is allowed to flow
inward with a small radial velocity
$$
v_r=-\delta \frac{3}{2}\frac{p}{\rho v_K(R_{\rm out})},
~~\delta=0.02,
$$
and
a poloidal magnetic field corresponding to the calculated magnetic
field at $r=R_{\rm out}$. The
remaining variables \textit{have} free boundary conditions. The
coronal boundary ($|z| > z_{\rm disk}$) has the same boundary
conditions as the top and bottom boundaries.

\subsection{Grid and code description}
\label{app:grid and code}

\paragraph*{Grid description:} The axisymmetric grid is in cylindrical ($r$,
$z$) coordinates with mesh compression towards
the equatorial plane and the $z$-axis, so that there is a larger
number of cells in the disk plane and near the star. In the models
presented here, we use a non-uniform grid with
$190 \times 306$ grid cells corresponding to a grid that is 43 by
82 stellar radii in size
At $r=20$, the number of grid cells that cover the disk in the
vertical direction is about 60.

\paragraph*{Code description:}
We use a Godunov-type numerical method with a five-wave Riemann
solver similar to the HLLD solver developed by
\citet{MiyoshiKusano2005}. The MHD variables are calculated in
four states bounded by five MHD discontinuities: the contact
discontinuity, two Alfv\'en waves and two fast magnetosonic waves.
Unlike \citet{MiyoshiKusano2005}, our method solves the equation
for entropy instead of the full energy equation. This
approximation is valid in cases (such as ours) where strong shocks
are not present.
We ensure that the magnetic fields are divergence-free by
introducing the $\phi$-component of the magnetic field potential,
which is calculated using the constrained transport scheme
proposed by \citet{GardinerStone2005}. The magnetic field is split
into the stellar dipole and the calculated components, $B=B_{\rm
dip} + B'$;
we omit the terms of the order $B_{\rm dip}^2$ which do not
contribute to the Maxwellian stress tensor \citep{Tanaka1994}. No
viscosity terms have been included in the MHD equations, and hence
we only investigate accretion driven by the resolved
MRI-turbulence. Our code has been extensively tested and has been
previously utilized to study different MHD problems (see
\citealt{KoldobaEtAl2016} for tests and some astrophysical
examples).

 Table \ref{tab-ref} shows sample reference values for three different
types of accreting stars: to apply the simulation results to a
particular class of star, multiply the dimensionless value by the
reference value. The dependence on $\mu$ is also shown.
\begin{table*}[ht!]
\centering
\begin{tabular}{llll}
\hline \hline & cTTs                                   & White Dwarf            & Neutron Star     \\
\hline {\bf initial} & & & \\ \hline
$M_*$ [\msun]                 & 0.8                    & 1                      & 1.4              \\
$R_*$                         & 2\rsun                 & 5000 km                & 10 km            \\
$B_*$ [G]                     & 1000                   & $1 \times 10^6$        & $1 \times 10^8$  \\
\hline {\bf derived}          &                        &                        &                  \\
\hline
$R_0$ [cm]                    & $1.40\times 10^{11}$   & $5\times 10^{8}$       & $1\times 10^{6}$        \\
$v_0$ [cm s$^{-1}$]           & $2.76 \times 10^{7}$   & $5.16\times 10^{8}$   &  $1.37 \times 10^{10}$ \\
$P_0$                         & 0.37 d                 & 6.08 s                 & 0.46 ms \\
$\nu_0$ [$s^{-1}$]            & $3.13 \times 10^{-5}$  & 0.16                  & $2.17 \times 10^3$ \\
$\Omega_0$ [$s^{-1}$]         & $1.97 \times 10^{-4}$  & 1.03                   & $1.37 \times 10^4$ \\
$T_0$ [K]                     & $9.17 \times 10^6$     & $3.21 \times 10^9$     & $1.13 \times 10^{12}$ \\
$T_{\rm disk}$ [K]            & $3.06 \times 10^3$     & $1.07 \times 10^6$     & $2.25 \times 10^{9}$ \\
$B_0$ [G]                     & $16.7 \mu_{60}^{-1}$            & $1.67 \times 10^4 \mu_{60}^{-1}$  &  $1.67 \times 10^6 \mu_{60}^{-1}$ \\
$\mu_0$ [G cm$^3$]            & $2.74 \times 10^{36}$  & $1.25 \times 10^{32}$  & $1.00 \times 10^{26}$ \\
$\rho_0$ [g cm$^{-3}$]        & $3.64 \times 10^{-13}\mu_{60}^{-2}$ & $1.04 \times 10^{-9}\mu_{60}^{-2}$  & $1.49 \times 10^{-8}\mu_{60}^{-2}$ \\
$\mdot_0$ [\msun\ yr$^{-1}$]  & $3.10 \times 10^{-9}\mu_{60}^{-2}$  & $2.11 \times 10^{-9} \mu_{60}^{-2}$  & $3.22 \times 10^{-12} \mu_{60}^{-2}$ \\
$\ldot_0$ [g cm$^2$ s$^{-2}$] & $7.61 \times 10^{35} \mu_{60}^{-2}$  & $3.46 \times 10^{34} \mu_{60}^{-2}$  & $2.80 \times 10^{30} \mu_{60}^{-2}$ \\
$\edot_0$ [g cm$^2$ s$^{-3}$] & $1.50 \times 10^{32} \mu_{60}^{-2}$  & $3.57 \times 10^{34} \mu_{60}^{-2}$  & $3.83 \times 10^{34} \mu_{60}^{-2}$ \\
\hline \hline
\end{tabular}
\caption{Reference values for three different types of accreting
stars. We use typical values of stellar mass $M_*$, radius $R_*$,
and magnetic field $B_*$ for each star, and the other reference
values are derived from these parameters. The dependence on the
dimensionless parameter $\mu$ is also shown, where the normalized
value $\mu_{60}= \mu/60$ is used.} \label{tab-ref}
\end{table*}

\subsection{Reference units}
\label{app:reference units}

The simulations are performed in dimensionless units and are
applicable to stars over a wide range of scales. There are four
free parameters: we choose the values of the stellar mass $M_*$,
radius $R_*$, magnetic field $B_*$ and dimensionless
magnetospheric parameter  $\tilde{\mu}$ and derive reference
values from these parameters. The magnetic moment
$\boldsymbol{\mu_\star} = \tilde{\mu}\mu_0\hat{z}$ is used to
initialize the stellar dipole field
\begin{equation}
\mathbf{B}_{\rm dip} =
\frac{3(\boldsymbol{\mu}\cdot\mathbf{R})\mathbf{R} -
\boldsymbol{\mu}R^2}{R^5}, \label{eqn_dipole}
\end{equation}
where $\mathbf{R}$ is the radius in spherical coordinates. In this
work, we take $\tilde{\mu}$ = 30, 60 and $100$.  The reference
units are as follows: length $R_0=R_*$, magnetic moment $\mu_0 =
B_0 R_0^3$, magnetic field $B_0 = B_*/\tilde{\mu} \times
(R_*/R_0)^3$ (the equatorial field dipole strength at $r = R_0$) ,
velocity $v_0=\sqrt{GM_*/R_0}$ (the Keplerian orbital velocity at
$r = R_0$), time $t_0=2\pi R_0/v_0$ (the Keplerian orbital period
at $r = R_0$), angular velocity $\Omega_0 = v_0/R_0$, pressure
$p_0 = B_0^2$, density $\rho_0=p_0/v_0^2$, temperature $T_0 =
p_0/\rho_0 \times m_H/k_B$ where $m_H$ is the mass of hydrogen and
$k_B$ is the Boltzmann constant, force per unit mass
$f_0=v_0^2/R_0$. Accretion rate $\mdot_0=\rho_0 v_0 R_0^2$,
angular momentum flux $\ldot_0 = \mdot_0 v_0 R_0$ and energy flux
$\edot_0 = \mdot_0 v_0^2$. We should stress out that in our
dimensionalization procedure, the reference magnetic field and
many other reference
 variables depend on the parameter $\tilde\mu$. Matter flux and other reference fluxes also depend on this parameter.
For practical purposes, we provide a useful form for reference
fluxes:
\begin{equation}\mdot_0=\rho_0 v_0 R_0^2 = ({B_\star}/{\tilde\mu})^2 (R_0^2/v_0)(R_\star/R_0)^6 ~,\label{eq:app-matter flux}\end{equation}
\begin{equation}\ldot_0=\mdot_0 v_0 R_0 = ({B_\star}/{\tilde\mu})^2 R_0^3 (R_\star/R_0)^6 ~,\label{eq:app-angmom flux}\end{equation}
\begin{equation}\edot_0=\mdot_0 v_0^2 = ({B_\star}/{\tilde\mu})^2 (R_0^2 v_0)(R_\star/R_0)^6 ~.\label{eq:app-energy flux}\end{equation}
All fluxes depend on parameter $\tilde\mu$ as $\sim
\tilde\mu^{-2}$. For example, in case of matter flux, this means
that at larger values of $\tilde\mu$, the matter flux is smaller,
and  (at fixed $B_\star$) the magnetospheric radius is expected to
be larger, because the general dependence $r_m\sim
(\mu_\star^2/\dot M)^{1/7}$ is approximately satisfied. That is
why in our model we use parameter $\tilde\mu$ to regulate the
dimensionless size $r_m/R_\star$ of the magnetosphere: the
magnetosphere is largest in case of $\tilde\mu=100$, and smallest
in case of $\tilde\mu=30$. In Tab. \ref{tab-ref}, we use the
normalized value $\mu_{60}=\mu/60$.

\section{Variation of the inner disk radius and matter fluxes in representative runs}
\label{appen:variation}

Left-hand panels of Figures \ref{app:fluxes-radii-d100},
\ref{app:fluxes-radii-d60}, and \ref{app:fluxes-radii-d30} show
temporal variation of the
 matter fluxes to the star $\dot M_\star$ and to the wind $\dot M_{\rm
 wind}$   in cases of
magnetospheres with different sizes (different parameter $\mu$)
and different corotation radii (different $r_{\rm cor}$). One can
see that in all cases accretion occurs in relatively brief bursts.
Matter flux to the wind also occurs in bursts.
 The dashed lines show the time-averaged values $\langle\dot
M_\star\rangle$ and $\langle\dot M_{\rm
 wind}\rangle$. Right-hand panels show temporal variation of the
 inner disk radius, $r_m$. The dashed lines show
the time-averaged values $\langle r_m \rangle$.

Left-hand panels of Figs. \ref{app:vmax-theta-d100},
\ref{app:vmax-theta-d60} and \ref{app:vmax-theta-d30} show
temporal variation of the normalized maximum velocity  $v_{\rm
max}/v_{\rm esc}$ in the matter-dominated component of the wind.
Dashed lines show the time-averaged values, $\langle v_{\rm
max}\rangle/v_{\rm esc}$. Figures show that the maximum velocity
rapidly decreases.hen $r_{\rm cor}$ increases.

Right-hand panels of Figs. \ref{app:vmax-theta-d100},
\ref{app:vmax-theta-d60} and \ref{app:vmax-theta-d30}  show
variation of the opening angle of the wind, $\Theta_{\rm wind}$,
with time. One can see that the opening angle systematically
increases, when $r_{\rm cor}$ increases. Dashed lines show
variation of the time-averaged values, $\langle\Theta_{\rm
wind}\rangle$. Figures show that the opening angle systematically
increases when $r_{\rm cor}$ increases.



\begin{figure*}
\centering
\includegraphics[height=5.8cm,clip]{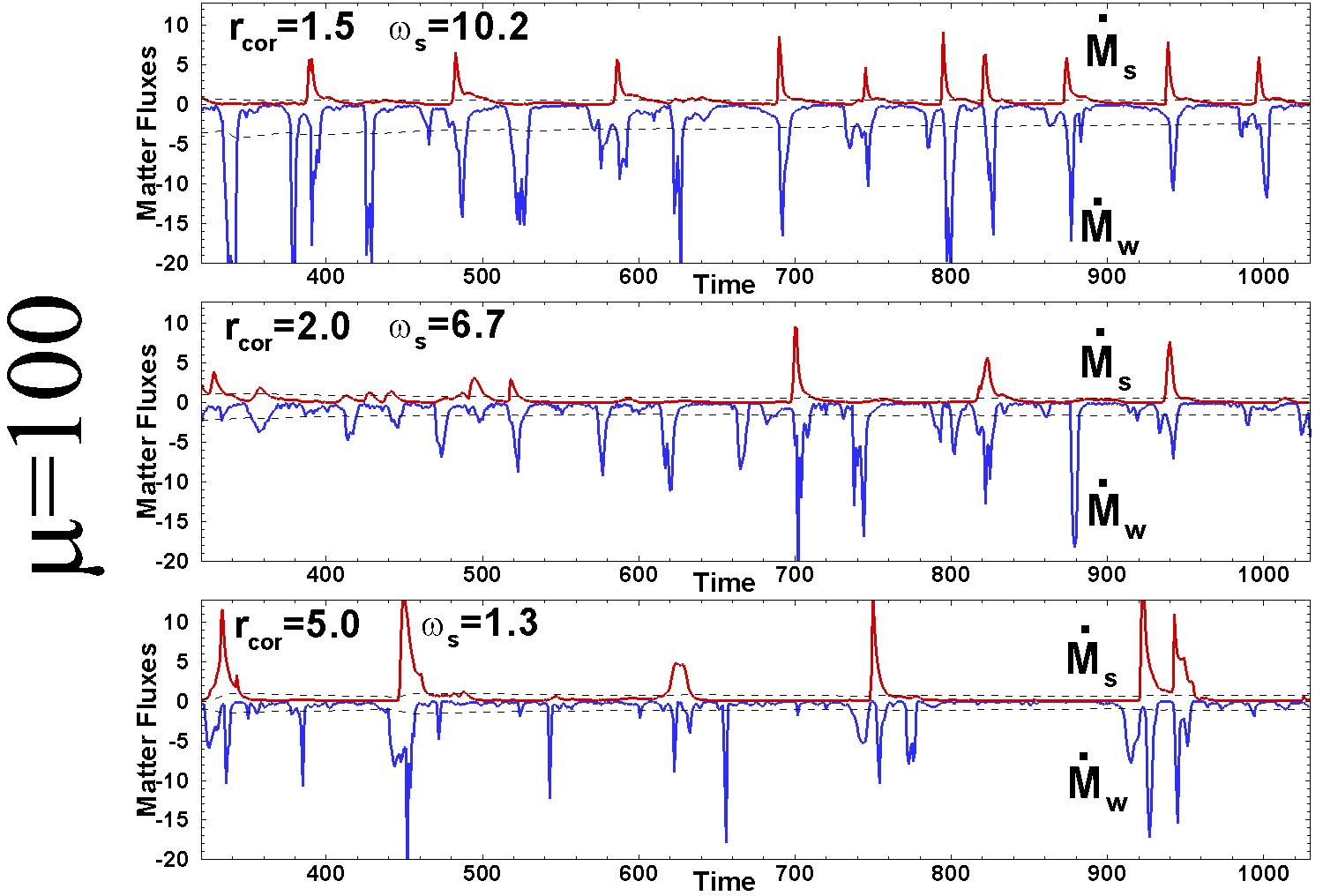}
\includegraphics[height=5.8cm,clip]{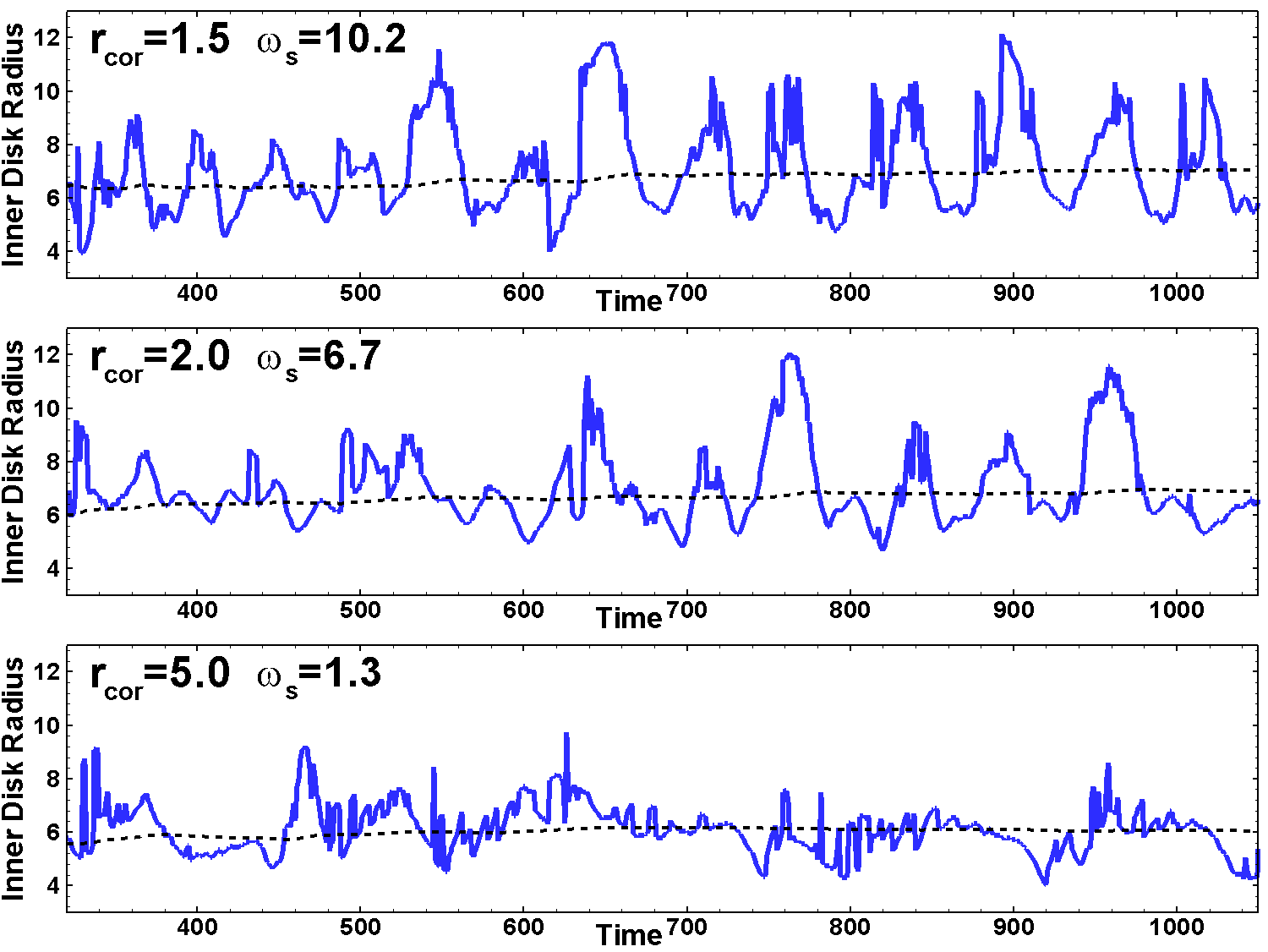}
\caption{\textit{Left Panels:} Variation of the inner disk radius
with time in case of $\mu=100$ and different corotation radii
$r_{\rm cor}$. \textit{Left Panels:} Matter fluxes to the star
$\dot M_s$ (red lines) and to the wind $\dot M_w$ (blue lines) for
same simulation runs. The matter flux to the wind has been
calculated through the surface $S(r=10,z=\pm10)$ at condition
$v>0.1 v_{\rm esc}$.} \label{app:fluxes-radii-d100}
\end{figure*}

\begin{figure*}
\centering
\includegraphics[height=5.8cm,clip]{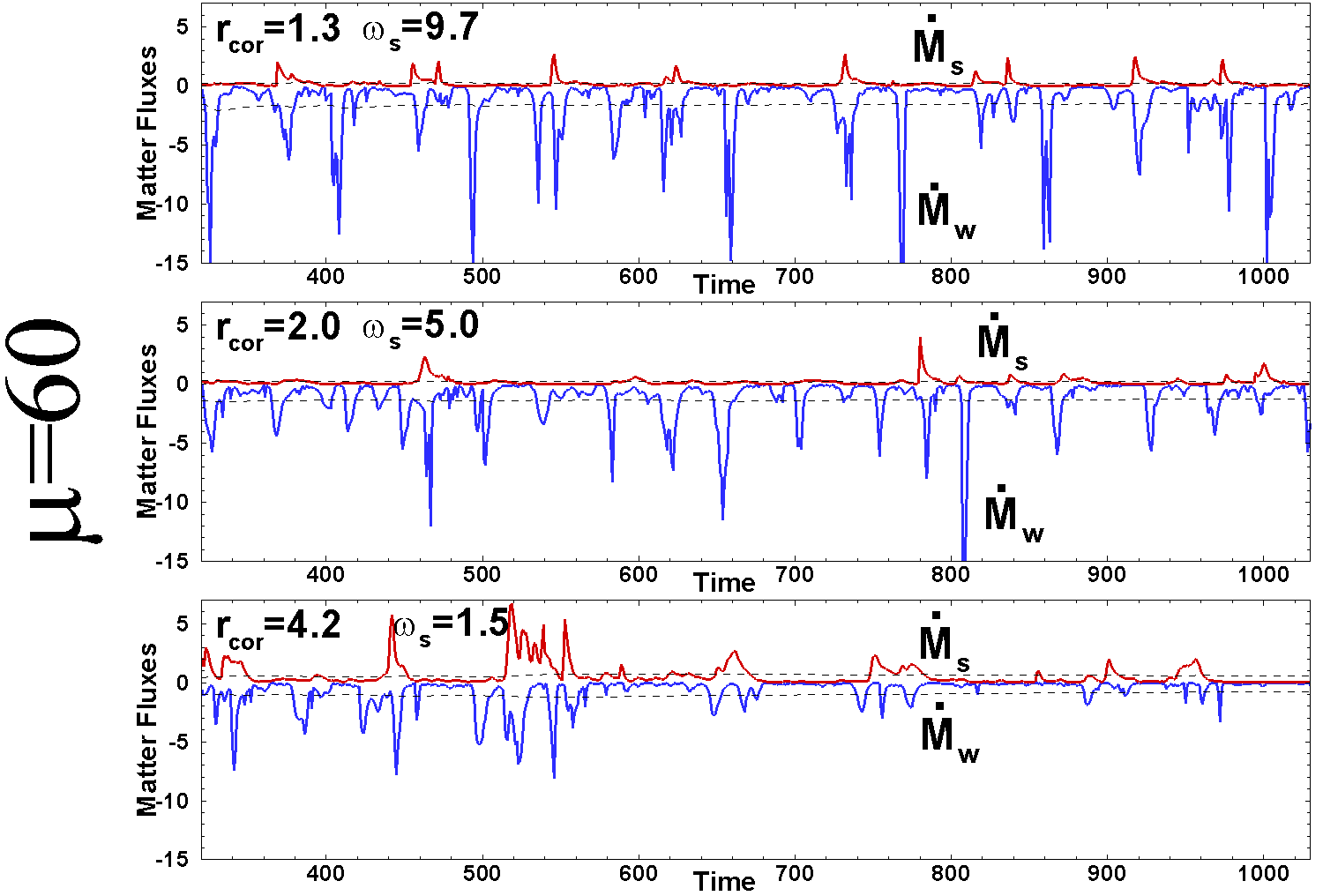}
\includegraphics[height=5.8cm,clip]{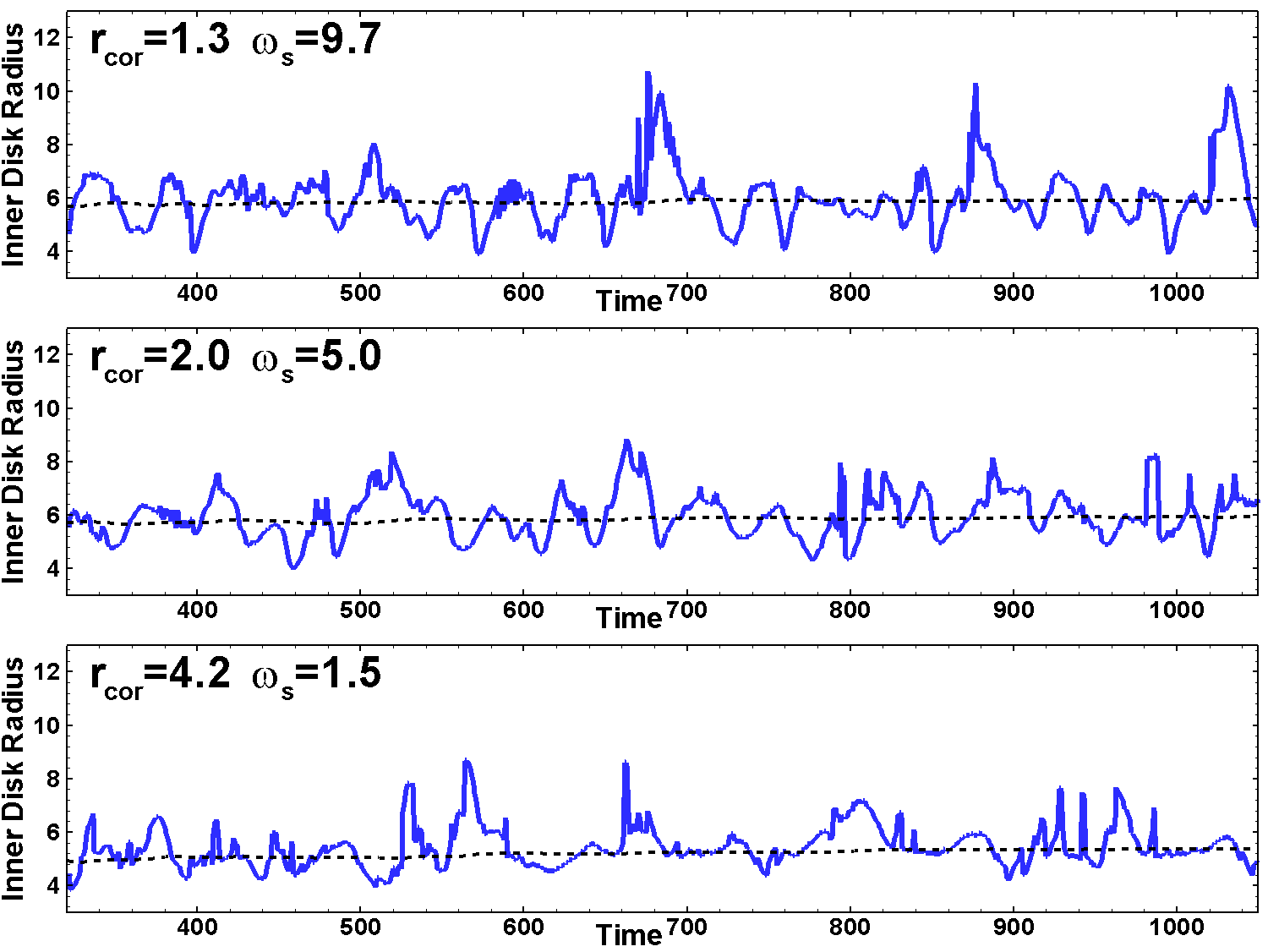}
\caption{Same as in Fig. \ref{app:fluxes-radii-d100} but for
$\mu=60$.} \label{app:fluxes-radii-d60}
\end{figure*}

\begin{figure*}
\centering
\includegraphics[height=5.8cm,clip]{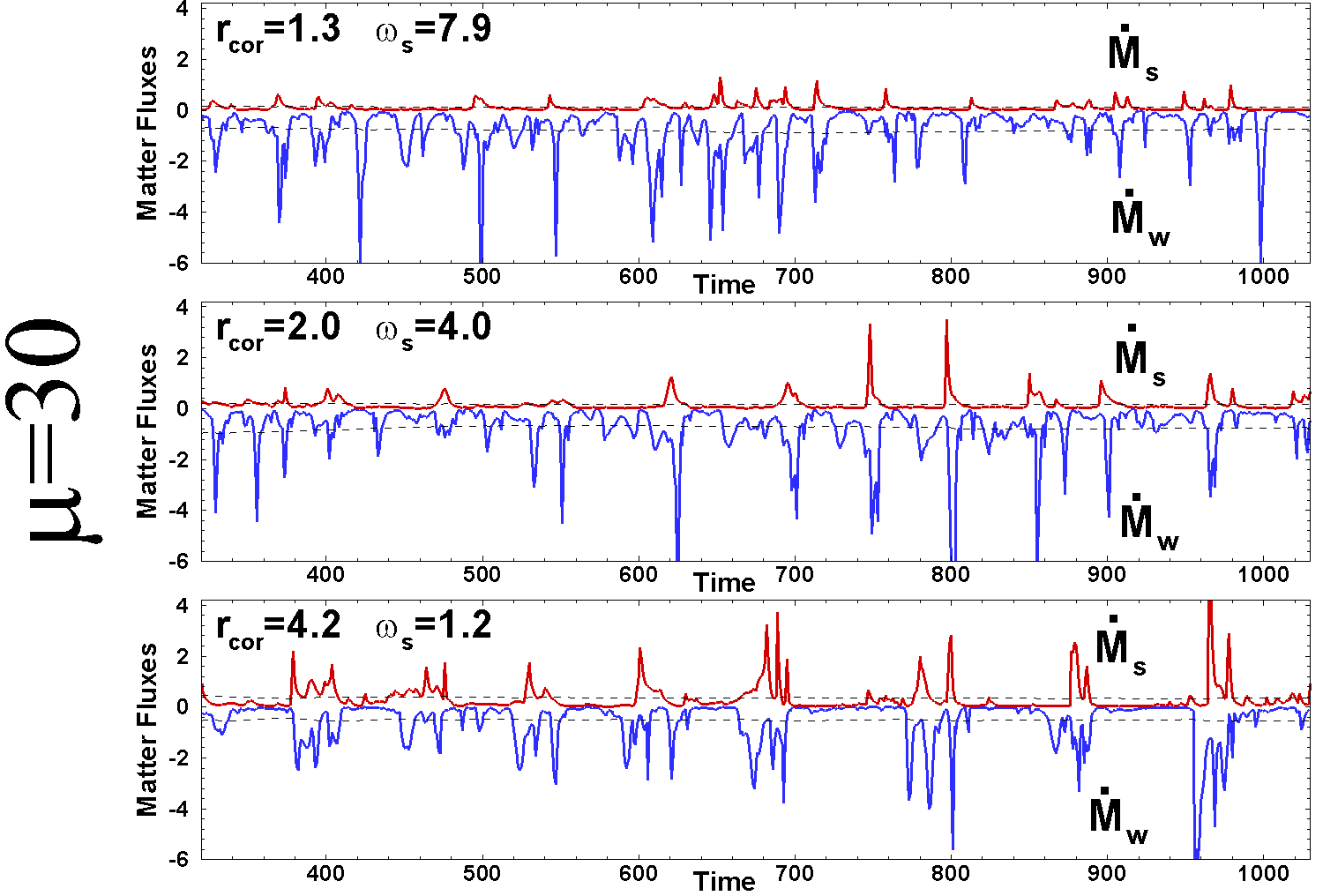}
\includegraphics[height=5.8cm,clip]{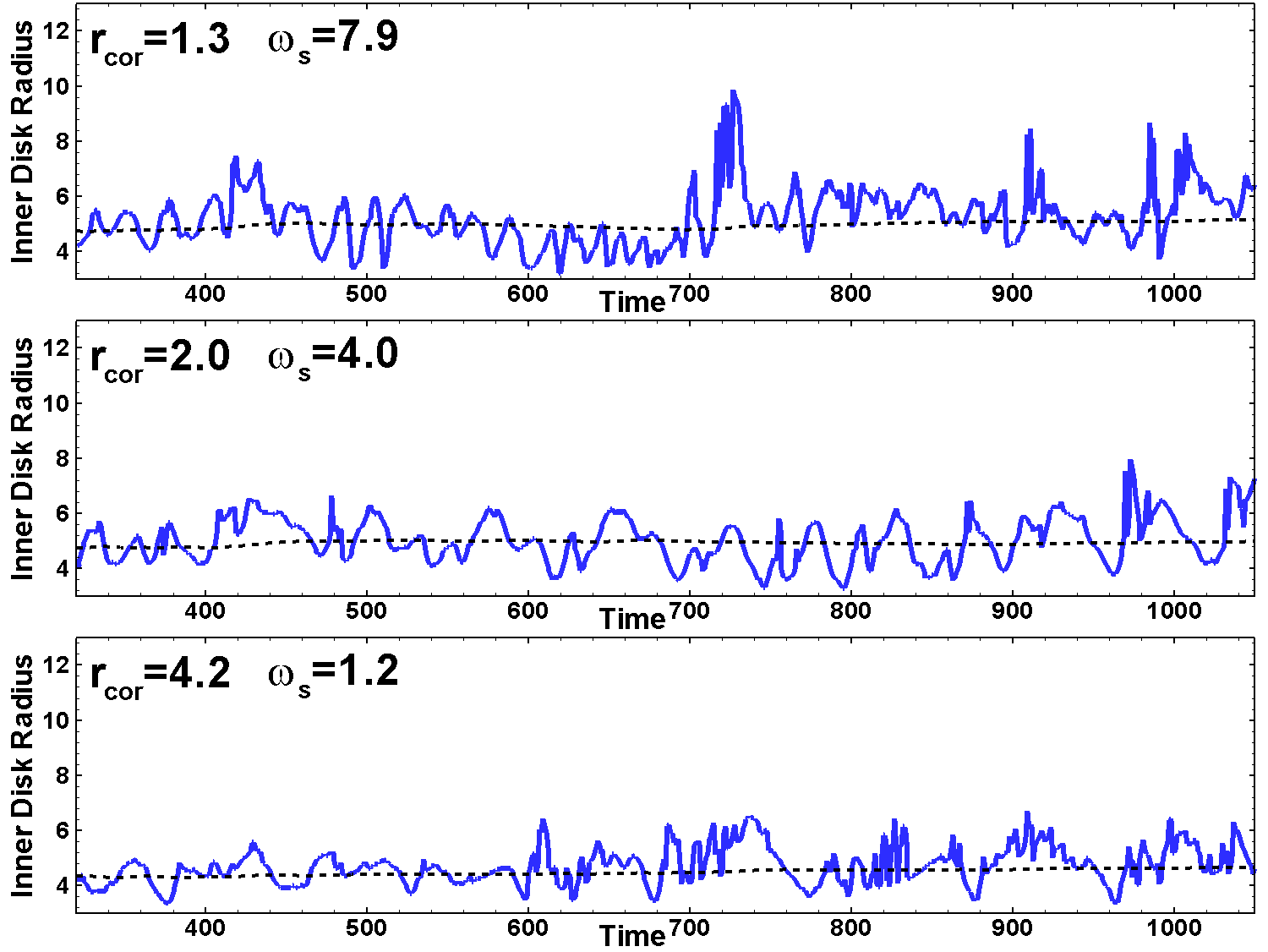}
\caption{Same as in Fig. \ref{app:fluxes-radii-d100} but for
$\mu=30$.} \label{app:fluxes-radii-d30}
\end{figure*}

\begin{figure*}[ht!]
\centering
\includegraphics[height=5.8cm,clip]{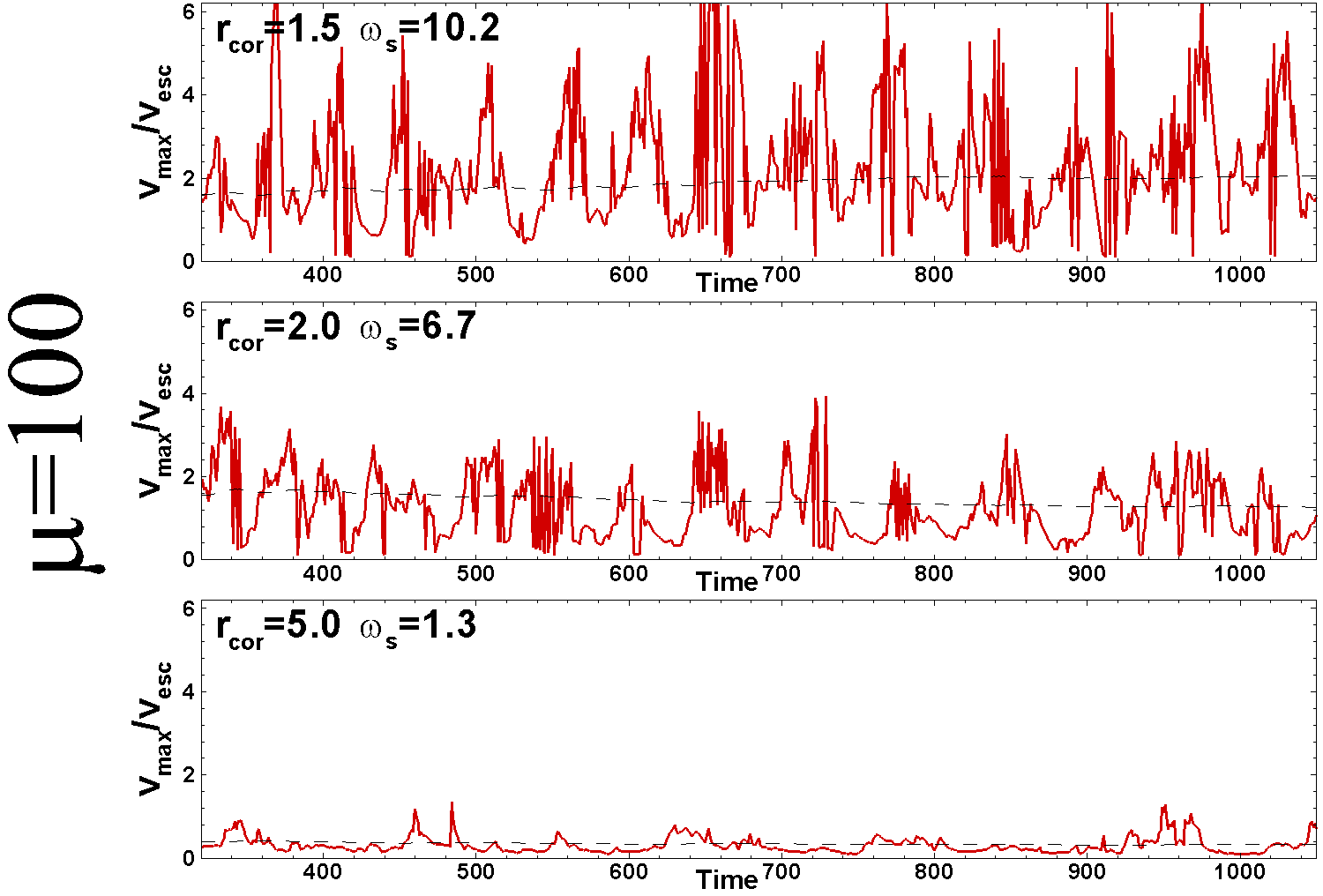}
\includegraphics[height=5.8cm,clip]{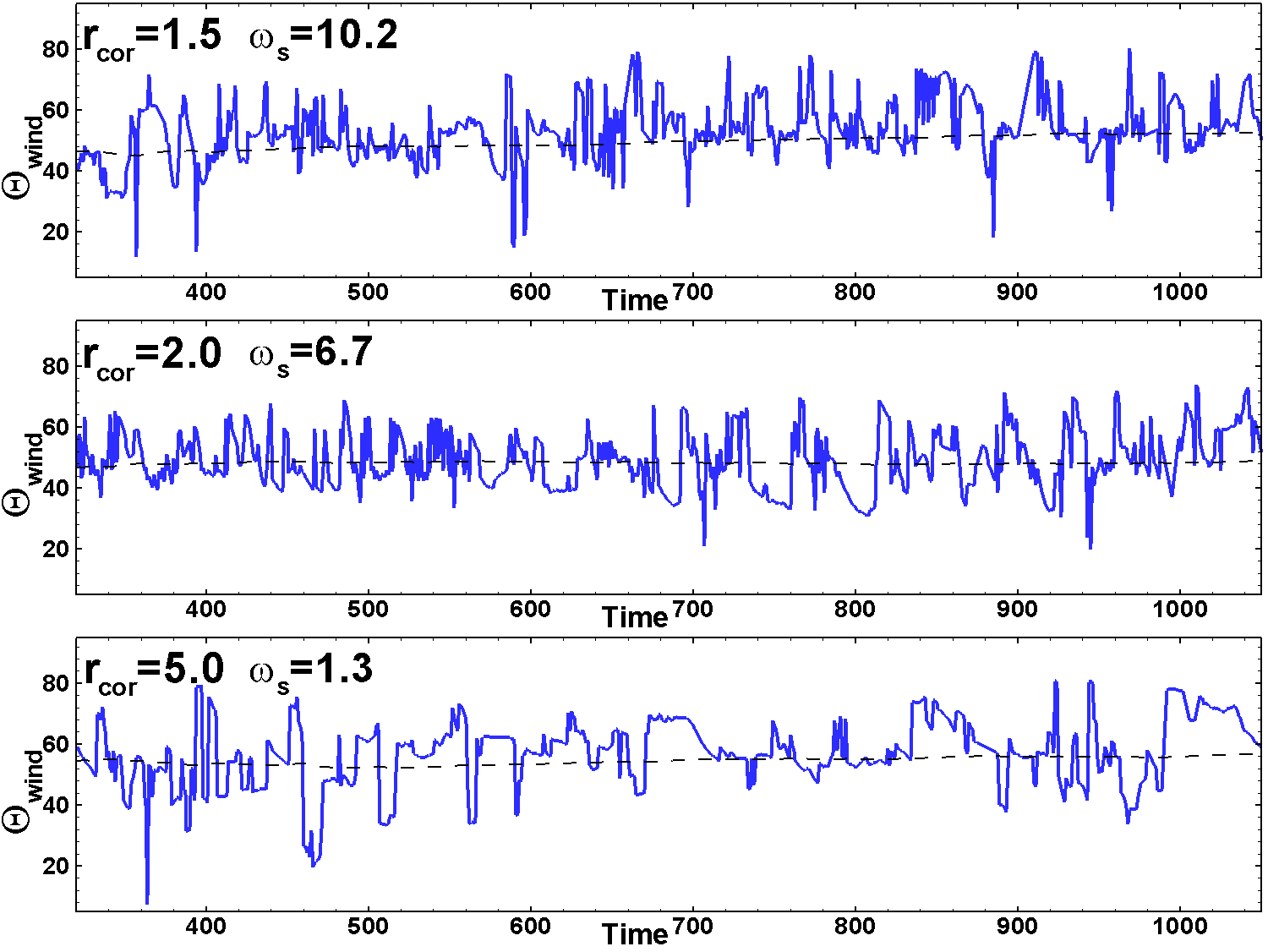}
\caption{\textit{Left Panels:} Variation of the maximum velocity
at the surface $S(r=10,z=\pm10)$ with time in case of $\mu=100$
and different corotation radii $r_{\rm cor}$. \textit{Right
Panels:} Variation of the opening angle of the wind, $\Theta_{\rm
wind}$, with time.} \label{app:vmax-theta-d100}
\end{figure*}

\begin{figure*}[ht!]
\centering
\includegraphics[height=5.8cm,clip]{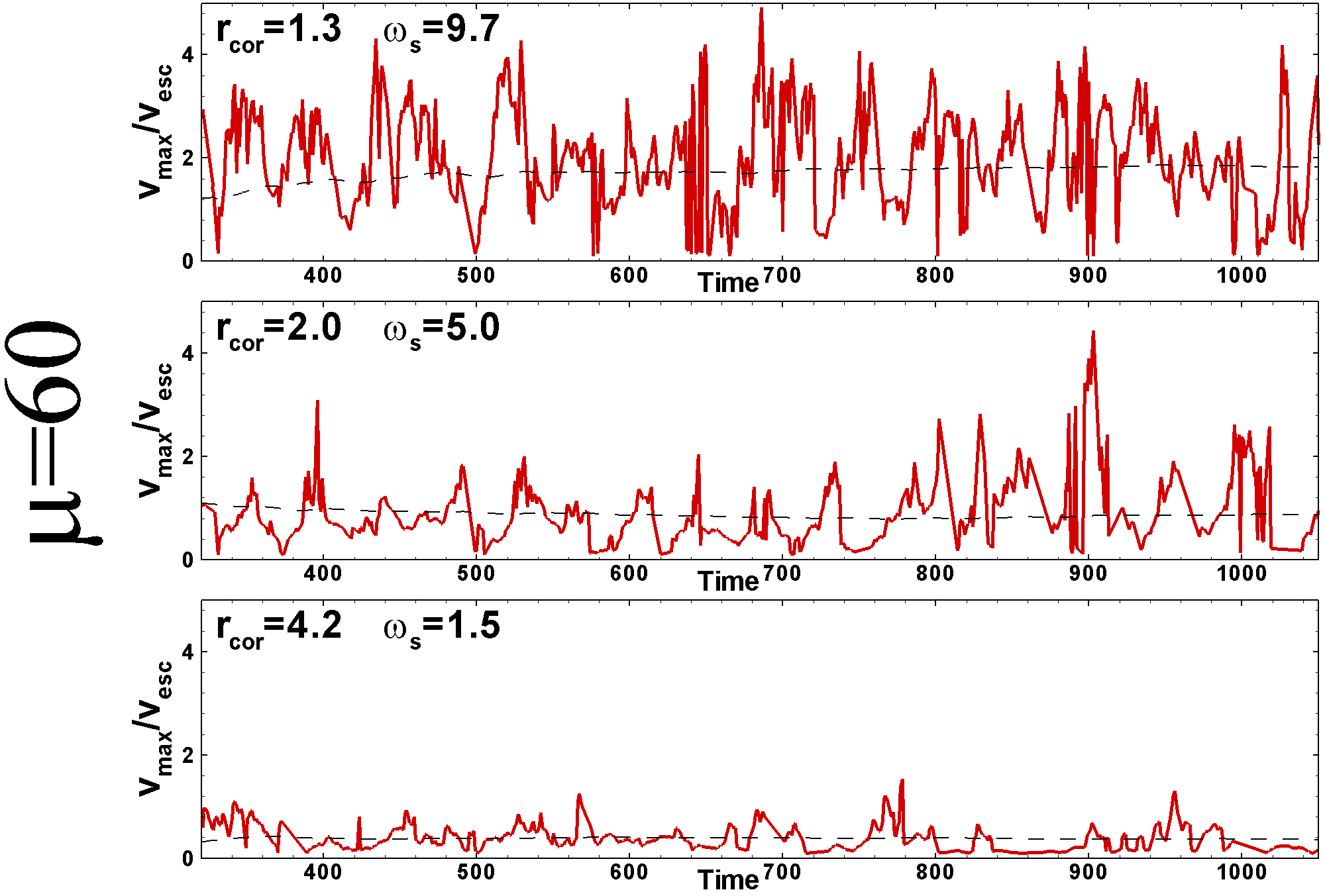}
\includegraphics[height=5.8cm,clip]{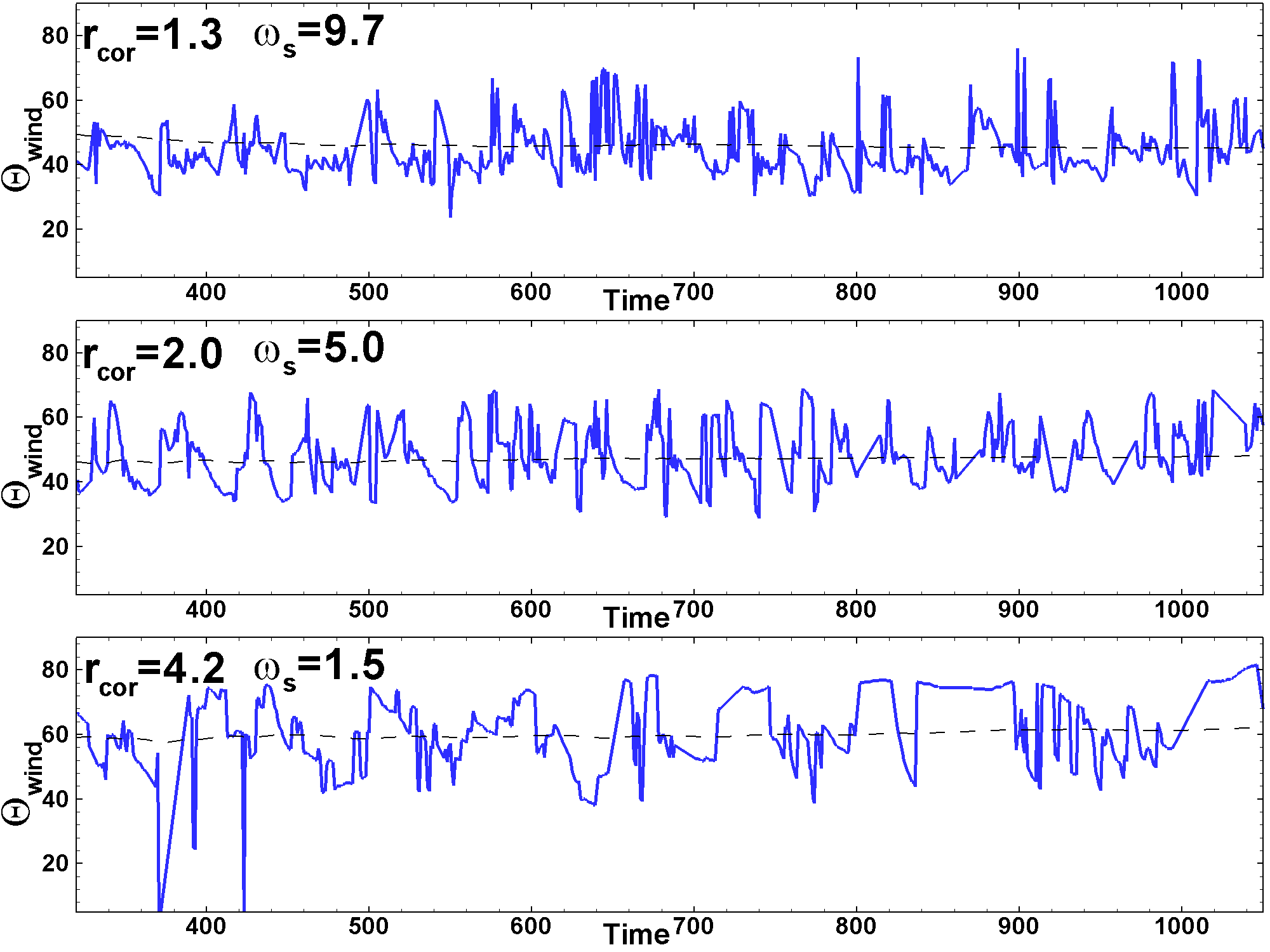}
\caption{Same as in Fig. \ref{app:vmax-theta-d100} but for
$\mu=60$. } \label{app:vmax-theta-d60}
\end{figure*}

\begin{figure*}[ht!]
\centering
\includegraphics[height=5.8cm,clip]{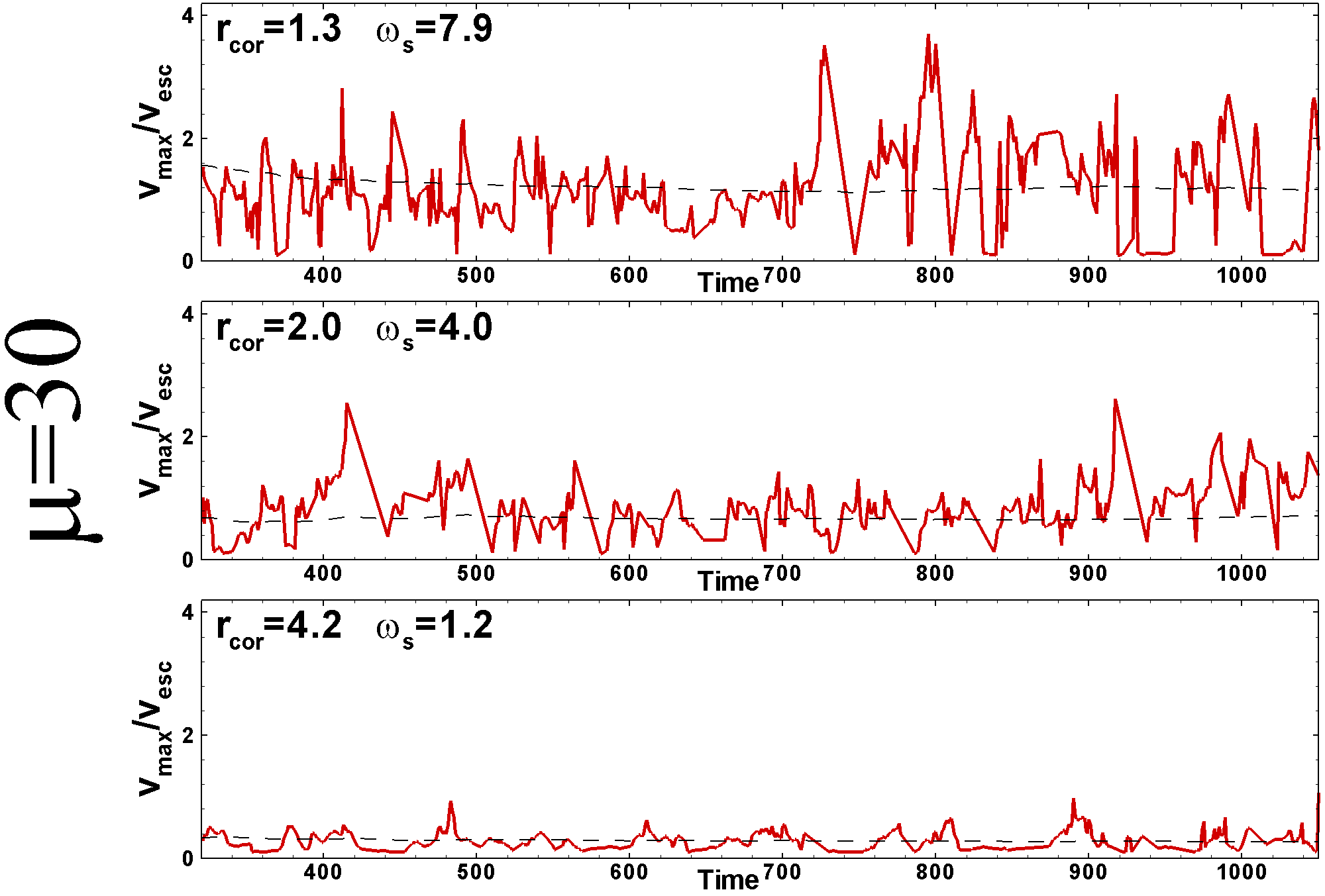}
\includegraphics[height=5.8cm,clip]{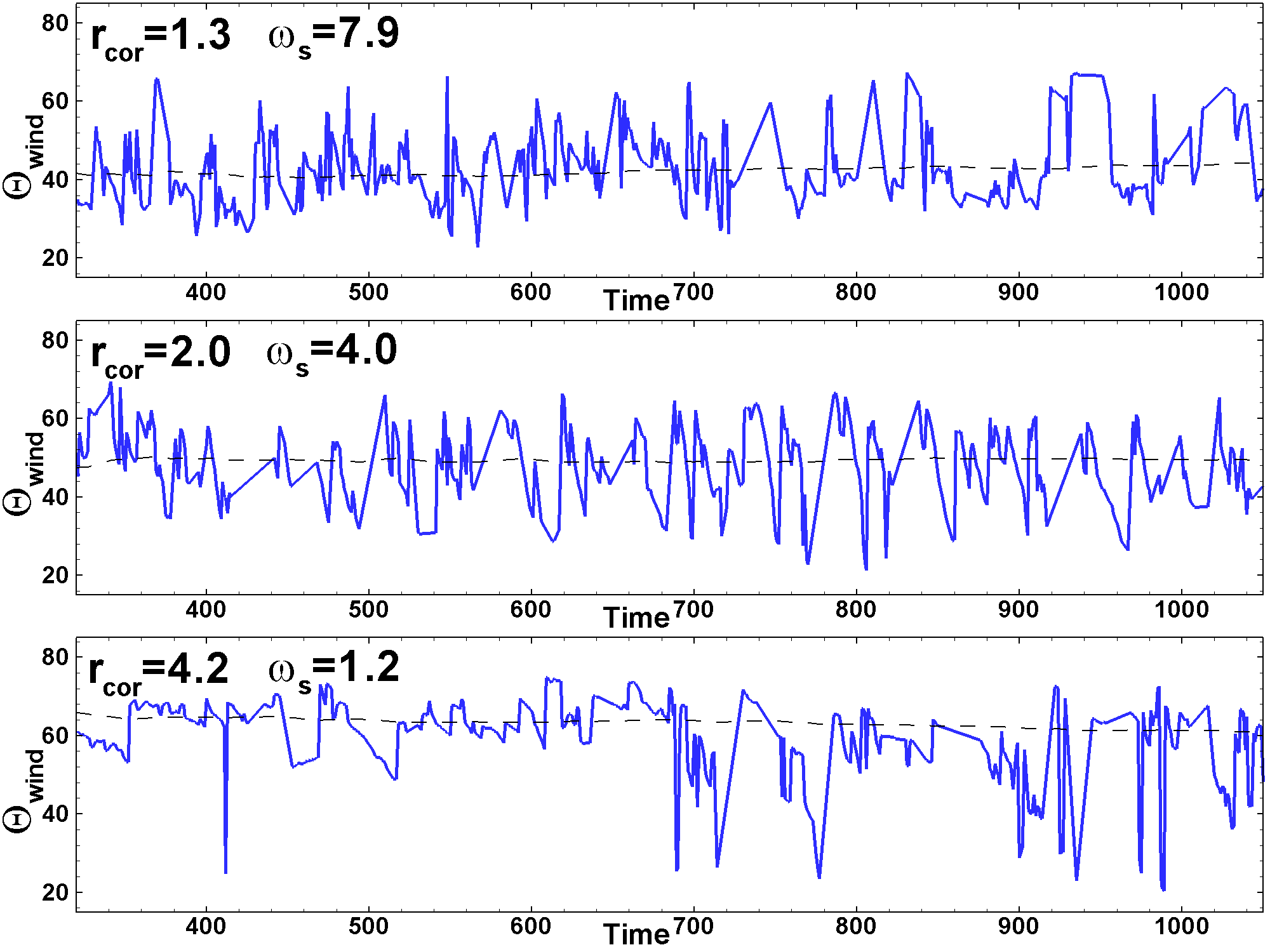}
\caption{Same as in Fig. \ref{app:vmax-theta-d100} but for
$\mu=30$.} \label{app:vmax-theta-d30}
\end{figure*}

\end{document}